\documentclass[sigconf,authorversion,screen]{acmart}
\AtBeginDocument{%
  }

\copyrightyear{2026}
\acmYear{2026}
\setcopyright{cc}
\setcctype{by}
\acmConference[CHI '26]{Proceedings of the 2026 CHI Conference on Human Factors in Computing Systems}{April 13--17, 2026}{Barcelona, Spain}
\acmBooktitle{Proceedings of the 2026 CHI Conference on Human Factors in Computing Systems (CHI '26), April 13--17, 2026, Barcelona, Spain}
\acmPrice{}
\acmDOI{10.1145/3772318.3791592}
\acmISBN{979-8-4007-2278-3/2026/04}

\usepackage[shortlabels]{enumitem}
\usepackage{subcaption}
\usepackage{multirow}
\usepackage{xcolor}
\newcommand{\highlight}[1]{\textcolor{black}{#1}}

\newif\ifrevdraft
\revdraftfalse

\newcommand{\maybeReviewer}[1]{%
  \if\relax\detokenize{#1}\relax
  \else\textcolor{magenta}{\textsuperscript{[#1]}}\fi
}

\ifrevdraft
  \newcommand{\newtext}[2][]{\maybeReviewer{#1}\textcolor{ForestGreen}{#2}}

  \newcommand{\deletedtext}[2][]{\maybeReviewer{#1}\textcolor{red}{\sout{#2}}}

  \newcommand{\replacedtext}[3][]{%
    \maybeReviewer{#1}
    \deletedtext[]{#2}\,\newtext[]{#3}
  }

  \newcommand{\emptext}[2][]{\maybeReviewer{#1}\textcolor{blue}{#2}}

  \newcommand{\reviewer}[1]{\textcolor{magenta}{\textsuperscript{[#1]}}}
\else
  \newcommand{\newtext}[2][]{#2}
  \newcommand{\deletedtext}[2][]{}
  \newcommand{\replacedtext}[3][]{#3}
  \newcommand{\emptext}[2][]{#2} 
  \newcommand{\reviewer}[1]{}
\fi

\begin{document}

\author{Anirban Mukhopadhyay}
\affiliation{%
 \institution{Virginia Tech}
 \city{Blacksburg}
 \state{Virginia}
 \country{USA}}

\author{Kevin Salubre}
\affiliation{%
 \institution{Honda Research Institute}
 \city{San Jose}
 \state{California}
 \country{USA}} 

\author{Hifza Javed}
\affiliation{%
 \institution{Honda Research Institute}
 \city{San Jose}
 \state{California}
 \country{USA}} 
 
\author{Shashank Mehrotra}
\affiliation{%
 \institution{Honda Research Institute}
 \city{San Jose}
 \state{California}
 \country{USA}} 

\author{Kumar Akash}
\affiliation{%
 \institution{Honda Research Institute}
 \city{San Jose}
 \state{California}
 \country{USA}} 

\title[Impact of GenAI Agent Roles on Collaborative Problem-Solving]{Exploring The Impact Of Proactive Generative AI Agent Roles In Time-Sensitive Collaborative Problem-Solving Tasks}

\renewcommand{\shortauthors}{Mukhopadhyay et al.}


\begin{abstract}
Collaborative problem-solving under time pressure is common but difficult, as teams must generate ideas quickly, coordinate actions, and track progress. Generative AI offers new opportunities to assist, but we know little about how proactive agents affect the dynamics of real-time, co-located teamwork. We studied two forms of proactive support in digital escape rooms: a facilitator agent that offered summaries and group structures, and a peer agent that proposed ideas and answered queries. In a within-subjects study with 24 participants, we compared group performance and processes across three conditions: no AI, peer, and facilitator. Results show that the peer agent occasionally enhanced problem-solving by offering timely hints and memory support; however, it also disrupted flow, increased workload, and created over-reliance. In comparison, the facilitator agent provided light scaffolding but had a limited impact on outcomes. We provide design considerations for proactive generative AI agents based on our findings.
\end{abstract}



\begin{teaserfigure}
  \centering
  \includegraphics[width=.8\textwidth]{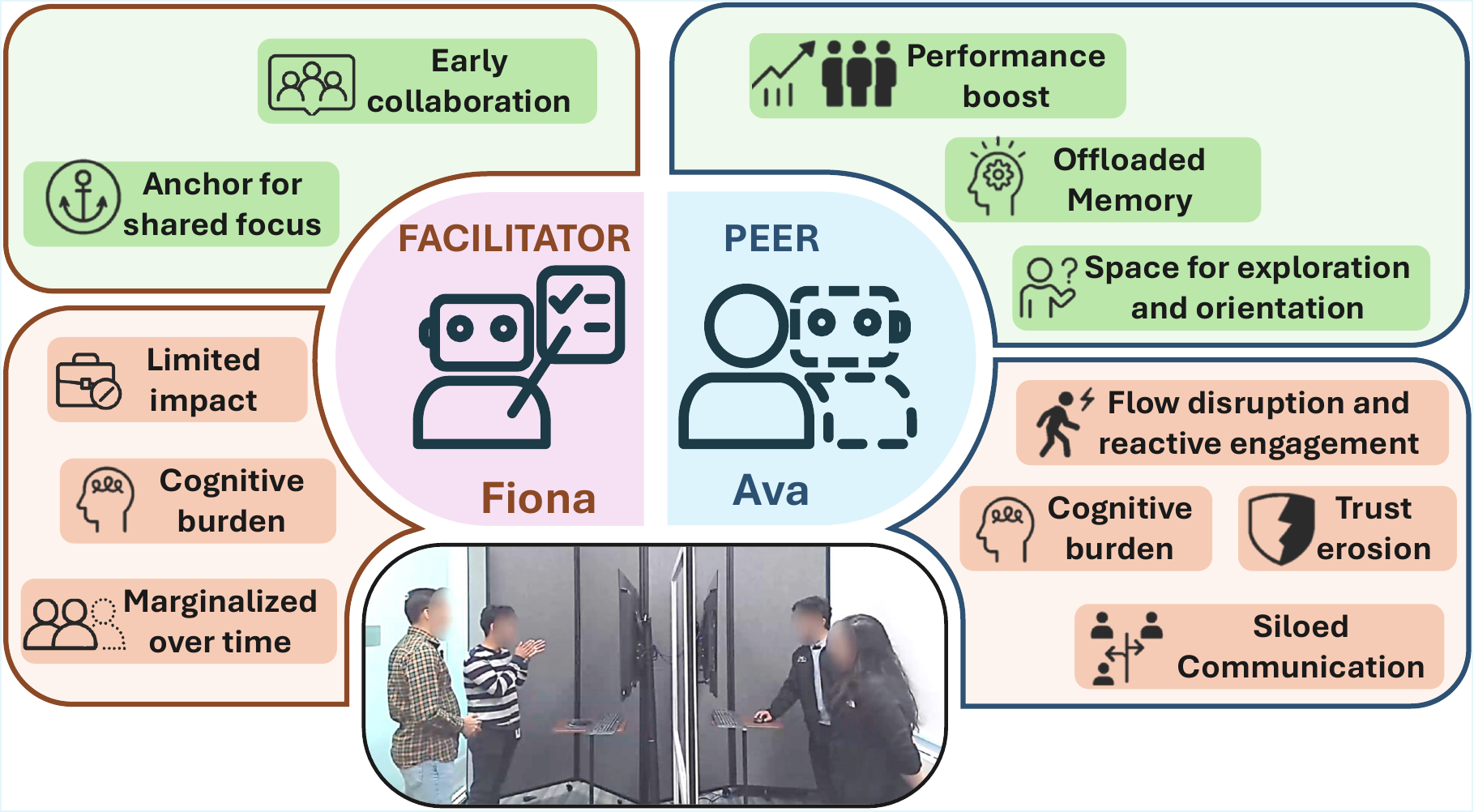}
    \caption{Perceived influence of proactive AI agents on group performance and processes. The top panel provides an overview of sub-themes from a reflexive thematic analysis comparing a Facilitator agent (“Fiona”) and a Peer agent (“Ava”). The benefits are highlighted in green, and the risks are highlighted in orange. The bottom image shows the study context, where a team is collaboratively solving a time-bounded digital escape-room task distributed across two screens.} 
    \label{fig:qualitative}
    \Description{The teaser image illustrates the perceived influence of proactive AI agents on collaborative, time-pressured puzzle-solving tasks. At the bottom, a group is shown working together on a digital escape-room style challenge, representing the study context. Above, two columns compare the roles of the Peer agent, Ava, and the Facilitator agent, Fiona. The green section highlights the advantages, with the peer agent offering offloaded memory and performance boosts, while the facilitator agent provides space for exploration, anchors shared focus, and supports early collaboration. The orange section captures the disadvantages, where the peer agent can add cognitive burden, disrupt flow, trigger reactive engagement, and erode trust, while the facilitator agent shows limited impact, becomes marginalized over time, and may encourage siloed communication. Together, the image conveys both the benefits and drawbacks of proactive AI roles in shaping group performance and teamwork processes.}
\end{teaserfigure}

\maketitle

\section{Introduction}


Complex and time-sensitive problem-solving in the real world is rarely an individual task. From disaster response teams coordinating under time pressure to cybersecurity analysts mitigating an attack, group collaboration is the norm. Co-located collaboration, where team members work together in the same location and at the same time, often enhances productivity in tasks that rely on frequent communication and joint efforts, such as brainstorming, knowledge building, and planning \cite{andrews_space_2010, olson_distance_2000, zhang_ladica_2025, hmelo-silver_facilitating_2008}. Prior research has shown that groups often outperform individuals because they can integrate diverse perspectives, share cognitive load, and adapt dynamically \cite{woolley_evidence_2010, wuchty_increasing_2007}. Team effectiveness depends not just on member ability but on the processes of coordination, communication, and shared attention \cite{ilgen_teams_2005}. However, despite decades of CSCW and HCI research, developing technology that supports group processes and outcomes remains difficult.

Two recent research trends offer new possibilities for augmenting teamwork: generative AI and proactive systems.
With advances in conversational and reasoning capabilities \cite{fui-hoon_nah_generative_2023, shi_understanding_2023, seymour_speculating_2024}, generative AI has become a powerful collaborator \cite{morris_design_2023, chiang_are_2023, van_den_broek_exploring_2024, shi_agent_2013, hwang2024whose}. Li et al. found that teams augmented with generative AI outperformed human-only groups across multiple performance measures \cite{li_generative_2024}. Groups can also help regulate appropriate reliance, as members have opportunities to challenge, refine, and set boundaries around AI contributions \cite{xu2023effectiveness, hagemann2017complex, johnson_exploring_2025}. At the same time, these systems are prone to persuasive but flawed outputs \cite{zercher2023ai}, leaving groups vulnerable to over-reliance, anchoring on AI suggestions, or deferring to them to avoid social conflict \cite{chiang_enhancing_2024}. 

In parallel, research has explored how AI and intelligent systems can act proactively \cite{samadi2024ai, houde_controlling_2025, johnson_exploring_2025}. Proactive behaviors have been shown to improve trust, situational awareness, and engagement across diverse contexts, from creativity to decision support \cite{zhang_investigating_2023, zhou2024understanding}. Hwang et al. found that people produced a higher number and quality of ideas when their AI partner behaved as an autonomous teammate \cite{hwang2021ideabot}. Teams often prefer AI that behaves as an active teammate, finding initiative-taking systems more supportive and peer-like. This line of work positions proactivity not just as a technical capability but as a design paradigm for how autonomous systems participate in collaboration.

Together, these trends converge in the growing CSCW and HCI framing of human-AI teams (HATs), where AI agents are understood not only as tools but as team members \cite{wang_adaptive_2025, chiang_enhancing_2024, ma_are_2024, kahr_trust_2024, calisto_assertiveness-based_2023}. O'Neill et al. define HATs as ``interdependence in activity and outcomes involving one or more humans and one or more autonomous agents, wherein each human and autonomous agent is recognized as a unique team member occupying a distinct role on the team, and in which the members strive to achieve a common goal as a collective.'' \cite{oneill_humanautonomy_2022}. This emphasizes the importance of the roles of AI agents; these could be task-focused, such as generating ideas or solving problems, or process-focused, such as facilitating communication or maintaining shared attention \cite{stevens_knowledge_1994, wang_adaptive_2025}. 
Generative AI has the potential to support both; its reasoning ability contributes task-specific guidance, while its conversational and summarization skills help facilitate coordination \cite{zhang_investigating_2023, morris_design_2023, hwang2024whose}. However, its limitations raise questions about how it should adopt such roles in practice.

While both generative AI and proactivity have shown promise independently, little is known about how they intersect in collaborative problem-solving. Should a proactive AI act as a \emph{Peer}, contributing ideas as an imperfect teammate, or as a \emph{Facilitator}, shaping group coordination and communication? How do these proactive roles influence not just team performance but also group processes? Addressing these questions is critical for designing AI that can effectively support collaboration in time-sensitive contexts.
To explore this further, we investigate the following research questions in the context of co-located collaboration in time-sensitive problem-solving tasks:
\begin{enumerate}[start=1,label={\bfseries RQ\arabic*:}]
    \item How do different AI agent roles (Peer vs. Facilitator) influence group performance? 
    \item How do these AI agent roles shape group processes such as workload, communication, and coordination?
\end{enumerate}

We explored these questions through the context of co-located teamwork in digital escape rooms. Escape rooms serve as a rich, high-pressure testbed for collaboration, requiring groups to solve interdependent puzzles under time constraints \cite{kleinman_untapped_2024, oszabo_anatomy_2022}. We then developed two functional technology probes \cite{hutchinson2003technology} in the form of generative AI agents: a facilitator, which provided discussion summaries and proposed group structures, and a peer, which contributed ideas as an imperfect teammate. The facilitator functionalities were based on previous work on such agents in group brainstorming and discussion contexts \cite{zhang_ladica_2025, earle-randell_how_2025, kuang_enhancing_2024}. For the peer agent, we ran a formative study with 6 participants to understand the preferences and develop the design features. 
Through a within-subjects mixed-methods study with 24 participants (6 groups with 4 participants each), we investigate how generative AI in the role of a peer versus a facilitator impacts both group processes and performance. 

Our findings highlight both the promise and pitfalls of proactive generative AI agents in group collaboration. The facilitator agent's summaries and collaboration cues initially captured attention, but its contributions were often sidelined when they became repetitive, lengthy, or poorly timed. The peer agent's thoughts and memory support sometimes boosted problem-solving, but they also increased workload, risked over-reliance, and disrupted flow. Importantly, teams followed varied trajectories: some relied on its thoughts before shifting to more reflective use; others' early enthusiasm led to dependence and later disillusionment; and in some, brief curiosity quickly turned into frustration and disengagement. Building on these insights, we provide design implications for tailoring facilitator and peer agents, and for supporting different trajectories of AI use in collaborative problem-solving.

In summary, we make the following three contributions:
\begin{enumerate}
    \item Functional technology probes of generative AI agents in facilitator and peer roles for group problem-solving tasks.
    \item An exploratory study of how proactive roles shape group processes and performance in time-sensitive collaboration.
    \item Design considerations for integrating proactive AI agents into group collaboration.
\end{enumerate}


\section{Background and Related Work}

In this section, we review co-located collaboration, with a focus on escape rooms as a research setting, which is used to study how groups coordinate and solve problems under time pressure (Section \ref{colocated}). Next, we examine generative AI in collaborative contexts, highlighting recent shifts from reactive to proactive support (Section \ref{proactive}). Finally, we consider research on the roles of AI agents, carving out the roles of facilitator and peer (Section \ref{roles}). By situating our work at this intersection, we show how embedding role-based proactive AI agents in escape-room settings offer new insights into both performance and group dynamics.

\subsection{Co-located Collaborative Problem-Solving}
\label{colocated}

Co-located synchronous collaboration describes situations where people work together in the same physical space and at the same time \cite{ellis_groupware_1991}. This mode of collaboration allows group members to share artifacts in a common workspace and benefit from subtle but important interactional cues \cite{olson_distance_2000}. For example, physical proximity enables coworkers to pick up on gestures, facial expressions, body posture, or shifts in attention that are often missed in remote settings. It also facilitates rapid feedback and turn-taking, helping groups quickly address misunderstandings or refine ideas as they arise. In addition, participants can easily co-reference objects in the shared environment, using gaze or pointing gestures to disambiguate expressions like ``this'' or ``that.'' At the same time, these benefits come with challenges. Social evaluation pressures can make people hesitant to share ideas, and production blocking may occur when turn-taking delays individuals from voicing contributions before they forget or overthink them \cite{nijstad_how_2006}. 

Escape rooms in particular offer a unique environment for studying co-located problem-solving, involving tightly-coupled interactions \cite{stuckel_effects_2008} and high synchronicity \cite{linebarger_benefits_2005}. They combine the control of a laboratory study with the ecological validity of a naturalistic group task \cite{cohen_using_2020}. Groups must search for clues, solve puzzles, and coordinate under strict time pressure, creating a setting that naturally elicits intense interaction. They have been used for training and education purposes \cite{fotaris2019escape}, and as platforms to study human behavior \cite{harteveld_teamwork_2019}. Escape rooms are well-suited for examining how groups work with AI under demanding conditions, also found in real-world settings such as disaster response, emergency medical teams, cybersecurity incident management, and air traffic control.  In this paper, we extend existing literature by examining collaboration in a novel escape room setting, where groups interact with generative AI in a fast-paced, synchronous environment that amplifies both the opportunities and the challenges of teamwork with AI.

\subsection{Generative AI Agents and Proactivity in Collaborative Contexts}
\label{proactive}

Generative AI has already demonstrated potential to enhance creativity and problem-solving across diverse stages, such as ideation, prototyping, deliberation, and decision-making \cite{shaer_ai-augmented_2024, cui_ai-enhanced_2024, he_ai_2024, chiang_enhancing_2024, hubert_current_2024}. Most of this research, however, has focused on one-to-one interaction between a single user and a generative AI system. HCI studies have shown that individual improvements in AI performance do not guarantee better team outcomes, highlighting the importance of designing for team-level dynamics \cite{schelble_lets_2022}. Only recently have studies begun to explore generative AI in multi-user, collaborative contexts. For example, researchers have investigated how groups use generative AI while co-designing \cite{deng_crossgai_2024, koch_imagesense_2020}, conducting qualitative analysis \cite{gao_coaicoder_2024}, or engaging in cooperative play \cite{sidji_human-ai_2024}. Others have examined how multiple participants jointly interact with tools like ChatGPT during group ideation \cite{he_ai_2024, shin_integrating_2023, shaer_ai-augmented_2024}, planning \cite{zhang_ladica_2025}, cybersecurity vulnerability assessments \cite{mukhopadhyay_osint_2025}, or creative activities such as music composition \cite{suh_ai_2021}.

Early explorations of generative AI in teamwork have often positioned these systems as enhanced ``AI-infused supertools'' rather than as genuine collaborators \cite{shneiderman2022human}. This framing emphasizes the centrality of human expertise, where AI's role is primarily to provide support upon request. However, findings from several studies suggest that users desire AI agents that can act more like teammates, with greater initiative and autonomy \cite{he_ai_2024, koch_imagesense_2020}. For example, when participants perceived an AI system as an autonomous partner, they generated more ideas and rated them as higher in quality \cite{hwang2021ideabot}. Other work has shown that groups often prefer AI agents with stronger decision-making roles and peer-like participation, especially in creative or cooperative tasks \cite{zhou2024understanding, zhang_investigating_2023, salikutluk_evaluation_2024}. Proactive communication from AI has also been found to enhance trust and situational awareness, underscoring the potential benefits of treating generative AI agents as active team members \cite{zhang_investigating_2023}.

Despite this promise, most current applications of generative AI remain fundamentally reactive \cite{han_when_2024, van_den_broek_exploring_2024, zamfirescu-pereira_why_2023}. They depend on users to issue prompts or instructions before producing output. While this model lowers barriers to use, it also creates friction in collaborative contexts. Non-experts may struggle to articulate their intentions effectively, and even skilled users must divide attention between crafting prompts and participating in group discussions \cite{han_when_2024}. This additional cognitive load can disrupt conversational flow, reduce efficiency, and diminish engagement in the shared workspace \cite{verheijden_collaborative_2023, van_den_broek_exploring_2024}. These limitations highlight the need for agents that can contribute more autonomously---responding to unfolding interactions without constant human direction.

Generative AI itself provides a foundation for building such proactive collaborators. Large language models (LLMs) in particular have demonstrated the capacity to simulate aspects of human cognition and social interaction, enabling them to participate more naturally in group exchanges \cite{park_generative_2023, park_social_2022, morris_design_2023}. Examples include LLM agents designed to play a devil's advocate role in deliberation \cite{chiang_enhancing_2024}, systems that surface relevant materials on shared displays to support discussion \cite{zhang_ladica_2025, imamura_serendipity_2024}, and AI teammates integrated into digital chat platforms \cite{samadi2024ai}. Evaluations of these systems show promise but also raise concerns. Collaborative agents may unintentionally bias groups, disrupt social dynamics, or reinforce existing power imbalances \cite{seymour_speculating_2024}. Findings show that people often treat AI as a secondary partner in group discussions, particularly when the system lacks the capacity to participate fully in social dynamics \cite{flathmann_empirically_2024}. Other work has highlighted a tendency for groups to over-rely on AI recommendations compared to individuals, raising concerns about dependence and reduced critical engagement \cite{chiang_are_2023, zercher2023ai}. Early explorations of co-located settings, including prototypes of generative AI teammates in mixed reality \cite{johnson_exploring_2025} or shared displays \cite{zhang_ladica_2025}, reveal both enthusiasm for their potential value and skepticism about their ability to navigate complex social interactions. Kraus et al. described proactivity as a ``double-edged sword,'' valuable when it aligns with user needs but problematic when it intrudes or misfires \cite{kraus_improving_2023}.

However, much of what we know about proactive AI in collaboration comes from speculative design or wizard-of-oz studies that do not fully capture the complexities of functional deployment. In this paper, we address this gap by developing and testing functional probes of proactive generative AI agents. We empirically examine how teams respond to and collaborate with proactive generative AI agents in complex, time-sensitive problem-solving scenarios.

\subsection{Roles of AI Agents in Collaborative Contexts}
\label{roles}

Roles are a critical lens for understanding how generative AI agents fit into collaborative contexts \cite{siemon_elaborating_2022, shi_agent_2013, zheng_roles_2019, wu_agent_2019}. In human teamwork, roles provide clarity, distribute responsibilities, and balance task-related and social demands. In the same way, when agents are introduced into groups, roles shape not only what it contributes but also how human members perceive and interact with them. Early studies have shown that people already apply social expectations to computers, treating them as legitimate teammates when interdependence exists between their actions \cite{stevens_knowledge_1994, robbins_title_nodate}. This highlights the need to carefully design and study the roles AI agents assume in group work.

Prior work has examined perceptions of AI in different social and functional roles. Kim et al. explored how people evaluated social versus functional AI, concluding that users tended to prefer functional systems, with usefulness acting as a key mediating factor \cite{kim_ai_2021}. However, their study was based on video demonstrations, leaving open questions about how people might respond when collaborating with functional and social agents in real tasks. Houde et al. argued that role specification could give users greater control and predictability in group brainstorming with AI, proposing roles such as responsive contributor, active reviewer, or conversation starter \cite{houde_controlling_2025}. Liu et al. studied peer roles in children's collaborative learning and showed that framing an AI as a teammate or moderator changes conversational dynamics \cite{liu_peergpt_2024}.

Bittner et al. provide a taxonomy of conversational assistant roles in collaborative work, grouping them into three categories: facilitator, peer, and expert \cite{bittner_where_2019}. Facilitators guide groups through structured processes, often using proactive or directive behavior grounded in scripts or models of the collaboration. They are common in contexts such as teaching, tutoring, or structured group interaction, where maintaining process flow is essential \cite{tegos2015promoting, dyke2013enhancing}. Peers, in contrast, blend into the group as equals, contributing socio-emotionally while offering knowledge that is ``enough but not too much.'' A well-designed peer agent avoids dominating discussions, encourages human contributions, and stays approachable \cite{dohsaka2009effects, porcheron2017animals}. Finally, expert roles emphasize domain-specific skill but remain largely reactive, providing help when prompted \cite{xu2017new}. Wang et al. further distinguish task capabilities (e.g., executing, planning, evaluating) from social capabilities (e.g., coordinating, resolving conflicts, building shared understanding), clarifying what proactive agents can target \cite{wang_adaptive_2025}.

Understanding the role of AI agents in collaborative problem solving also requires attention to group processes, not only outcomes \cite{oneill_humanautonomy_2022, zhang_investigating_2023}. Group processes are the interdependent acts that transform individual inputs into collective results \cite{ilgen_teams_2005}. Communication, coordination, and workload distribution can be less visible than final performance, yet they are central to explaining team effectiveness, especially in time-sensitive contexts such as escape rooms. \newtext[R2]{While prior work has predominantly examined human-AI collaboration in human-AI teams or human-human-AI teams \cite{mcneese2021my, schelble_lets_2022, flathmann_empirically_2024, munyaka2023decision}, there is limited empirical understanding of how AI agents shape group processes in larger teams. We situate our study in four-member teams to capture the interaction dynamics that characterize more realistic collaborative settings.}

In this work, we draw on these insights to focus on two roles, facilitator and peer, that occupy distinct parts of the design space. The facilitator allows us to examine how an agent can structure and guide collaborative problem solving. The peer enables us to explore what happens when the agent positions itself as an equal sparring partner. Unlike prior studies that place AI outside the group, assume perfect knowledge, or consider them as tools, we embed agents directly within co-located activity. By instantiating these roles as functional probes, we study how design features influence both group outcomes and the processes that mediate them.

\section{Task Environment}

Our study was designed around escape-room–style puzzles that serve as co-located collaborative problem-solving tasks \cite{kleinman_untapped_2024, cohen_using_2020}. 
In designing the environment, we had three motivations: (1) tasks needed to require active communication and coordination among multiple group members, not passive or individual effort; (2) each puzzle had to accommodate four co-located participants; and (3) puzzles had to be adequately difficult to be engaging for at least 20 minutes for the group while remaining unsolvable by state-of-the-art multimodal generative AI models when viewed in isolation. \newtext[R2]{We selected four-person groups because time-sensitive problem solving with bigger team sizes requires teams to divide work, coordinate across sub-tasks, and maintain shared situational awareness. This setting allows us to observe how human-AI collaboration shifts as these group processes take shape. 
}

\paragraph{Puzzle Design}

We created three puzzles (Puzzles 1--3), each consisting of three interconnected sub-puzzles. Sub-puzzles were distributed across two screens, such that solving them required integrating information from both displays. While individuals could walk between screens, this was slower and more effortful than communicating with teammates stationed at the other screen, thus encouraging interdependence.

There were nine unique sub-puzzles across all conditions. This avoided learning effects while maintaining a consistent ``two-screen'' theme. Sub-puzzles could be solved in parallel and each relied on exclusive puzzle elements, opening up possibilities for division of labor. We aimed for puzzles to last 15–25 minutes, though exact difficulty varied because solutions often depended on sudden ``Aha!'' moments when participants connected multiple pieces of information. Sub-puzzle designs were inspired by cooperative online puzzle games such as Acorn Cottage \cite{noauthor_acorn_nodate} and Alone Together \cite{team_enchambered_nodate}.

We tested the puzzles against state-of-the-art multimodal and reasoning generative AI models (e.g., GPT-4.1, o3, o4) to evaluate model performance. While these models generated partial ideas by linking elements across screens, they consistently failed to produce full solutions. This reinforced our goal of designing tasks that were challenging for AI alone but could benefit from AI as a teammate, sharing partial reasoning with human collaborators.



\paragraph{Implementation}

The puzzles were implemented using HTML, CSS, and JavaScript, and hosted on a Django server. Each puzzle's two screens corresponded to separate webpages. The countdown timer was synchronized through a backend SQLite database that stored the puzzle's status (START/STOP). Puzzle elements included both static images and interactive components. Here's an example of a sub-puzzle: in Puzzle 1 (Figures \ref{fig:facilitator_shot} and \ref{fig:peer_shot}), Screen 2 contained green and yellow buttons. When pressed in the order shown by the Color Strip on Screen 1, these buttons triggered an ``@'' symbol to appear in the slot above. This symbol then became a clue for the next sub-puzzle, which required linking the Symbol and Buttons element on Screen 1 with the Paper element on Screen 2.

\begin{figure*}
    \centering
    \includegraphics[width=.8\linewidth]{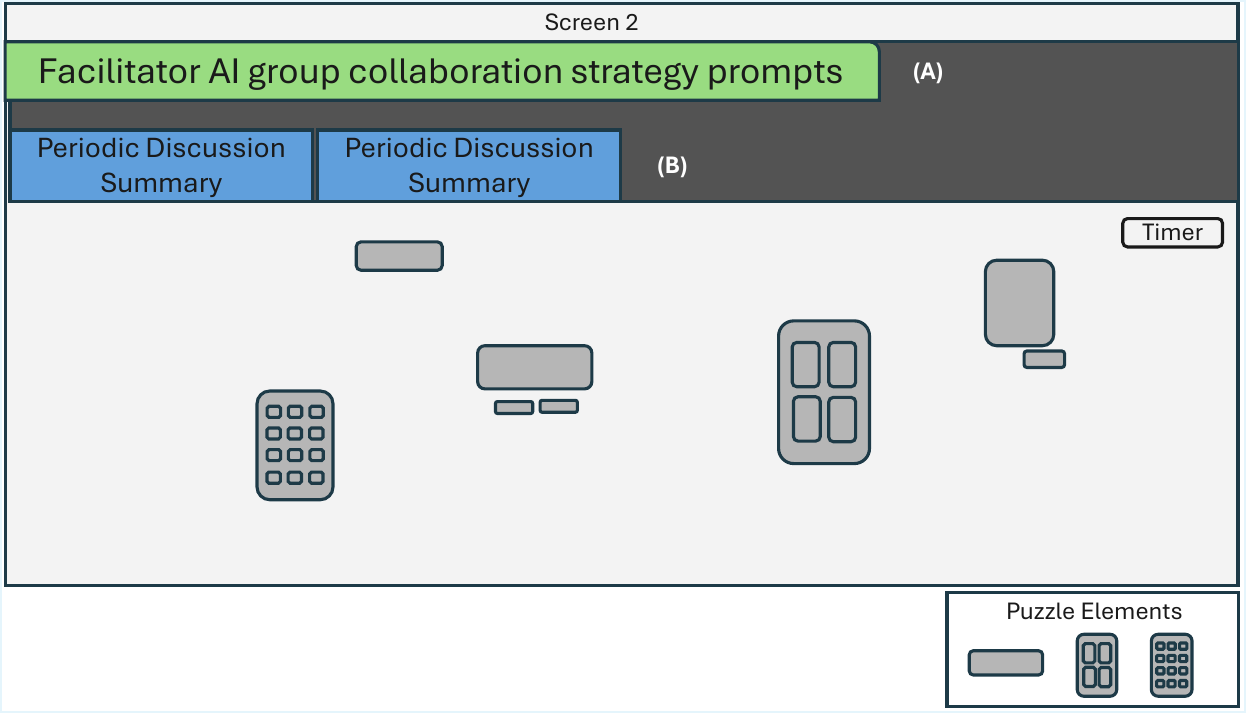}
    \caption{Diagram of Screen 2, Puzzle 1 with the facilitator agent condition. The gray boxes represent the puzzle elements. There are two main features of the facilitator agent (black box at the top): (A) The green text field shows where Fiona suggested structured collaboration strategies like the 1-2-4-All liberating structure \cite{mccandless_liberating_2020}, provided time reminders, and asked groups to divide up unsolved parts of the puzzle; (B) The blue text field displays Fiona's summary of ideas discussed by group, presented every three minutes.}
    \label{fig:facilitator}
    \Description{A schematic interface showing the Facilitator AI system. At the top, a green banner labeled “Facilitator AI group collaboration strategy prompts” spans the width of the screen. Below it is a dark gray bar with two blue buttons labeled “Periodic Discussion Summary.” The main area displays abstract representations of puzzle pieces spread across a white workspace. A small “Timer” label appears in the top right. The bottom portion shows a preview strip with miniature puzzle representations.}
    
\end{figure*}

\begin{figure*}
    \centering
    \includegraphics[width=.8\linewidth]{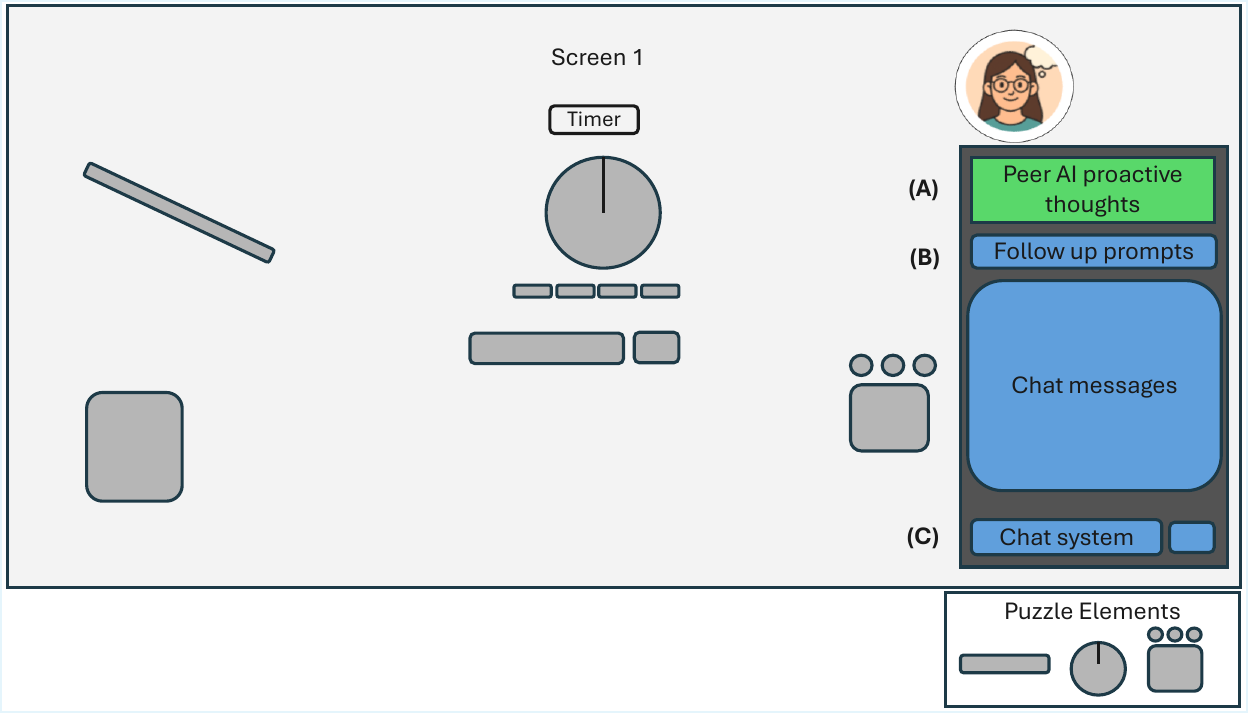}
    \caption{Diagram of Screen 1, Puzzle 1, with the peer agent condition. The gray boxes represent the puzzle elements. There are three main features of the peer agent (black box on the right): (A) Ava proactively shared brainstorming thoughts every 3 minutes (displayed in the green text field), based on puzzle screenshots and contextualized by group conversations; (B) The blue text field indicates that groups could follow up by asking Ava to explain or vary its ideas; and (C) Ava was available as a chat-based partner on each puzzle screen, responding to user queries.}
    \label{fig:peer}
    \Description{A schematic interface showing the Peer AI system. On the right side, a vertical panel includes a character avatar, a green box labeled “Peer AI proactive thoughts,” a blue button labeled “Follow up prompts,” and a large blue chat window titled “Chat messages,” with a chat input bar labeled “Chat system” at the bottom. The main white workspace displays abstract puzzle pieces, while a circular timer appears near the top center.}
\end{figure*}

\section{Design of Generative AI Agents}
\label{sec:design_agent}

We approached the agents as functional technology probes \cite{hutchinson2003technology}, designed to explore qualities of proactivity and interdependence that are central to human-AI teams \cite{oneill_humanautonomy_2022}. Given the vast design space of proactive agents and the rapid evolution of generative AI, our probes are not intended as final or optimal solutions. Instead, they serve as design instances that help us investigate how different role configurations shape group dynamics and problem-solving processes.

To address our research questions, we set out three design goals for the agents: (1) they should work with multiple participants and take on active roles within the group, rather than acting from the outside; (2) they should not rely on perfect or pre-defined solutions, but collaborate with humans to construct answers in real time; and (3) they should act proactively, stepping in without waiting for explicit prompts. These goals align with what's currently possible with generative AI, including summarizing complex dialogue, contextualizing responses within group discussions, and supporting image-based puzzle solving. They also highlight the limitations of generative AI, including its inaccuracy and lack of social and cultural awareness.

Based on these goals, we designed two probes: \textit{Fiona}, a process-focused facilitator; and \textit{Ava}, a task-focused peer. Together, they represent two distinct regions of the design space for collaborative AI agents. \newtext[R1]{Our goal in designing the two agents was not to isolate each design feature but to instantiate representative probes of the facilitator and peer roles. The specific features we implemented focused process-focused and task-focused mechanisms through which these roles typically manifest. We used prior CSCW/HCI literature \cite{earle-randell_how_2025, kuang_enhancing_2024, zhang_ladica_2025, hubert_current_2024, suh_ai_2021, houde_controlling_2025} and a formative study to identify features which are characteristic of each role. We refined them through a pilot to ensure that the two agents remained usable and meaningfully differentiated.} 

\newtext[2AC]{We avoided adding further variations (e.g., agent gender \cite{duan2025gender} or communication styles \cite{zhang_investigating_2023}), as these would introduce additional interpretive ambiguity in comparing the two agent conditions. While real-world AI teammates may blend multiple roles, testing facilitator and peer behaviors separately allowed us to first understand their distinct impact before exploring adaptive or hybrid role configurations. By examining facilitator-like and peer-like support separately, we wanted to explore how each type of intervention shaped group processes and outcomes.}


The pilot study was conducted with four participants outside of the research team. Across three sessions, the group solved two-screened puzzles under different conditions: without AI support, with a facilitator agent duplicated across both screens, and with a peer agent duplicated across both screens. Each session lasted 20 minutes and was followed by a focus group interview to gather feedback on puzzle difficulty, room setup, and experiences with the agents, especially around usability. We iterated on the designs based on this feedback. In the following sections, we describe the agents' features and implementation details. 

\subsection{Fiona: The Facilitator Agent}

Meta-cognition, or ``cognition about cognition,'' enables groups to reflect on and regulate how they process information, approach problems, and coordinate efforts \cite{thompson_metacognition_2012}. Prior work shows that groups with strong metacognitive skills are better able to monitor progress, adapt strategies, and leverage diverse perspectives, resulting in improved outcomes \cite{nonose_effect_2012}. Two processes in particular can benefit groups: task monitoring, where groups regularly evaluate progress toward goals and adjust their approach \cite{kim_effect_2018}, and metacognitive prompting, where questions or reminders encourage reflection on decision-making and collaboration \cite{wiltshire_training_2014}. Previous HCI research has supported these functions in contexts of group discussion and brainstorming \cite{earle-randell_how_2025, kuang_enhancing_2024, zhang_ladica_2025}.

Building on these insights, we designed the facilitator agent (Fiona) to scaffold groups' metacognitive processes during co-located problem-solving. Rather than replacing human judgment, we designed the facilitator agent to scaffold the group's own reflective capabilities, allowing them to remain aligned and adaptive to emerging ideas. The \newtext[2AC]{schematic} user interface of the facilitator is presented in Figure~\ref{fig:facilitator} along with the features. \newtext[]{The screenshot of the facilitator embedded within Puzzle 1 screen is shown in Figure~\ref{fig:facilitator_shot}}. Fiona was implemented with two core features:
\begin{enumerate}
    \item \emph{Group collaboration strategy prompts, time reminders, and coordination support: }
The facilitator provided periodic reminders to regroup and prompted groups to divide up tasks if necessary. For example, Fiona started the session by suggesting the 1-2-4-All liberating structure, a well-established facilitation technique where individuals first reflect independently, then pair up, and finally synthesize as a group \cite{mccandless_liberating_2020}. This approach is relevant for small groups working across two shared displays, as it balances individual contributions with collective integration. Fiona also sent time reminders along with encouragement to keep communicating.
    \item \emph{Periodic discussion summaries: }
Every few minutes, Fiona generated concise summaries of the last segment of discussion and displayed them as idea cards. These summaries were designed to help groups step back and evaluate what had been covered, reinforcing task monitoring and ensuring shared awareness of progress.
\end{enumerate}

\subsubsection{Iteration based on Pilot}

We found that frequent interruptions and long speech output from the facilitator disrupted the puzzle-solving process. The initial implementation used screen overlays to display up to four summaries grouped by puzzle elements, which blocked puzzle elements. It also relied on rigid countdowns to suggest collaboration structures that constrained the group's natural pacing.

In response, we removed strict timing enforcement to allow groups more flexible pacing during initial stages, eliminated screen overlays so task elements remained visible, and added a short demo/tutorial to set expectations. Fiona's language was made more concise, and the frequency of summaries was reduced to every three minutes. Instead of reading out summaries, Fiona briefly mentioned that new ideas had been added.

\subsubsection{Implementation}

\begin{figure*}
    \centering
    \includegraphics[width=1\linewidth]{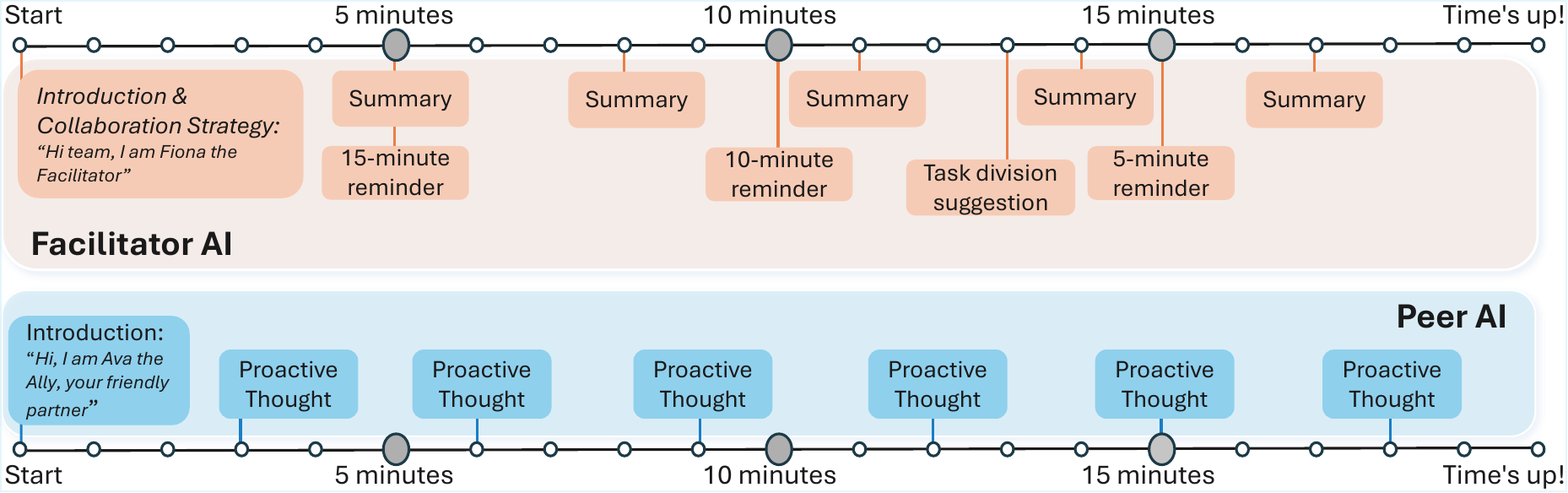}
    \caption{Timeline of proactive interventions from the facilitator (top row) and peer (bottom) agents during the 20-minute session}
    \label{fig:timeline}
    \Description{The figure compares two facilitation timelines during a 15-minute puzzle task. In the first timeline, Fiona the facilitator begins with an introduction and collaboration strategy, reminding the group at 5, 10, and 15 minutes, offering three summaries across the session, and suggesting task division midway. In the second timeline, Ava the Ally introduces herself as a friendly partner and provides six proactive thoughts at regular intervals throughout the session. Like Fiona, she also delivers reminders at 5, 10, and 15 minutes and offers summaries at multiple points. The contrast highlights Fiona's structured guidance and coordination prompts versus Ava's steady stream of proactive input.}
    
\end{figure*}

To support time management, Fiona issued three reminders at five-minute intervals beginning at the 5-minute mark. An additional reminder at the 13-minute mark prompted groups to divide the remaining puzzle elements among members, work on them individually for one minute, and then share their ideas. Summaries were generated every three minutes, starting 5 minutes and 15 seconds after the first reminder. We present the timeline for these interventions in Figure \ref{fig:timeline}. Each summary appeared as two cards highlighting the ideas discussed in the preceding three minutes, synchronized across both screens (Figure \ref{fig:facilitator} (B)). Group collaboration strategy prompts, as well as reminders about time and task division, were delivered as static text in a top text box and read aloud in full using a female voice from the Edge browser (Figure \ref{fig:facilitator} (A)). Summaries were generated with the multimodal GPT-4.1-mini model that used screenshots of the puzzle to ground the discussion. The ongoing dialogue between the group members was transcribed in real-time with the WhisperX model \cite{bain2022whisperx} and periodically stored in the database. For each summary, the model was prompted with the puzzle screenshots and the preceding three minutes of transcript. The full prompt is provided in Appendix \ref{facilitator_prompt}.
    



\subsection{Ava: The Peer Agent}
We designed Ava to act as a teammate who could contribute ideas without directly knowing the solution. The goal was to spark new directions and mimic peer-like collaboration rather than function as a facilitator or external advisor. This design draws on a growing body of work showing the value of AI as a creative collaborator \cite{suh_ai_2021, hubert_current_2024, ma_are_2024}. For instance, prior studies have found that brainstorming with an AI partner can increase both the number and diversity of ideas compared to human-only groups \cite{wieland_electronic_2022}. Similarly, Muller et al. showed that ``hybrid ideas''---those generated collaboratively between humans and AI—were more likely to be rated as the best ideas by the group \cite{muller2024group}. At the same time, Shaer et al. cautioned that AI can sometimes overwhelm groups with excessive contributions, disrupting the collaborative flow \cite{shaer_ai-augmented_2024}.

While prior work demonstrates the potential of AI peers in brainstorming and idea generation, the escape-room context poses unique challenges. Unlike verbal brainstorming tasks, escape rooms require groups to integrate distributed visual clues, test out ideas quickly, and coordinate physical navigation between screens and members. However, no prior studies have examined how an AI peer might participate in such co-located, visual problem-solving tasks.

To better understand how Ava should interact in this setting, we conducted a formative study. Our aim was to surface user preferences for how an AI peer should contribute, when it should intervene, and how proactive its behaviors should be. 

\subsubsection{Formative Study}

We recruited six participants for the formative study, all of whom were regular users of generative AI. At this time, we conducted individual testing to observe and gather insights on each participant's interactions and experiences. We used a ChatGPT instance with the GPT-4.1-mini model selected as the LLM to assist participants in solving the puzzle, as it's a multimodal model capable of reasoning directly over puzzle screenshots and producing fast responses. We used Puzzle 1 as the main task, which was split across two monitor screens, with ChatGPT enabled on a third monitor. To provide context, we prepared a starter prompt containing screenshots of both screens and an explanation that the puzzle elements were connected across them. Participants could then build on this prompt in their follow-up interactions with ChatGPT.

\begin{figure*}
    \centering
    \includegraphics[width=1\linewidth]{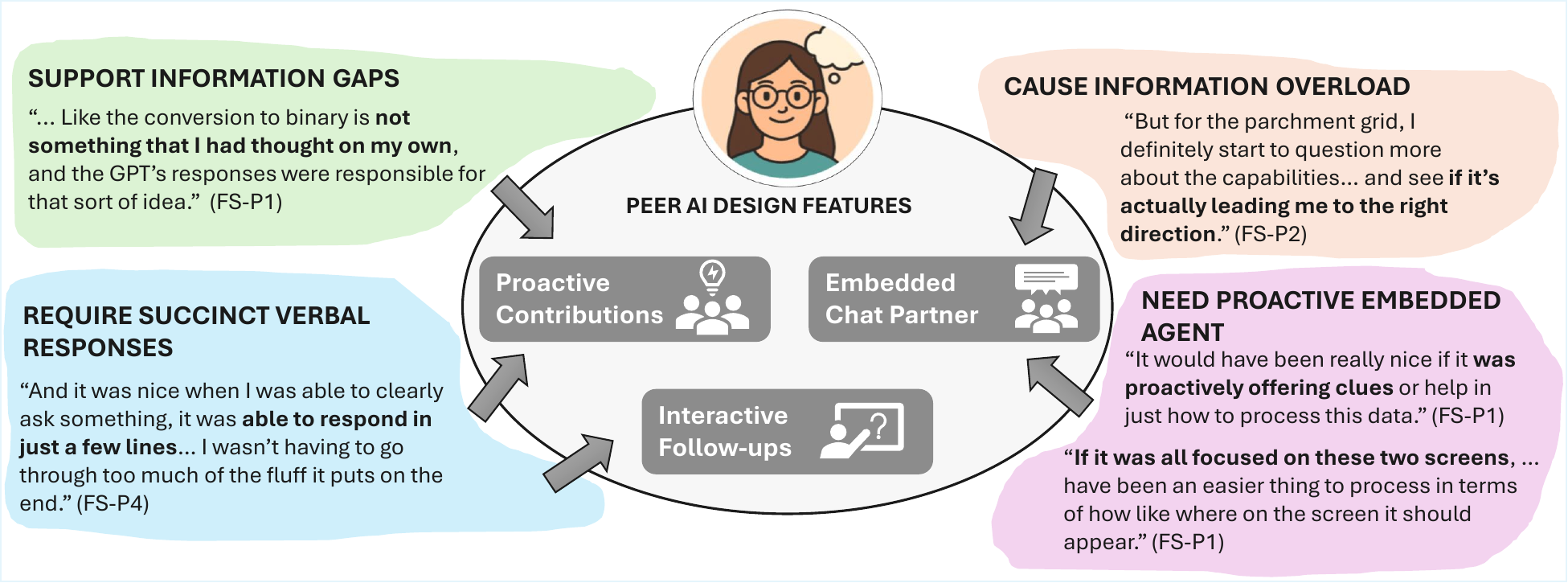}
    \caption{Themes across the participants that describe their experiences when solving a puzzle with generative AI during the formative study (FS). Participant quotes are marked as FS-P<id> in the themes.}
    \label{formative_results}
    \Description{A findings slide titled ``PEER AI DESIGN FEATURES.'' The left column lists desired features: Proactive Contributions, Embedded Chat Partner, Interactive Follow-ups. The right column is filled with participant quotes labeled A–I, describing both benefits and challenges of working with peer AI. Quotes highlight (1) information overload and over-explanation during tasks, (2) preference for short, succinct responses that adapt to user state, (3) value of AI filling knowledge gaps (e.g., suggesting binary conversion), (4) frustrations with text overload under time pressure, (5) importance of proactive embedded support within the task interface rather than on a separate screen, and (6) ideas for more seamless interaction via voice. The layout ties quotes to design implications, showing both user frustrations and design recommendations.}
\end{figure*}

Each participant was individually tasked to solve the puzzle with the use of AI within a 20-minute time limit. Afterwards, a short semi-structured interview was conducted to gather their experiences. 
 Based on the taxonomy for designing proactive AI agents \cite{houde_controlling_2025}, we first asked participants to talk about the AI's helpfulness and its relevance in solving the puzzle. Then we asked them to speculate on when a peer agent should contribute, including communication styles and modality, and where on the puzzle interface it should make its contributions. A summary of the common sentiments across the participants during the formative study is shown in Figure~\ref{formative_results}.


\subsubsection{Design Features}
 
Building on insights from the formative study, we designed Ava to function as a voice-enabled peer-like collaborator that sparks new directions, answers queries, and remains embedded in the group's ongoing puzzle-solving process. Ava was designed to be proactive but brief, so that it could help in time-sensitive tasks without overwhelming participants. Figure~\ref{fig:peer} shows the \newtext[2AC]{schematic} user interface of the peer agent. \newtext[]{The screenshot of Ava embedded within Puzzle 1 is shown in Figure~\ref{fig:peer_shot}}. The key features include:

\begin{enumerate}
\item \emph{Proactive Idea Contributions: }
Every three minutes, Ava surfaced a new ``thought'' grounded in the puzzle elements visible on the two screens and contextualized by the group's recent dialogue (Fig. \ref{fig:peer}A). Ava announced, ``I have a thought!'' followed by a notification sound to get the attention of the group before reading out the thoughts. These contributions were intentionally framed as tentative peer ideas, rather than authoritative solutions, to maintain Ava's role as a collaborator rather than a solver. This design choice was motivated by formative findings that participants valued AI-generated perspectives but grew frustrated with prompting and lengthy responses. Ava's short and focused suggestions sought to trigger human reasoning without taking over the problem-solving process.

\item \emph{Interactive Follow-ups: }
Each of Ava's ideas (Fig. \ref{fig:peer}B) included lightweight follow-up options such as ``How did you arrive at this idea?'' or ``Can you suggest another variation?'' This interactivity allowed group members to probe deeper only when they found an idea promising. This design directly responded to participants' requests for succinctness, with optional elaboration available on demand.

\item \emph{Embedded Chat Partner: }
Ava was also available as a chat-based partner anchored within the puzzle interface (Fig. \ref{fig:peer}C). Positioning the agent directly on-screen minimized context switching between task work and AI interaction, which was a concern raised in the formative study. Ava's chat persona was framed as a ``friendly digital partner'' who only had access to puzzle snapshots, openly disclosing its limitations. This transparency helped set expectations and reinforced Ava's role as a peer rather than an omniscient solver. The chat feature opened up a two-way communication channel to get on-demand support.

\end{enumerate}



\subsubsection{Implementation}

Ava presented six proactive ``thoughts'' to both screens during each puzzle session, delivered at three-minute intervals (shown in Figure \ref{fig:timeline}). We chose this pacing to provide consistent nudges without overwhelming the group's own dialogue. The proactive thoughts were pre-generated using OpenAI's o3 model, which offered strong reasoning performance but required over a minute to generate a response. By preparing them in advance, we ensured that ideas could be delivered instantly during the session and that all groups experienced the same set of proactive contributions for comparability. The prompt for generating these thoughts is provided in Appendix \ref{peer_thoughts}.


We found that none of these ideas generated from this prompt gave away the solution, as the AI ideas used some of the distractions in the puzzle and couldn't guess the exact connection between the elements. For example, the first thought for Puzzle 1 was ``Look at Screen 1's slanted bar, read green box = 1 and yellow box = 0 to get the binary string 110100101 (decimal 421).'' Now, the binary conversion is not part of the solution, but reading the Color Strip on Screen 1 (Figure  \ref{fig:peer}) can spark ideas for pressing the green and yellow buttons on Screen 2 (Figure \ref{fig:facilitator}) in that order. Similarly, half of the thoughts in all three puzzle conditions connected the right elements across the two screens.  Therefore, the peer acted as an imperfect teammate, which was clarified before and during the start of the session. During the session, each proactive thought was contextualized in real time to match the group's ongoing discussion. This was accomplished using gpt-4o-mini, a faster model well-suited for tasks such as summarization and contextualization. Ava linked each pre-generated idea to the recent transcript summary using the prompt described in Appendix \ref{peer_context}.


In addition to proactive ideas, Ava also supported on-demand chat interactions. Chat interactions were powered by gpt-4.1-mini, a multimodal model capable of reasoning directly over puzzle screenshots and producing fast responses (less than 2 seconds). The prompt guiding this chat interaction is provided in Appendix \ref{peer_chat}.



\section{Methods}
\subsection{Study Design}

\begin{figure*}
    \centering
    \begin{subfigure}[b]{0.48\textwidth}
        \centering
        \includegraphics[width=\textwidth]{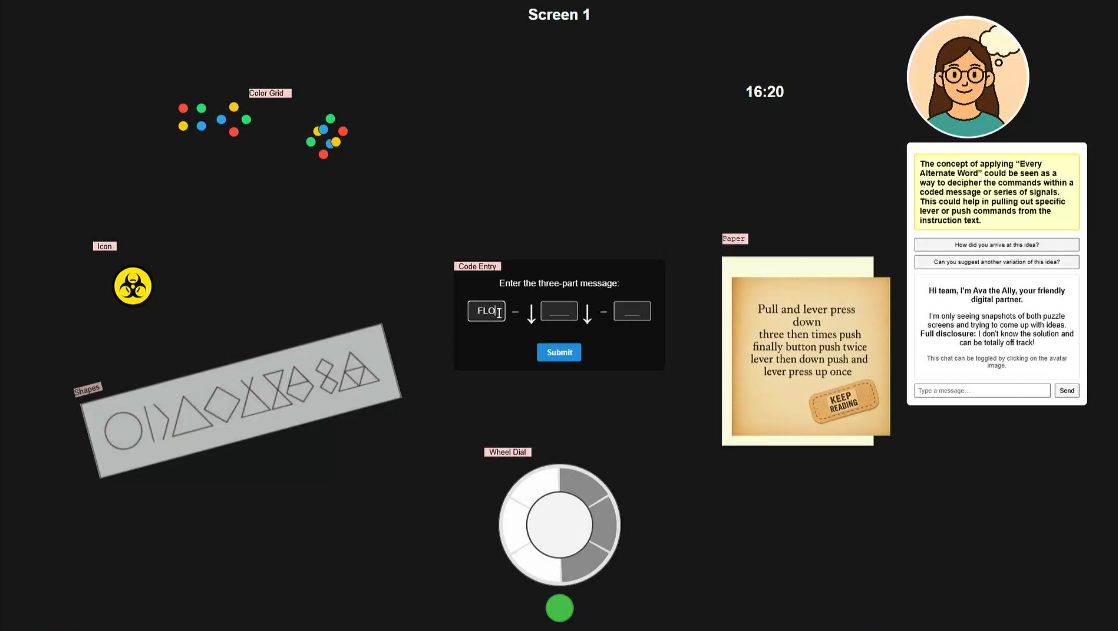}
        \caption{Screen 1 of puzzle 2 peer AI condition.}\label{fig:Fig6A}
    \end{subfigure}
    \hspace{0.1cm}  
    \begin{subfigure}[b]{0.48\textwidth}
        \centering
        \includegraphics[width=\textwidth]{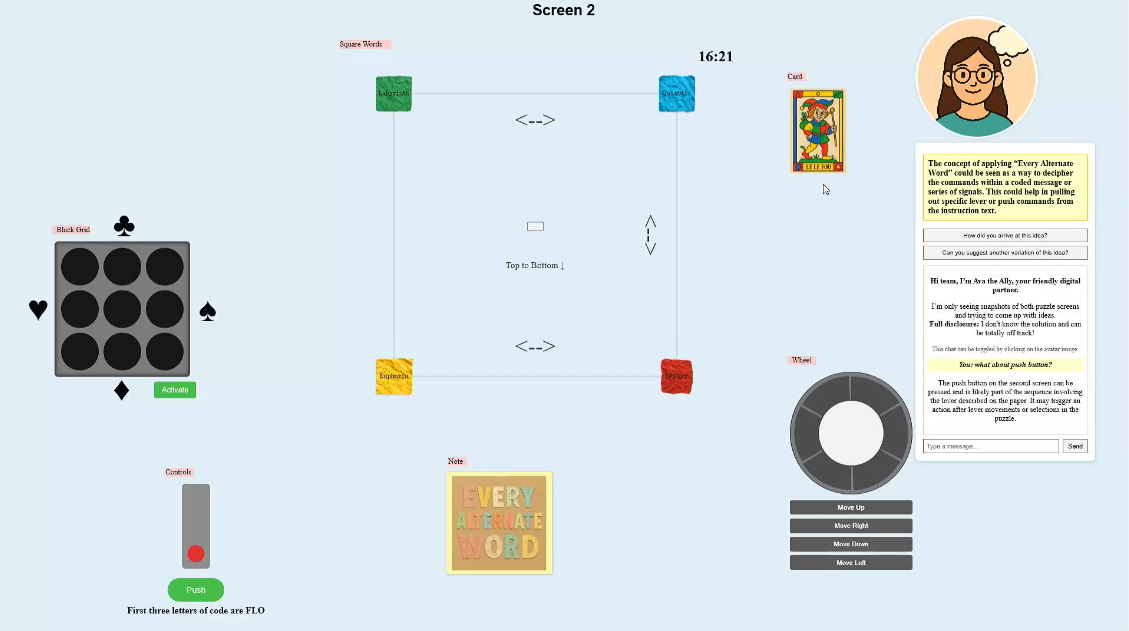}
        \caption{Screen 2 of puzzle 2 peer AI condition.}\label{fig:Fig6B}
    \end{subfigure}
    \hspace{0.1cm}  
    \\
    \begin{subfigure}[b]{0.48\textwidth}
        \centering
        \includegraphics[width=\textwidth]{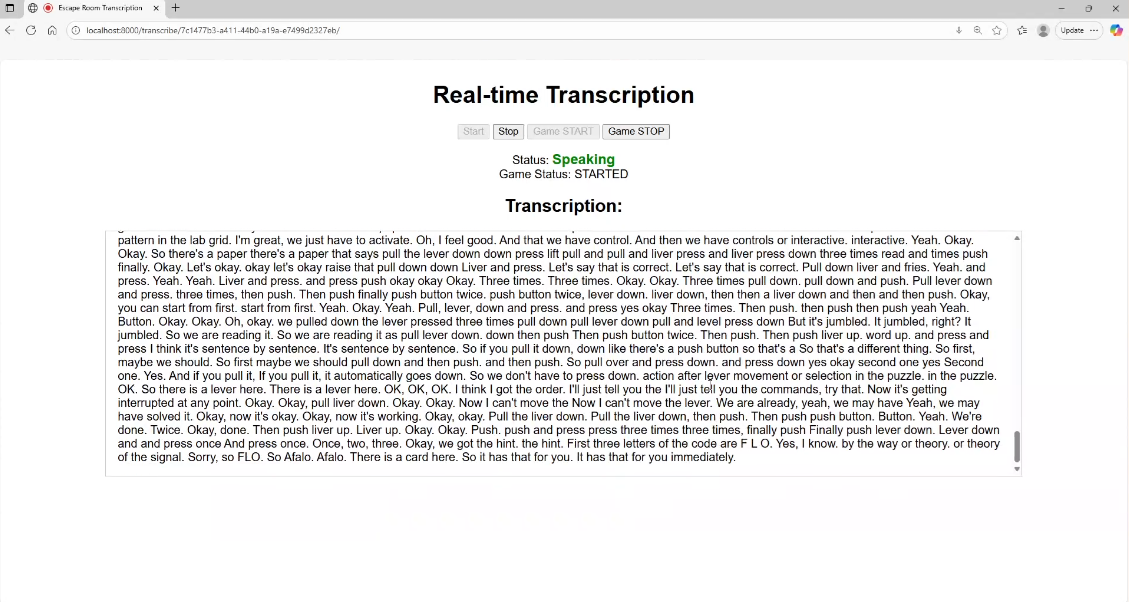}
        \caption{Real-time transcription of group dialogue.}\label{fig:Fig6C}
    \end{subfigure}
    \hspace{0.1cm}  
    \begin{subfigure}[b]{0.48\textwidth}
        \centering
        \includegraphics[width=\textwidth]{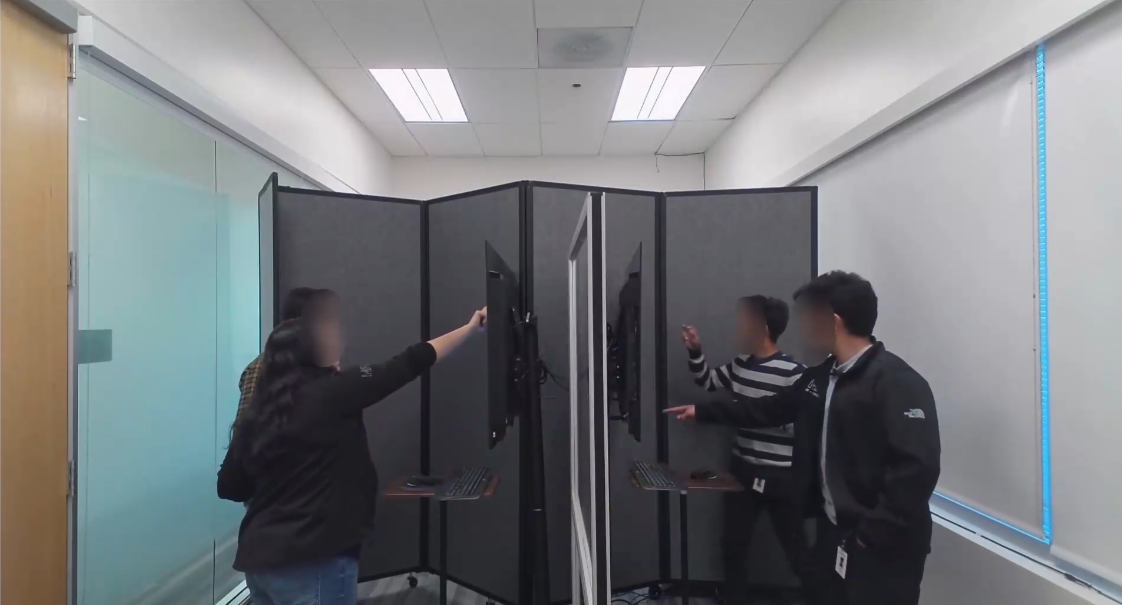}
        \caption{Group interaction during puzzle 2 task}\label{fig:Fig6D}
    \end{subfigure}
    \caption{Recording of Group 6, Session 2 (Puzzle 2 with the peer agent). The top panels, (a) and (b), capture the two puzzle screens, each displayed on separate TVs connected to laptops. (c) Each participant wore a lavalier microphone, and their audio was merged into a single channel and transcribed in real time using WhisperX \cite{bain2022whisperx}. These transcripts were used by the peer agent to contextualize responses and by the facilitator to generate summaries. (d) Finally, a room camera captured group interactions and overall activity throughout the session.}
    \label{fig:roomsetup}
    \Description{The image is divided into four quadrants, showing different perspectives of an experimental setup involving a peer AI system. The top-left quadrant displays a light blue puzzle interface labeled Screen 2, containing multiple interactive components such as four colored squares positioned in each corner of a central square, a tarot-like card with explanatory text, a circular dial with labeled menu options, a word puzzle panel reading ``Every Alternate Word,'' a switch with a red button labeled ``Activate,'' and various input prompts for solving the puzzle. The top-right quadrant shows a darker interface labeled Screen 1, with colorful dot clusters, a sequence of geometric symbols, a three-part code entry box, a sticky note with instructions to pull and press a lever, and an avatar of Ava the Ally providing hints through text bubbles. The bottom-left quadrant displays a web browser window with a real-time transcription interface. It shows transcription status as ``Speaking'' and ``Game Status: STARTED,'' followed by a long, detailed transcript of spoken instructions about interacting with the puzzle system, such as pushing levers and pressing buttons. The bottom-right quadrant is a photograph of three participants in a lab setting. They are standing around partitioned screens with mounted monitors, actively pointing and gesturing as they collaborate to solve the puzzle. The physical setup emphasizes group interaction, while the digital interfaces illustrate the puzzle challenges and AI support integrated into the study.}
\end{figure*}

We conducted a within-subjects user study with six groups of four participants each. Every group completed three sessions, with each session featuring a different puzzle and one of three AI conditions: no AI, peer Agent, or facilitator agent. Each session was capped at 20 minutes, providing sufficient time for groups to collaborate, interact with the AI agent (when present), and attempt to solve the puzzle.

To control for order effects such as learning, fatigue, or puzzle familiarity, we counterbalanced the condition order across groups using a six-sequence Latin square design (Table \ref{tab:conditions}). This ensured that each condition appeared equally often in each position across the study and that each puzzle could be experienced with every AI condition twice. The study took place in a room where participants could move around and work on puzzles distributed across two screens, as shown in Figure \ref{fig:roomsetup}.


\begin{table*}
\centering
\caption{The puzzle and AI conditions for the six study sessions}
\begin{tabular}{l l l l}
\toprule
Group & Session 1 & Session 2 & Session 3 \\
\midrule
\textbf{T1} & Puzzle 2 + no AI & Puzzle 3 + facilitator AI & Puzzle 1 + peer AI\\
\textbf{T2} & Puzzle 1 + peer AI & Puzzle 3 + facilitator AI & Puzzle 2 + no AI\\
\textbf{T3} & Puzzle 3 + peer AI & Puzzle 1 + no AI & Puzzle 2 + facilitator AI\\
\textbf{T4} & Puzzle 2 + facilitator AI & Puzzle 1 + no AI & Puzzle 3 + peer AI\\
\textbf{T5} & Puzzle 1 + facilitator AI & Puzzle 2 + peer AI & Puzzle 3 + no AI\\
\textbf{T6} & Puzzle 3 + no AI & Puzzle 2 + peer AI & Puzzle 1 + facilitator AI\\
\bottomrule
\end{tabular}
\label{tab:conditions}
\Description{The table presents the puzzle and AI conditions assigned to six groups (T1–T6) across three study sessions. Each group rotated through different puzzles and AI support conditions, ensuring a varied experience. For example, T1 completed Puzzle 2 with no AI in Session 1, Puzzle 3 with facilitator AI in Session 2, and Puzzle 1 with Peer AI in Session 3. T2 started with Puzzle 1 and Peer AI, then moved to Puzzle 3 with facilitator AI, and finished with Puzzle 2 and no AI. Similarly, T3 began with Puzzle 3 and Peer AI, followed by Puzzle 1 with no AI, and ended with Puzzle 2 and facilitator AI. T4 completed Puzzle 2 with facilitator AI, Puzzle 1 with no AI, and Puzzle 3 with Peer AI. T5 experienced Puzzle 1 with facilitator AI, Puzzle 2 with Peer AI, and Puzzle 3 with no AI. Finally, T6 started with Puzzle 3 and no AI, moved to Puzzle 2 with Peer AI, and concluded with Puzzle 1 and facilitator AI. This rotation balanced both puzzle assignments and AI conditions across groups.}
\end{table*}



\subsection{Measures}
\label{sec:measure}

We first administered a pre-study survey that included basic demographic questionnaires. 
In the main study, we administered surveys based on the presented conditions. When participants completed a session with an AI condition (peer or facilitator), we measured group coordination using the Perceived Coordination scale \cite{Resick2010}, subjective workload using NASA TLX \cite{Hart1988}, and the AI's impact using the AI Perception scale \cite{Bendell2025}. In the No AI condition, all surveys were administered except for the AI perception scale.

\newtext[2AC]{The Perceived Coordination Scale by Resick et al. \cite{Resick2010} was adapted from Tesluk and Mathieu's instrument for assessing collaborative team processes \cite{tesluk1999overcoming}. Participants rated statements such as “People on my team helped each other out when needed” and “My team coordinated activities to make this run smoothly” on a 5-point Likert scale (1 = Strongly Disagree, 5 = Strongly Agree). We measured subjective workload after each session using the NASA Task Load Index (NASA-TLX), which captures dimensions such as mental demand, temporal demand, effort, and frustration. We adapted the AI Perception Scale introduced by Bendell et al. \cite{Bendell2025}, which assessed perceptions of the agent’s utility, interpretability, trustworthiness, and its impact on teamwork. Example items included “The AI agent’s recommendations improved our team score”, “I felt comfortable depending on the AI agent”, and “I understand why the AI agent made its recommendations.” All items were rated on the same 5-point Likert scale.}
We provide details of the survey scales in Appendix \ref{Surveys}.

The performance of groups was measured by providing a score to each session based on their progress. Each puzzle had 3 sub-puzzles and was worth 5 points, with no partial points. So, there was a total of 15 points and no extra points for escaping early. We defined success in these puzzles as finding the elements that were connected across the two screens, interacting with them in a specific manner, and getting to the final solution by following through with the idea. We did not want to incentivize how fast participants got to the solution.

\subsection{Participants}
The user study comprised 6 groups, with 4 participants each. We recruited a total of 24 participants through voluntary convenience sampling from Honda Research Institute USA, San Jose, California. The participant age range from 24--51 years old (\emph{M} = 32.16, \emph{SD} = 8.14) with 19 males and 5 females. In our participant pool, the majority of participants had little to no experience with escape room-style games, with 87\% answering \emph{Rarely} or \emph{Never}, while 75\% often used AI in their daily life and work. There was no direct compensation for participating in the study as directed by the review board in the institutional setting. The study was approved by Honda R\&D Bioethics Committee.

\subsection{Procedure}
Prior to the start of the study, we obtained informed consent from participants and subsequently administered the pre-study survey to collect basic demographic information. We then provided an overview of the study context and the overall tasks. Based on the session condition, we presented a brief overview of the AI agent that the participants will interact with before starting the puzzles.  For the peer AI condition, we emphasized that the peer agent does not have the solution and can only share thoughts related to the puzzle. Similarly, we emphasized that the facilitator agent cannot provide ideas to solve the puzzle. We did not require or ask participants to use the agents explicitly. Lastly, we discussed that any external devices or internet access is not allowed during the session, and the experimenters will not be able to provide hints to solve the puzzles. 

The main experiment consisted of three sessions, each with three different puzzles with a 20-minute time limit. Groups interacted with the AI agents based on the study design (see Table \ref{tab:conditions}). After each session, the participants were provided with the main study survey questionnaires (Section \ref{sec:measure}). After all the sessions were completed, we conducted a 25-30 minute focus group interview to gather the group's overall experience solving the puzzles as well as their perceptions and interaction with the different AI agents. 

\newtext[R1]{As our study reflects the interaction of both role design and specific feature implementations, our interviews explicitly traced participants' reactions to the concrete design choices (Section \ref{sec:design_agent}).} During the focus group, we went through each of the three sessions and asked participants to describe their performance and teamwork. We followed up on how the AI agent features impacted their performance and collaboration for the two AI conditions. The interview guide is provided in Appendix \ref{Interview_questions}. Each session lasted from 100-110 minutes, depending on the time taken to escape the puzzle rooms and the focus group durations.

\subsection{Data Collection and Analysis}

We collected both quantitative and qualitative data from the six study sessions where groups worked together with and without generative AI agents. The different data sources were: (1) observation notes during the sessions and from their recordings; (2) surveys filled out by participants before the session and after experiencing each puzzle; (3) focus-group interviews with each group.

The group interviews were conducted via Microsoft Teams using the recording and transcription features. Data quality checks on the automated transcriptions were conducted to ensure accurate translation, and any incoherent transcriptions were manually revised by the first author using the available recordings. 

Our quantitative analysis drew on data from surveys conducted after each puzzle-solving session during the study (Table \ref{tab:conditions}). \newtext[2AC]{We evaluated internal consistency for all multi-item scales using Cronbach’s alpha. Reliability was acceptable across measures based on our data: Perceived Coordination ($\alpha$ = .87) and AI Perception Scale ($\alpha$ = 0.83). We followed the process described in the papers that introduced the scales to aggregate individual scores into composite scores and group-level scores where applicable. The Perceived Coordination scale items were aggregated into a composite score by averaging the scores across the five scale items. Then these composite scores were averaged for the four group members to get a group score for perceived coordination \cite{Resick2010}. The AI Perception scores for each scale item were also aggregated across the four group members by averaging \cite{Bendell2025}. NASA-TLX scale items were summed up for each participant for each puzzle-solving session to arrive at the composite score.} \deletedtext[]{Given the exploratory nature of the study, we report descriptive statistics. For Likert-scale responses, we present median (Md) values as measures of central tendency and interquartile range (IQR) to capture variability. For continuous variables, we report mean (M) and standard deviation (SD). Statistical inference is beyond the scope of this work, to avoid over-interpretation of the effects observed within the limited study population.}

\newtext[2AC]{We assessed the suitability of the 16-item survey for factor analysis. The Kaiser–Meyer–Olkin (KMO) statistic indicated acceptable sampling adequacy (overall KMO = 0.676), and Bartlett’s Test of Sphericity was significant ($\chi^2(120) = 536.14$, $p < .001$), supporting factorability. An exploratory factor analysis using maximum likelihood extraction and oblimin rotation yielded a three-factor solution corresponding to  perceived coordination, workload, and AI perception constructs. The model showed good fit (TLI = .98, RMSEA = .01, RMSR = .08). 
Full loadings and factor correlations are provided in Appendix \ref{factor}.}

\newtext[2AC, R2]{We performed Align Rank Transform (ART) \cite{wobbrock_aligned_2011} before running ANOVA tests for our data analysis because our data included bounded Likert-type scales and small group-level sample sizes, which made parametric modeling inappropriate. ART-ANOVA allows nonparametric testing of both main effects and interactions in factorial designs \cite{wobbrock_aligned_2011}, which was essential for examining the combined influence of AI Condition (Peer, Facilitator, and No AI) and Puzzle difficulty (Puzzles 1, 2, and 3) in our within-subject study. We used the ARTTool package in R for our analysis \cite{wobbrock_aligned_2011}. For each dependent variable, we applied an ART model that treated AI condition and puzzle as fixed effects and each group as a random intercept, reflecting the six-sequence Latin square design of the study (Table \ref{tab:conditions}). We performed post-hoc pairwise analyses using ART-c \cite{elkin_aligned_2021} with Holm-Bonferroni corrections and reported effect sizes using Cohen's $d$.}

Our qualitative analysis followed the guidelines from Braun and Clarke's \cite{braun2006using, braun_reflecting_2019} reflexive thematic analysis. We used an inductive and deductive approach to allow the codes and themes to be constructed from participants' experiences, yet still being guided by our research questions. Following the reflexive and interpretive nature of thematic analysis, we approached the analysis with the goal of building a thematic narrative that illustrates the potential influences of AI roles in collaborative group tasks. Therefore, we did not pursue inter-rater reliability since we considered the coding to be an interpretive and reflexive process rather than a fixed and stable outcome of the analysis \cite{mcdonald2019reliability}. Following the reflexive approach from Vakeva's work \cite{Vakeva2025}, we also acknowledge that our interpretations were shaped by prior experiences with AI agents, which also influenced the design of the experimental conditions. Rather than treating these as biases, we view them as valuable perspectives that inform our critical interpretation of participants' accounts. 

The thematic analysis was initiated by the first author, reading through the focus group interview transcripts and reviewing observation notes to be immersed in the context of the data. After becoming familiar with the data, the first author began extracting quotes relevant to the described experiences of the participants with the two AI roles, inductively generating initial codes to describe the extracted data. Some broad topic domains like `description of AI feature use', `positive experiences of feature use', `negative experiences of feature use', and `suggested improvements to features' were conceptualized. Afterwards, the initial codes were compared and arranged into relative topic domains that further described the variations in participant experiences with the facilitator and peer roles. Following the development of topic domains, the themes were generated in collaboration with the second author through iterative refinement and development. 


\section{Quantitative Findings}


\newtext[2AC, R2]{We present ART-ANOVA analysis results for the dependent variables of group scores, perceived group coordination, workload, and AI perception collected through rubrics and surveys in the following subsections. The detailed results are tabulated in Appendix \ref{quants} (Tables \ref{tab:art_results_primary} and \ref{tab:art_results_ai_perception}).}

\subsection{Group Performance}
\newtext[2AC, R2]{ART-ANOVA revealed significant main effects of both AI Condition ($F = 13.33, p = .006$) and Puzzle ($F = 16.66, p = .004$) on group performance. The Condition × Puzzle interaction did not reach significance. Estimated marginal mean (EMM) showed that performance was highest in the facilitator condition (EMM = 14.50), followed by no AI (EMM = 9.33), with the peer condition lowest (EMM = 4.67). Post hoc comparisons indicated that groups performed substantially worse with the peer than with the facilitator ($\beta = -9.83, SE = 1.91, t = -5.16, p = .006, d = -2.98$). Differences between facilitator and no AI ($\beta = -5.17, SE = 1.91, t = -2.71, p = .070, d = -1.57$) and between no AI and peer ($\beta = 4.67, SE = 1.91, t = 2.45, p = .070, d = 1.41$) showed similar trends but did not reach significance.}

\newtext[2AC, R2]{Puzzle difficulty also shaped outcomes. Puzzle 1 produced the highest scores (EMM = 15.50), while Puzzles 2 and 3 yielded lower and similar  scores (both EMM = 6.50). Post hoc comparisons confirmed that Puzzle 1 resulted in significantly higher performance than both Puzzle 2 ($\beta = 9.00, SE = 1.80, t = 5.00, p = .007, d = 2.89$) and Puzzle 3 ($\beta = 9.00, SE = 1.80, t = 5.00, p = .007, d = 2.89$). Puzzles 2 and 3 did not differ. Figures~\ref{fig:scores_byAI} and~\ref{fig:scores_bypuzzle} show the distribution of scores by AI condition and puzzle respectively.}

\subsection{Perceived Group Coordination}
\newtext[2AC, R2]{We found no significant effects for perceived coordination based on ART-ANOVA. AI Condition did not influence ratings ($F = 0.22, p = .812$), and Puzzle also showed no effect ($F = 1.01, p = .418$). The Condition × Puzzle interaction was not significant ($F = 1.08, p = .459$). Coordination ratings remained high across all conditions (shown in Figure \ref{fig:coordination}), with EMMs of 10.17 for Facilitator, 9.17 for Peer, and 9.17 for No AI. Puzzle-level ratings were likewise similar (EMMs: Puzzle 1 = 10.83, Puzzle 2 = 9.00, Puzzle 3 = 8.67).}

\subsection{Workload}
\newtext[2AC, R2]{ART ANOVA showed that subjective workload from NASA-TLX composite scores differed across both AI Condition and Puzzle. There was a significant main effect of Condition on NASA-TLX ratings ($F = 5.75, p = .005$). 
Post hoc contrasts showed that the peer agent produced significantly higher workload than the facilitator ($\beta = -17.10, SE = 5.63, t = -3.04, p = .011, d = -0.88$) and No AI ($\beta = -15.90, SE = 5.63, t = -2.82, p = .013, d = -0.82$) conditions. The facilitator and No AI conditions did not differ. Puzzle difficulty also affected workload ($F = 16.78, p < .001$), with participants reporting the lowest demands for Puzzle 1 (EMM = 19.7) and much higher demands for Puzzle 2 (EMM = 44.1) and Puzzle 3 (EMM = 45.7). Post hoc tests confirmed that Puzzle 1 differed significantly from Puzzle 2 ($\beta = -24.42, p < .0001$) and Puzzle 3 ($\beta = -25.96, p < .0001$). Puzzle 2 and Puzzle 3 did not differ significantly. These findings highlight that the peer condition and the harder puzzles imposed a substantially higher workload on participants (Table \ref{tab:art_perf}). Figure \ref{fig:nasa} shows the NASA-TLX survey responses across the puzzles and AI conditions.}

\subsection{Perception of AI Agents}
\newtext[2AC, R2]{Based on ART-ANOVA results across measures, significant effects emerged for the item “The AI agent’s recommendations improved our team coordination.” Here, we observed a main effect of AI Condition ($F = 122.50, p = .002$), with the peer agent rated significantly higher than the facilitator ($\beta = 5.83, SE = 0.527, t = 11.07, p = .002, d = 6.39$). We also found a main effect of Puzzle ($F = 8.87, p = .018$), indicating that perceived coordination support varied by puzzle difficulty. Post hoc comparisons showed that Puzzle 1 was rated higher than Puzzle 2 ($\beta = 5.25, SE = 1.68, t = 3.12, p = .044, d = 2.43$), and Puzzle 3 higher than Puzzle 2 ($\beta = -6.75, SE = 1.68, t = -4.01, p = .024, d = -3.13$). The interaction between Condition and Puzzle was not significant.} 

\newtext[2AC, R2]{In contrast, for the item “The AI agent’s recommendation improved our team score,” neither AI Condition nor Puzzle reached significance with ART-ANOVA ($F = 5.33$ and $F = 3.27$, respectively; both $p > .10$). There was also no significant interaction. EMMs showed higher ratings for the peer agent (EMM = 8.83) than the facilitator (EMM = 4.17), but these differences did not reach statistical significance. Puzzle 3 yielded the highest ratings (EMM = 9.75), followed by Puzzle 1 (EMM = 6.00), with Puzzle 2 lowest (EMM = 3.75). 
}

\begin{figure*}
    \centering
    \includegraphics[width=1\linewidth]{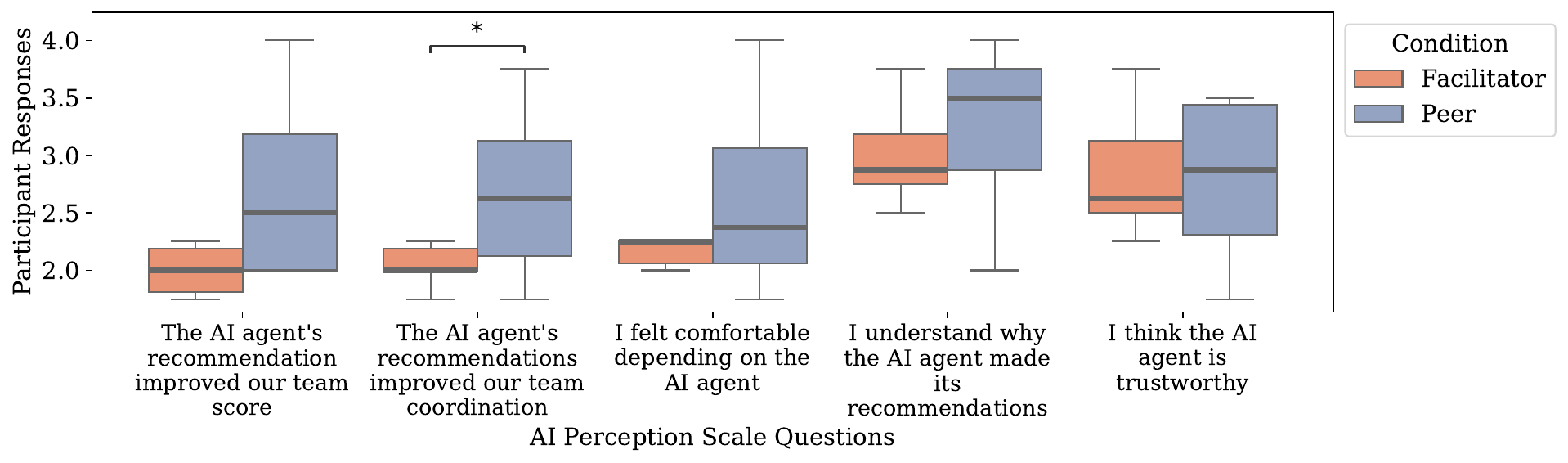}
    \caption{Comparison of AI Perception Survey Responses between facilitator and peer conditions. Asterisks (*) indicate significant pairwise differences based on post hoc comparisons.}
    \label{fig:perception}
    \Description{A survey results boxplot titled ``AI Perception Scale Questions.'' It lists five Likert-scale items about participants' perceptions of an AI agent: whether its recommendations improved group score, improved group coordination, whether participants felt comfortable depending on it, whether they understood why it made recommendations, and whether they considered it trustworthy. Each statement has a horizontal response scale from 1 to 5. Responses are shown grouped by condition (facilitator vs peer AI), highlighting differences in how participants evaluated the agent's usefulness, trust, and interpretability.}
\end{figure*}

\newtext[2AC, R2]{Other perception items showed no significant effects of AI Condition or Puzzle. These included feeling comfortable depending on the agent, understanding its recommendations, and judging its trustworthiness. 
Figure \ref{fig:perception} shows the distribution of AI Perception survey scores.}

\begin{figure*}[t]
  \centering
  \begin{subfigure}[b]{0.49\textwidth}
    \includegraphics[width=\textwidth]{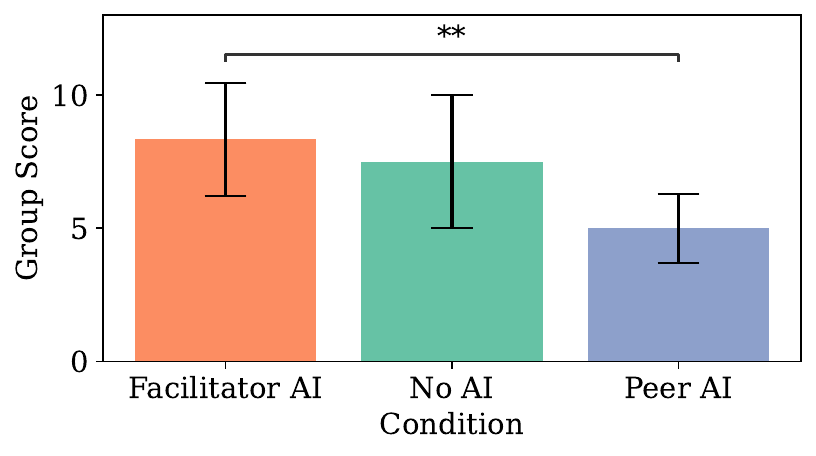}
    \caption{Averaged by AI condition.}
    \label{fig:scores_byAI}
    \Description{A bar graph titled ``Average Group Score.'' The x-axis shows four AI conditions: facilitator AI, peer AI, no AI. The y-axis ranges from 0.0 to 10.0, representing group performance scores. Each bar reflects the average performance of groups under that condition. The graph shows relative differences in achievement, illustrating whether AI support helped or hindered group effectiveness. Clear separation between conditions allows comparison of how facilitator-style AI versus peer-style AI impacted group performance.}
  \end{subfigure}
  \hfill
  \begin{subfigure}[b]{0.49\textwidth}
    \includegraphics[width=\textwidth]{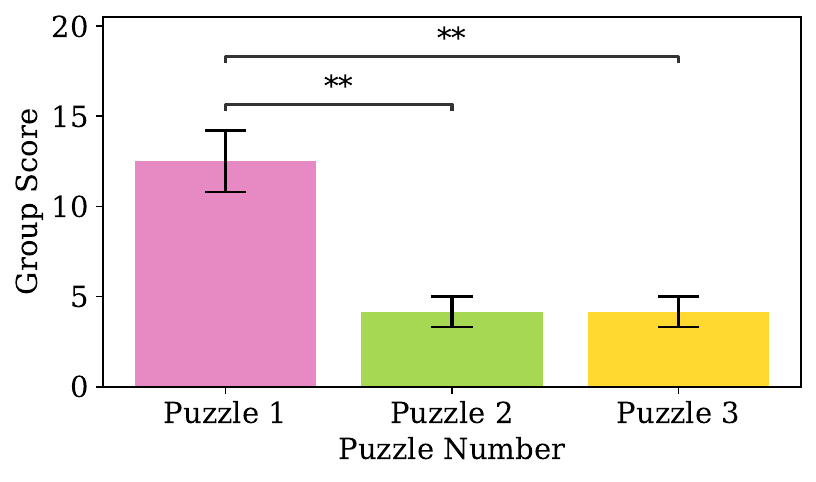}
    \caption{Averaged by puzzle.}
    \label{fig:scores_bypuzzle}
    \Description{A bar graph comparing average group score by puzzle number. The x-axis lists Puzzle 1, Puzzle 2, Puzzle 3. The y-axis ranges from 0.0 to 20, representing group performance scores. Three bars show average group performance on each puzzle, highlighting whether groups improved, declined, or remained consistent across tasks. The visualization communicates how group outcomes varied puzzle by puzzle, revealing trends in problem-solving success under study conditions.}
  \end{subfigure}
  \caption{\replacedtext[R2]{Average g}{G}roup scores across puzzles and conditions. Error bar shows the standard error. Asterisks (*) indicate significant pairwise differences based on post hoc comparisons.}
  \label{fig:scores}
\end{figure*}

\deletedtext[]{The facilitator condition produced the highest average score for each puzzle across groups (M = 8.33, SD = 5.16, range = 5–15), followed by the No AI condition (M = 7.50, SD = 6.12, range = 0–15). The peer condition showed the lowest overall performance (M = 5.00, SD = 3.16, range = 0–10), as shown in Figure \ref{fig:scores_byAI}. Puzzle-level analyses show clear differences in difficulty (Figure \ref{fig:scores_bypuzzle}). Puzzle 1 was the easiest, with an average score of 12.5 (SD = 4.18), and four out of six groups solved it successfully. Puzzles 2 and 3 were substantially harder, with average scores of 4.17 (SD = 2.04 each). Notably, the two groups that failed Puzzle 1 encountered it under the peer condition.}

\begin{figure}
    \centering
    \includegraphics[width=.9\linewidth]{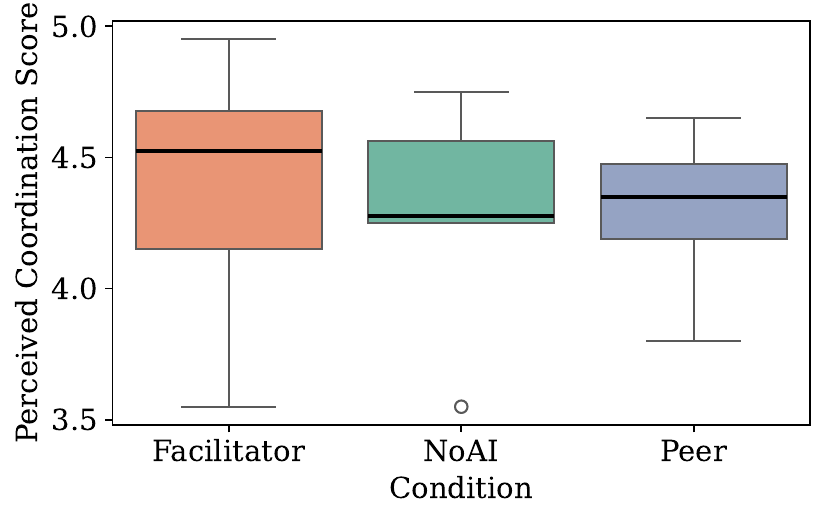}
    \caption{Comparison of Perceived Coordination Survey Responses across Conditions.}
    \label{fig:coordination}
    \Description{}
\end{figure}

\begin{figure*}[t]
  \centering
  \begin{subfigure}[b]{0.49\textwidth}
    \includegraphics[width=\textwidth]{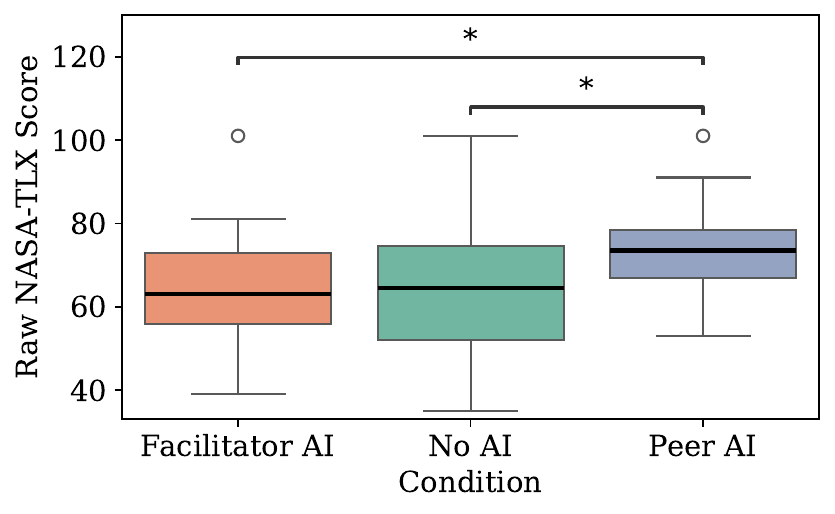}
    \caption{Averaged by AI condition.}
    \label{fig:nasatlx_byAI}
    \Description{A boxplot showing raw NASA-TLX workload scores across three conditions: Facilitator AI, No AI, and Peer AI. Scores range roughly from 40 to 120. The Peer AI condition shows visibly higher median workload than both the Facilitator AI and No AI conditions, with asterisks indicating significant differences.}
  \end{subfigure}
  \hfill
  \begin{subfigure}[b]{0.49\textwidth}
    \includegraphics[width=\textwidth]{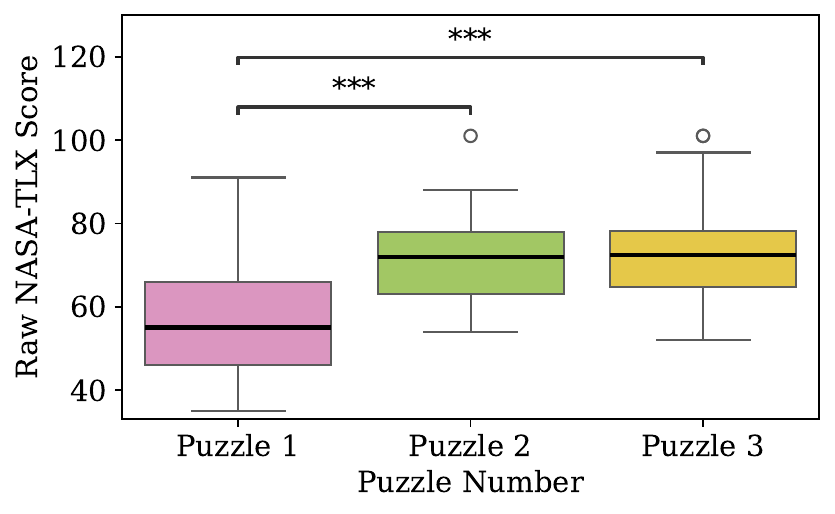}
    \caption{Averaged by puzzle.}
    \label{fig:nasatlx_bypuzzle}
    \Description{A boxplot showing NASA-TLX workload scores for Puzzle 1, Puzzle 2, and Puzzle 3. Puzzle 1 has notably lower workload, while Puzzles 2 and 3 show higher and similar workload distributions. Asterisks mark significant differences between Puzzle 1 and the other puzzles.}
  \end{subfigure}
  \caption{Raw NASA-TLX scores across conditions and puzzles. Asterisks (*) indicate significant pairwise differences based on post hoc comparisons.}
  \label{fig:nasa}
\end{figure*}

\deletedtext[]{We found high perceived group coordination scores throughout the sessions (Figure \ref{fig:coordination}). Across conditions, participants consistently reported that teammates helped each other and coordinated smoothly, with median ratings of 4.0–5.0 on a 5-point scale. In terms of workload, average scores from the NASA-TLX survey revealed higher workload for the peer condition (M = 12.19, SD = 1.96) compared to both the facilitator (M = 10.60, SD = 2.39) and no AI conditions (M = 10.72, SD = 2.68) (Figure \ref{fig:nasa}).}

\deletedtext[]{Based on the AI perception survey results, we found a lower rating for the facilitator compared to the peer for improving overall group score (Md = 2.0, IQR = 1.0–3.0 vs. Md = 2.5, IQR = 2.0–4.0; Figure \ref{fig:perception}). The facilitator also had a lower rating for improving group coordination (Md = 2.0, IQR = 1.75–2.25) compared to the peer (Md = 3.0, IQR = 2.0–3.25). For the peer, groups expressed widely varying perceptions, with ratings spanning from strong disagreement to agreement on whether it improved overall group score. On other dimensions, such as trustworthiness, the differences between facilitator (Md = 3.0, IQR = 2.0–3.25) and peer (Md = 3.0, IQR = 2.0–3.0) were smaller, suggesting that perceptions diverged more strongly around impact on outcomes and coordination.}

\section{Qualitative Findings}
In the following sections we present our findings from our group interviews and observations that describe their experiences with the two generative AI agents during collaborative problem-solving tasks. We conceptualize our results into themes that illustrate the influence of each agent role on their group performance and group processes. An overview of the sub-themes is presented in Figure \ref{fig:qualitative}.


\subsection{Facilitator Agent}


\subsubsection{Theme 1: Provided Early Guidance and Subtle Anchors}

\paragraph{Structuring Early Collaboration}

At the beginning of collaborative tasks, participants approached the facilitator agent with curiosity. Some groups \newtext[2AC]{(3/6)} followed its early suggestions, such as ``look at other screen'', even when these reiterated behaviors they were already engaged in. When the facilitator AI was introduced in the very first session (e.g., groups 4 and 5), its guidance strongly shaped how groups organized themselves. Participants treated its interventions almost like ``rules'', adopting practices such as rotating screens after short intervals and dividing puzzle elements among members. While these structures were not actively reinforced by the AI in later sessions, they often persisted with that workflow. Even participants \newtext[2AC]{(3/24)} in groups that did not have the facilitator in their first session mentioned that such guidance could have helped them orient more quickly. As P4 explained, ``In session one, we totally had no idea about what is the setting\ldots if it could provide something like ‘split the group and focus on screen one,' that would help.''

\paragraph{Anchors for Shared Focus}

Once groups \newtext[2AC]{(6/6)} had begun to establish their own rhythm, the facilitator shifted into a quieter role: not directing the group, but anchoring its focus and maintaining structure when needed. For participants with prior experience in escape room puzzles \newtext[2AC]{(3/24)}, the facilitator's suggestions on work division and team structure were not novel, but were valued as reminders. The agent functioned as a grounding presence as P18 described, ``It was more like a grounding element, like a teacher in the room or like a moderator in an exam room. Like you do your thing, but I'm there.'' This sense of quiet oversight was appreciated, with P18 further noting, ``It gives us clear, you know, what we did, what we talked about, instead of giving us something that can lead us to somewhere else, I think this is the right amount of AI.''

A typical facilitator summary looked like this: ``The team discussed converting given values into a time format (hour, minute, second) and trying different button-press patterns according to arrow directions.'' (Group 5) Sometimes such summaries surfaced useful details that drew collective attention. P14 recalled, ``I remember that moment when the AI cue with the offset 30 popped out and then everyone focused on that and we were able to solve.'' Subtle prompts to divide tasks also offered helpful nudges. As P22 reflected, ``One time in the middle it kind of reminded us to split the task...that actually helped because we were going back and forth together.''


\subsubsection{Theme 2: Misaligned Support Led to Declining Use and Disengagement Over Time}

\paragraph{Limited Impact}

Across groups \newtext[2AC]{(6/6)}, participants felt that the agent had little effect on puzzle performance. Summaries were frequently described as redundant, overly lengthy, or poorly timed—factors that limited their usefulness under time pressure. As P9 put it, ``It was just saying what we already said… So I don't really think the summary helped at all.'' Similarly, P17 explained, ``When I saw the summarization, probably that was like couple minutes ago… by then I already knew which parts I should work on.'' Participants \newtext[2AC]{(10/24)} often emphasized that their existing communication made the AI's inputs unnecessary. 

Many groups \newtext[2AC]{(4/6)} relied more on human teammates than on the agent to coordinate work, with some comparing its role to an unneeded manager. P1 explained, ``I only want to take hints from the AI or use it to remember what I've been saying. But…from a team process level, my teammates are more reliable.'' Once communication patterns solidified, the AI's contributions became increasingly redundant. As P15 noted, ``We already had a good enough communication and collaboration from the first puzzle. So we kind of already solved that facilitation issue.''

Groups \newtext[2AC]{(3/6)} also described their interactions as primarily brainstorming-driven, which clashed with the agent's retrospective style of summarization. P23 summarized this disconnect: ``Our communication was brainstorming, not strategic. The AI summaries didn't spark anything, they just restated what was already there.'' While some \newtext[2AC]{(4/24)} acknowledged that short, targeted summaries might enhance coordination, the implementation was misaligned with the group's fast-paced, improvisational style.

\paragraph{Added Cognitive Burden}

Building on participants' critiques of redundancy, some \newtext[2AC]{(6/24)} described the agent as not just unhelpful but disruptive. In groups 1, 4, and 6, interventions sometimes interfered with momentum rather than supporting it. As P2 reflected, ``Maybe if we were all standing around and doing nothing, then that might have helped. But…I think that actually disrupted our flow.''

Rather than lightening the workload, the agent introduced a subtle distraction when Fiona spoke up to introduce summaries. The two-line summaries required extra attention at moments when participants already felt pressed for time. As P11 explained, ``The time pressure gets me going, not wanting to read the AI… I would rather interact with the person who said that.''

\paragraph{Marginalized Over Time}

We observed that initial curiosity about Fiona's summaries quickly faded as the task progressed. While some participants \newtext[2AC]{(7/24)} glanced at outputs early on, most \newtext[2AC]{(20/24)} reported forgetting or ignoring the agent as time pressure mounted. As P7 described it, ``At some point, I just forget about its existence. I kind of just ignored it.'' By the latter half of puzzles, the agent was largely sidelined, with participants only occasionally glancing at the output if it appeared concise or relevant. This decline was amplified by prior negative experiences with the peer agent, which reduced trust: ``My trust in the AI system was reduced because of the previous round…so I discouraged people from even looking at it.'' (P22)

\subsection{Peer Agent}

\subsubsection{Theme 1: Enhanced Problem Solving by Offering Timely Hints, \replacedtext[R2]{Cognitive}{Memory} Offloading, and Exploratory Support}

\paragraph{Timely Hints Boosted Group Performance}

Unlike the facilitator agent, Ava was remembered for its ability to steer groups toward solutions, especially at moments of impasse. Participants in Group 2 attributed much of their puzzle progress directly to its thoughts that were perceived as hints. As P5 explained, ``Most of the tasks we solved were hints given by the agent… it basically directed us towards correct answers.''  

The usefulness of the peer agent was closely tied to the timing of its contributions. Participants \newtext[2AC]{(9/24)} valued nudges most when momentum had stalled. As P10 summarized, ``We were pursuing an idea for some time and maybe we weren't going anywhere, and that's exactly when it popped up… so it was useful.'' Some participants \newtext[2AC]{(4/24)} also wished that the peer agent had been available in all sessions, ideally in a form that could be invoked on demand. As P22 explained, ``I would have preferred an invokable AI agent…because we were stuck on a bunch of things and it was like maybe just, ‘Do you think that photo frame having three sides is relevant?''' 

\paragraph{Offloaded Memory and Calculation Tasks}

Beyond offering hints, the peer agent was valued as a dependable support for \deletedtext[R2]{lower-level cognitive tasks, helping groups offload} offloading memory and calculation work. Participants used it to handle number-to-letter conversions, decode Morse code, recall prior details, and combine information across screens. As P2 explained, ``A lot of those tasks have a long context… we had to rely on her to remember those things, but humans you probably make mistakes.'' Similarly, P4 emphasized its reliability in arithmetic: ``I was trying to decode...doing the math, but I do it wrong… she do a really good job on remembering.''

Participants \newtext[2AC]{(5/24)} also described how the peer agent supported problem-solving by confirming their thoughts about puzzle elements and providing targeted clarifications. For instance, P12 recalled, ``It did have us identify that it is Morse code and what Morse code was saying.'' P11 noted its usefulness in keeping the group oriented: ``One thing that helps is it actually reminds you some details… guides you back to what you want to focus.'' The system's responses also preserved parallel streams of thought by recording questions and ideas that might otherwise be forgotten. As P4 reflected, ``Maybe everyone has different ideas and it's hard to memorize everyone's idea… with AI probably… that sort of records the idea [when you type out a question].''

\paragraph{Chat Provided Space for Orientation and Exploration}

\newtext[R2]{We examined system log data from the embedded chat interface of the peer agent to understand how often groups engaged with it. Engagement levels varied across groups (M = 11.5, SD = 5.5), with some teams using the peer chat and follow-up prompts frequently while others interacted with it only occasionally.  We further reviewed the recorded study sessions to count how many times individual participants engaged with the chat features. Most participants exhibited low engagement, with 16/24 participants recording only 0–3 interactions, while only 5/24 participants showed high engagement (6–9 interactions). There was an average of 2.83 interactions (SD = 2.53). The dedicated follow-up buttons of “How did you arrive at this idea?” and “Can you suggest another variation?” were used only twice across all six sessions, suggesting that participants primarily engaged with the agent using their own prompts.}

Groups engaged with Ava through text chat in varied ways, using it to orient themselves and probe puzzle elements. Groups 1 and 3 began by asking high-level questions, starting with ``What are we trying to solve?'' In fact, half of the groups interacted with Ava through chat even before its first proactive thought appeared around the three-minute mark.

Most queries \newtext[2AC]{(42/67)} focused on specific puzzle elements and their functions, often phrased with reference to their labels. Participants asked questions like, ``What does the symbol between veg and triad mean?'', ``What are the blinking lights for?'', or ``Can you count the colors for us?'' Groups \newtext[2AC]{(2/6)} occasionally debated an open question before typing it into the chat, sometimes duplicating the same query across both screens.

Chat was also a channel for exploration, where participants sought additional ideas after uncovering new puzzle information. For example, after manipulating grid elements in Puzzle 2 to reveal a heart symbol, P21 asked, ``What do I do with a heart symbol?''

\subsubsection{Theme 2: Misaligned Interactions Disrupted Flow, Added Effort, and Fragmented Communication}

\paragraph{Disrupted Flow and Reactive Engagement}

Although the peer agent sometimes provided useful hints, participants \newtext[2AC]{(14/24)} also described how its outputs disrupted the natural flow of collaboration. Suggestions were often vague, mistimed, or irrelevant to the group's immediate focus. P14 noted, ``Sometimes I see the answers from AI kind of confuse me… it points out which might be related to which, but not in a very specific way.'' Some wished for clearer, more directive cues, with P15 reflecting, ``Whatever hints it gave, we need to interpret… probably it's better if AI just tells you like ‘focus on this part.''' In some cases, verbose input even made tasks feel harder than they were: ``It made it look harder than what it actually was. We fell in the loop… I kind of was disappointed.'' (P14)

The timing of interventions often compounded this disruption, with unsolicited input breaking group momentum. P18 recalled, ``Ava would come out of nowhere and be like, so if you look at this and this means this… and we're like, not right now.'' In several groups \newtext[2AC]{(3/6)}, this dynamic pushed groups into a reactive mode, working collectively in response to AI suggestions rather than dividing tasks or generating independent hypotheses. As a result, collaboration became less self-directed and more tethered to interpreting Ava's outputs.

\paragraph{Cognitive Burden}

Participants \newtext[2AC]{(10/24)} emphasized that interacting with the peer agent often introduced more effort than it saved. Typing prompts and parsing lengthy responses added friction in a fast-paced setting. As P4 explained, ``Thinking in my head is faster than consolidating that and putting it in the prompt.'' Others echoed concerns about the format: ``The outputs right now were very long to process in a time-constrained setup.'' (P8) 

Participants \newtext[2AC]{(6/24)} also struggled to develop a clear mental model of the peer agent, citing unclear roles and capabilities. P9 mentioned, ``Sometimes we don't know how to ask a good question. We don't know this AI, we don't understand its capability.'' Several \newtext[2AC]{(4/24)} noted that their limited experience with the system constrained its usefulness. P6 reflected, ``If we had the AI maybe on the third one, we could interact in a better way, asking more correct questions that could be more efficient.''

\paragraph{Trust Erosion Through Over-Reliance and Unmet Expectations}

Ava's interventions sometimes undermined trust within groups. Participants \newtext[2AC, R1]{(8/24)} described how its confident outputs encouraged reliance without offering reasoning, which in turn reduced group-led problem solving. As P14 admitted, ``It gave us information… but the issue is there was no reasoning behind it. It just gave us the final output, and we didn't know if it was right. I just trusted her completely---I thought we were dumb to understand.'' When such outputs proved unhelpful, the result was disappointment. P23 reflected, ``I expected it to come up with something that we have not thought about... but it didn't.'' The gap between expectations of an ``intelligent'' peer and the reality of inconsistent support eroded participants' confidence in the system. As P22 summarized, ``I think since most people didn't trust it a lot, so it didn't really add to collaboration.''

\paragraph{Siloed Communication and Fragmented Awareness}

Compounding issues of cognitive burden and trust, the peer agent sometimes reshaped communication in ways that fragmented teamwork. As chat-based exchanges were not voiced aloud, it resulted in confusion and redundancy. As P1 described, ``There was one time where… people had individually been talking to Ava… but without like talking out loud… then we were like, ‘oh wait, what did Ava say?''' These private interactions created parallel conversations that left some participants out of the loop.

Several participants \newtext[2AC]{(7/24)} admitted that engaging with the AI reduced their group involvement. P16 reflected, ``For me personally, I think this was the one that I was less engaged with the others… I spoke very less during this puzzle than during any of the other.'' Others noted that typing to Ava pulled them away from shared context. This dynamic was described as akin to interacting with a ``fifth teammate,'' but one that fragmented rather than enriched group collaboration. P3 explained, ``It definitely changes the dynamic… you have this fifth person that you interact with kind of alone.'' 

\subsubsection{Theme 3: Varied Trajectories of Agent Use Showing Contrasting Patterns of Enthusiasm, Reliance, and Disengagement}


\paragraph{Reliance on Hints Giving Way to Positive Reflections} In the first trajectory, we observed that groups 2 and 3 began with strong reliance on Ava's hints. They treated its suggestions as decisive, turning to its ``thoughts'' when stuck. Ava often directed them toward answers, and the group followed its lead. They continued to engage with the peer throughout the session. In the focus group, these groups reflected that more deliberate engagement might have helped them collaborate effectively. As P6 recalled, ``\ldots we could have used the agent more, and the conclusion after the first experiment was that we should use the agent more. But the problem was in the next two [sessions] we didn't end up having agent.'' 

\paragraph{From Early Enthusiasm to Dependence and Disillusionment} In the second trajectory, groups 1 and 4 eagerly embraced the AI as a helpful partner, treating its early suggestions as breakthroughs. But this excitement soon shifted into dependence, with participants waiting for AI thoughts and query responses and interacting less with one another. When the AI's reasoning later fell short, the over-reliance turned into frustration and disappointment. For example, P3 described how, when Ava's first proactive thought surfaced, it seemed strikingly clever compared to the group's reasoning, prompting him to shift his attention toward the AI. But when a teammate solved the same sub-puzzle with a simpler approach, he lost trust in Ava altogether.

\paragraph{From Initial Curiosity to Frustration and Eventual Disengagement} The third trajectory reflected a sharper arc from curiosity to disengagement. At first, participants from groups 5 and 6 framed the AI as a potential teammate and actively explored its capabilities. Their expectations were high, and they experimented with asking questions to probe its usefulness. However, unfamiliarity with its limits soon led to confusion, and poorly timed or repetitive outputs disrupted the flow of collaboration. For example, in Group 6, P23 initially paid close attention to Ava's proactive thoughts, but when several suggestions repeated information the group had already used to solve an earlier sub-puzzle, she gradually stopped engaging with them. As such frustrations mounted, these groups redirected their attention to one another, gradually sidelining the AI and disengaging from it.

\section{Discussion}

\newtext[2AC]{In this section, we synthesize our quantitative and qualitative findings to answer the research questions and provide design considerations for future AI agents. First, we examine how the facilitator and peer roles shaped performance and participants’ perceptions. Next, we describe how these roles influenced coordination, communication, and workload during the task. Finally, we outline concrete design considerations grounded in the features that shaped teams’ experiences.}

\subsection{RQ1: How did different AI agent roles (peer vs. facilitator) influence group performance in co-located, time-sensitive problem-solving tasks?}

\newtext[R2]{Our findings reveal a disconnect between how participants perceived the AI agents’ contributions and the actual performance outcomes across conditions. Objectively, groups performed best with the facilitator agent, followed by the no AI condition, and worst with the peer agent. AI Condition had a significant effect on performance, and post hoc comparisons showed that the facilitator condition outperformed the peer condition. Puzzle difficulty also impacted scores, with groups scoring significantly higher in Puzzle 1 than Puzzles 2 and 3.}

\newtext[R2]{Interestingly, in the focus group interviews, participants never credited the facilitator agent’s (Fiona’s) features for their higher scores. Groups often described its periodic summaries and coordination nudges as redundant or poorly timed under pressure.  The AI Perception survey reflected a similar pattern: participants did not significantly favor either agent for improving their team’s score, though the peer agent was rated more positively on average. The peer agent (Ava) actually evoked polarized reactions. Some groups thought that they advanced because Ava offered timely ideas, memory support, and quick calculations, while others felt its unsolicited ``thoughts'' disrupted flow or slowed progress.}

\deletedtext[]{Across sessions, the facilitator condition produced the highest average puzzle scores (Fig.\ref{fig:scores_byAI}), yet participants never credited Fiona's features for this improvement (Fig.\ref{fig:perception}). Groups often described its periodic summaries and coordination nudges as redundant or poorly timed under pressure. By contrast, the peer agent (Ava) evoked polarized reactions. Some groups advanced because Ava offered timely ideas, memory support, and quick calculations, while others felt its unsolicited ``thoughts'' disrupted flow or slowed progress. Puzzle difficulty (Fig.~\ref{fig:scores_bypuzzle}) and group familiarity with the escape-room format also shaped how both agents were received.}

\newtext[R2]{This misalignment arises from how differently the two agents’ contributions surfaced during the task. The peer agent’s messages were immediate and easy to notice, sometimes intervening when groups felt stuck. This made its help feel direct and impactful, even when the suggestions were imperfect or added to participants’ workload. However, the facilitator’s summaries and coordination cues blended into the ongoing discussion. They supported groups in indirect ways like helping maintain orientation, but were rarely experienced as actionable help in the moment. As a result, participants overlooked the facilitator’s role in their success, while the peer felt more influential despite its lower overall performance outcomes.}

The results also reflect the nature of our task environment: short, co-located, and tightly interdependent tasks under strict time pressure. In such conditions, groups must quickly test ideas and converge on promising ones. As prior work shows, AI suggestions are more likely to be adopted when decision time is longer \cite{cao_how_2023}. Time pressure made participants less receptive to global summaries or rigid scaffolds, which may be more effective in open-ended ideation or distributed work \cite{imamura_serendipity_2024, clark_supersizing_2010}.

The peer role both helped and hindered. It \highlight{enabled progress} but also \highlight{anchored groups on its suggestions}. Prior work shows that groups often defer to AI more than individuals do \cite{chiang_are_2023}, with some members defending its advice or using it as a tie-breaker under load. At the same time, groups can push back when at least one member has strong contrary evidence \cite{chiang_enhancing_2024}. This explains our mixed results: Ava sometimes catalyzed sensemaking, but at other times derailed parallel problem-solving when no one challenged its ideas.

The contrast between roles highlights different challenges. Fiona's metacognitive scaffolds plausibly \highlight{supported coordination} \cite{reicherts_ai_2025, nussbaum_technology_2009} and may explain higher scores, but participants often experienced them as \highlight{invisible or repetitive}, since groups already shared knowledge aloud in a co-located setting. Previous research has also found that facilitation around group structures and discussion scaffolds lacks authority, which can cause them to be overlooked \cite{liu_peergpt_2024}. Ava’s contributions were more visible, resembling the kinds of behaviors expected from a teammate. However, when timing or topical fit was off, Ava risked \highlight{crowding out conversation}, echoing findings that proactive AI teammates can overwhelm group discussions \cite{memmert_towards_2023, wieland_electronic_2022}.


\subsection{RQ2: How did the AI agent roles shape group processes such as workload, communication, and coordination in co-located, time-sensitive problem-solving tasks?}

\newtext[R2]{Workload measured using NASA-TLX was significantly higher in the peer condition than in both the facilitator and no AI conditions, which did not differ from each other. This suggests that any offloading the peer provided was outweighed by the additional interaction and monitoring demands it introduced.}
\deletedtext[]{NASA-TLX scores were higher with the peer than with the facilitator or no-AI conditions (Fig. \ref{fig:nasa}), suggesting that any offloading it provided was outweighed by interaction and monitoring costs.} The facilitator sat lightly on the conversation, offering time prompts and summaries that many groups ignored but did not find disruptive. By contrast, the peer behaved more like a ``fifth teammate,'' injecting ideas that sometimes aided sensemaking but also diverted attention into an AI-centered side channel. Because its contributions had to be read, checked, and often queried under time pressure, the peer added to participants' workload. Prior work similarly shows that proactive, talkative agents can overwhelm groups unless their initiative is tightly governed \cite{shaer_ai-augmented_2024}; Houde et al. also found that frequent or lengthy posts distorted discussion and argued that groups should be able to control when, what, and where an agent contributes \cite{houde_controlling_2025}. Our findings mirror these concerns: unmanaged initiative increased attention switching and cognitive load.

Communication patterns diverged by role. The facilitator rarely intruded. Its summaries sometimes \highlight{helped as an anchor}, but when too long or mistimed, they \highlight{felt redundant to fast and ad hoc brainstorming}. Systems like LADICA explain this tension \cite{zhang_ladica_2025}. They aim to foster mutual awareness on shared displays while avoiding dominance of the human–human discussion, and they caution against features that over-clutter or steer the flow. By contrast, the peer often redirected attention to private chats with the agent, creating silos. Johnson et al. surface this as a design tension around social prominence and engagement: groups want augmentation, but they worry that agent channels will split attention and disrupt the shared workspace \cite{johnson_exploring_2025}. Our findings match the generative AI-centered interaction pattern found by Feng et al., where students engaged considerably more with the chatbot than with their peers during the collaborative problem-solving process \cite{feng_group_2025}.

Groups reported strong coordination across all conditions, possibly because members knew each other from before (Fig. \ref{fig:coordination}). Coordination remained human-led in most sessions. Groups divided work and set rhythm with each other, \highlight{treating agent input as optional}. However, early facilitator nudges sometimes stuck. When the facilitator condition appeared first, its suggestions (for example, divide-and-rotate or brief individual reflection before group synthesis) helped groups set a structure that persisted. \newtext[R2]{However, participants perceived the peer agent as contributing significantly more to their coordination, likely because its proactive thoughts and query responses pulled the team together to interpret and act on them.} It sometimes shifted coordination by nudging groups away from dividing tasks and toward more joint problem-solving. In several cases, \highlight{groups worked reactively around the peer's responses, staying together rather than splitting work}. 

\subsection{Design Considerations for Proactive Generative AI Agents in Colocated Time-Sensitive Problem-Solving Tasks}

Our study revealed several patterns of user interaction with the AI agents, highlighting both opportunities and risks. \newtext[R1]{Before turning to design considerations, we note that our findings reflect the interaction of both role design and the specific features used to instantiate those roles. To avoid overgeneralizing about “roles”, our design implications focus on the features that shaped these experiences and that can inform the design of future agents.} Building on participants' suggestions and prior work, we outline concrete design considerations for proactive AI agents that can support colocated, time-sensitive, collaborative problem solving.


\emph{Summaries} and \emph{coordination cues} offered by an AI agent to guide the collaborative problem-solving task emerged as key design features. However, these same outputs risked marginalization when they became overly long, repetitive, or poorly timed. This suggests that AI agent contributions must be designed to remain concise, progress-aware, and embedded in the shared workspace rather than appearing as standalone text. Their utility increases when paired with actionable next steps grounded in the group's ongoing discussion, which participants described as critical for sustaining progress. In addition to text-based support, participants emphasized the importance of visual features that act as ``external memory'', such as progress indicators or expandable cards that align with the task flow while minimizing reading overhead. These preferences resonate with prior findings on large-display systems, where visual artifacts effectively scaffold group awareness without disrupting conversation \cite{zhang_ladica_2025}. 



Another important consideration was the design of \emph{ thoughts or suggestions} from the agent. For proactive input to be useful in time-bounded collaboration, suggestions should be kept short and accompanied by clear rationales to build user trust. Additionally, when appropriate, it may be helpful to offer multiple alternatives rather than a single idea. Such designs may help mitigate over-anchoring, promote comparative reasoning, and support more deliberate group decision-making. This also supports prior arguments that systems should intervene when users show signs of over-reliance while still respecting group autonomy and pace \cite{johnson2025collaboration}. 

An agent's \emph{social presence} and \emph{influence} also warrant careful design. Maintaining a moderate presence: visible but peripheral, and offering support without disrupting conversation or dominating the group's attention, can be helpful for sustaining engagement.
Participants also emphasized the value of user controls for agent initiative. Adjustable mechanisms, such as rate limits, silence thresholds, or options like ``never speak first'', ``pause'', or a quick ``volume'' dial, could give groups flexibility in determing how and when the agent participates. Such features align with broader human-centered AI principles calling for transparency and controllability in agent interjections \cite{houde_controlling_2025, wang_adaptive_2025}. This is also line with theories of social influence suggest that preventing AI from becoming the de facto authority helps preserve critical engagement and reduces risks of social loafing or conformity \cite{kelman2006interests}.

Finally, \emph{timing} and \emph{relevance} of AI agent's contributions proved crucial, with poorly timed or off-topic input often leading groups to abandon its support. Participants suggested that timing could be improved through activity-based triggers that respond to cues such as silence, stalls, or bursts of talk, ensuring interventions occur at appropriate moments. Relevance, in turn, could be strengthened through tighter integration into the workspace, such as anchoring suggestions to specific visual elements. Together, these features could reduce the cognitive cost of shifting attention between the agent and the task, making contributions easier to interpret in context.




\section{Limitations and Future Work}

While our study provides encouraging insights into how proactive generative AI can participate in real-time teamwork, several limitations also point to valuable directions for future research. 
First, we recruited participants through a convenience sample pool who already knew and worked closely with one another. While this familiarity likely influenced group communication and coordination, such dynamics are common in many real-world group settings, for example, in emergency response units, clinical care teams, and workplace project groups. Studying these established teams provided insight into how proactive AI agents integrate into pre-existing social dynamics. Future work can extend this approach to ad hoc teams or cross-organizational collaborations where roles and norms are less established.

\emptext[R1]{Second, our work focused on a specific collaborative setting---digital escape-room puzzles with small groups in co-located, time-sensitive conditions. While this testbed offered a controlled yet engaging way to observe group communication and interaction with AI agents, the puzzles themselves consisted of relatively simple, self-contained sub-tasks. Such tasks may not fully reflect the nature of real-world collaborative work. High-stakes domains such as healthcare, emergency response, or crisis management involve complex workflows, domain knowledge, and longer time horizons than the discrete puzzle elements used in our study. As a result, our findings may not directly generalize to these settings. Expanding into these more representative environments will be important for testing the robustness of our results and understanding how proactive AI adapts to more varied and interdependent teamwork.}

\newtext[R1]{Third, the effects we observed came not only from the facilitator and peer roles but also from the specific features used to represent them. The facilitator mainly gave summaries and reminders, while the peer offered short ideas, and both agents intervened at fixed times. These design choices shaped how teams perceived and used each agent. Future work could refine and examine each design feature on its own so that its specific effect on teamwork can be understood. These insights can then guide the development of effective facilitator and peer agents that better support co-located teamwork.}

\emptext[2AC]{Finally, each agent performed a single, non-adaptive role in our study. Because neither agent adapted to the group’s progress, needs, or experience level, their influence on teamwork was shaped by these fixed behaviors. Our participants themselves noted that agents should adapt their role to the session phase and group experience: early on, offering more process guidance like a facilitator, and later, shifting to on-demand, minimal cues like a peer. Prior work has shown the importance of such signals for detecting disengagement, over-reliance, or social loafing \cite{kohn2011collaborative}. Future work can focus on developing such context-aware and adaptive mechanisms that will move proactive agents closer to being integrated teammates who support evolving group needs \cite{wang_adaptive_2025}.}


\section{Conclusion}
Our work presents an early but promising step toward understanding how proactive generative AI agents can enrich real-time, co-located teamwork. By comparing two distinct roles, a facilitator that provided summaries and team structure cues, and a peer that contributed ideas and memory support, we examined how proactive AI influences not only task performance but also group processes such as workload, coordination, and communication.
Our findings show that facilitators initially captured attention but were often sidelined when their input became lengthy or poorly timed, while peer agents generated more varied trajectories. Some groups used peer contributions to move from reliance toward more reflective engagement; others shifted from enthusiasm to dependence and disillusionment. These patterns reveal both the promise and the fragility of proactive support in high-pressure collaboration.
Taken together, our results highlight that the value of proactive generative AI lies not in static roles but in the ability to adapt---providing the right kind of support at the right moment. Designing such adaptable agents opens a pathway toward AI that participates as a trusted teammate, flexibly balancing task and process contributions to strengthen human collaboration in diverse and time-sensitive domains.


\begin{acks}
We thank Dr. Teruhisa Misu and Dr. Kurt Luther for their thoughtful feedback and guidance throughout this work. The first author was partially supported by the Virginia Commonwealth Cyber Initiative. We also sincerely thank the study participants for sharing their time and valuable insights.
\end{acks}

\bibliographystyle{ACM-Reference-Format}
\bibliography{sample-base,mylibrary,chi26-paper, chi26-related-work, chi26-revision}

@article{mcneese2021my,
  title={Who/what is my teammate? Team composition considerations in human--AI teaming},
  author={McNeese, Nathan J and Schelble, Beau G and Canonico, Lorenzo Barberis and Demir, Mustafa},
  journal={IEEE Transactions on Human-Machine Systems},
  volume={51},
  number={4},
  pages={288--299},
  year={2021},
  publisher={IEEE}
}

@article{tesluk1999overcoming,
  title={Overcoming roadblocks to effectiveness: Incorporating management of performance barriers into models of work group effectiveness.},
  author={Tesluk, Paul E and Mathieu, John E},
  journal={Journal of applied Psychology},
  volume={84},
  number={2},
  pages={200},
  year={1999},
  publisher={American Psychological Association}
}

@article{kim_ai_2021,
	title = {{AI} as a friend or assistant: {The} mediating role of perceived usefulness in social {AI} vs. functional {AI}},
	volume = {64},
	issn = {0736-5853},
	shorttitle = {{AI} as a friend or assistant},
	url = {https://www.sciencedirect.com/science/article/pii/S0736585321001337},
	doi = {10.1016/j.tele.2021.101694},
	urldate = {2025-08-26},
	journal = {Telematics and Informatics},
	author = {Kim, Jihyun and Merrill Jr., Kelly and Collins, Chad},
	month = nov,
	year = {2021},
	pages = {101694},
}

@article{zercher2023ai,
  title={When AI joins the team: a literature review on intragroup processes and their effect on team performance in team-AI collaboration},
  author={Zercher, D{\'e}sir{\'e}e and Jussupow, Ekaterina and Heinzl, Armin},
  year={2023}
}

@inproceedings{muller2024group,
  title={Group brainstorming with an ai agent: Creating and selecting ideas},
  author={Muller, Michael and Houde, Stephanie and Gonzalez, Gabriel and Brimijoin, Kristina and Ross, Steven I and Moran, Dario Andres Silva and Weisz, Justin D},
  booktitle={International conference on computational creativity},
  pages={10},
  year={2024}
}

@inproceedings{johnson2025collaboration,
  author    = {Janet G. Johnson and Steven R. Rick},
  title     = {The Promise and Peril of Collaboration: Fostering Appropriate Reliance When Problem-Solving with GenAI},
  booktitle = {CHI '25 Workshop on Tools for Thought: Research and Design for Understanding, Protecting, and Augmenting Human Cognition with Generative AI},
  year      = {2025},
  address   = {Yokohama, Japan}
}

@article{kelman2006interests,
  title={Interests, relationships, identities: Three central issues for individuals and groups in negotiating their social environment},
  author={Kelman, Herbert C},
  journal={Annu. Rev. Psychol.},
  volume={57},
  number={1},
  pages={1--26},
  year={2006},
  publisher={Annual Reviews}
}

@article{kohn2011collaborative,
  title={Collaborative fixation: Effects of others' ideas on brainstorming},
  author={Kohn, Nicholas W and Smith, Steven M},
  journal={Applied Cognitive Psychology},
  volume={25},
  number={3},
  pages={359--371},
  year={2011},
  publisher={Wiley Online Library}
}

@book{shneiderman2022human,
  title={Human-centered AI},
  author={Shneiderman, Ben},
  year={2022},
  publisher={Oxford University Press}
}

@inproceedings{hwang2024whose,
  title={In whose voice?: examining AI agent representation of people in social interaction through generative speech},
  author={Hwang, Angel Hsing-Chi and Siy, John Oliver and Shelby, Renee and Lentz, Alison},
  booktitle={Proceedings of the 2024 ACM Designing Interactive Systems Conference},
  pages={224--245},
  year={2024}
}

@article{samadi2024ai,
  title={The AI collaborator: Bridging human-AI interaction in educational and professional settings},
  author={Samadi, Mohammad Amin and JaQuay, Spencer and Gu, Jing and Nixon, Nia},
  journal={arXiv preprint arXiv:2405.10460},
  year={2024}
}

@article{cao_how_2023,
	title = {How {Time} {Pressure} in {Different} {Phases} of {Decision}-{Making} {Influences} {Human}-{AI} {Collaboration}},
	volume = {7},
	issn = {2573-0142},
	url = {https://dl.acm.org/doi/10.1145/3610068},
	doi = {10.1145/3610068},
	language = {en},
	number = {CSCW2},
	urldate = {2025-08-27},
	journal = {Proceedings of the ACM on Human-Computer Interaction},
	author = {Cao, Shiye and Gomez, Catalina and Huang, Chien-Ming},
	month = sep,
	year = {2023},
	pages = {1--26},
}

@article{oneill_humanautonomy_2022,
	title = {Human–{Autonomy} {Teaming}: {A} {Review} and {Analysis} of the {Empirical} {Literature}},
	volume = {64},
	issn = {0018-7208},
	shorttitle = {Human–{Autonomy} {Teaming}},
	url = {https://doi.org/10.1177/0018720820960865},
	doi = {10.1177/0018720820960865},
	number = {5},
	urldate = {2025-08-27},
	journal = {Human Factors},
	author = {O’Neill, Thomas and McNeese, Nathan and Barron, Amy and Schelble, Beau},
	month = aug,
	year = {2022},
	note = {Publisher: SAGE Publications Inc},
	pages = {904--938},
}

@inproceedings{imamura_serendipity_2024,
	address = {New York, NY, USA},
	series = {{AHs} '24},
	title = {Serendipity {Wall}: {A} {Discussion} {Support} {System} {Using} {Real}-time {Speech} {Recognition} and {Large} {Language} {Model}},
	isbn = {9798400709807},
	shorttitle = {Serendipity {Wall}},
	url = {https://doi.org/10.1145/3652920.3652931},
	doi = {10.1145/3652920.3652931},
	urldate = {2025-08-27},
	booktitle = {Proceedings of the {Augmented} {Humans} {International} {Conference} 2024},
	publisher = {Association for Computing Machinery},
	author = {Imamura, Shota and Hiraki, Hirotaka and Rekimoto, Jun},
	month = may,
	year = {2024},
	pages = {237--247},
}

@article{oszabo_anatomy_2022,
	title = {The anatomy of social dynamics in escape rooms},
	volume = {12},
	copyright = {2022 The Author(s)},
	issn = {2045-2322},
	url = {https://www.nature.com/articles/s41598-022-13929-0},
	doi = {10.1038/s41598-022-13929-0},
	language = {en},
	number = {1},
	urldate = {2025-08-27},
	journal = {Scientific Reports},
	author = {O'Szabo, Rebeka and Chowdhary, Sandeep and Deritei, David and Battiston, Federico},
	month = jun,
	year = {2022},
	note = {Publisher: Nature Publishing Group},
	pages = {10498},
}

@article{cohen_using_2020,
	title = {Using {Escape} {Rooms} for {Conducting} {Team} {Research}: {Understanding} {Development}, {Considerations}, and {Challenges}},
	volume = {51},
	issn = {1046-8781, 1552-826X},
	shorttitle = {Using {Escape} {Rooms} for {Conducting} {Team} {Research}},
	url = {https://journals.sagepub.com/doi/10.1177/1046878120907943},
	doi = {10.1177/1046878120907943},
	language = {en},
	number = {4},
	urldate = {2025-08-27},
	journal = {Simulation \& Gaming},
	author = {Cohen, Tara N. and Griggs, Andrew C. and Keebler, Joseph R. and Lazzara, Elizabeth H. and Doherty, Shawn M. and Kanji, Falisha F. and Gewertz, Bruce L.},
	month = aug,
	year = {2020},
	pages = {443--460},
}

@article{ilgen_teams_2005,
	title = {Teams in {Organizations}: {From} {Input}-{Process}-{Output} {Models} to {IMOI} {Models}},
	volume = {56},
	issn = {0066-4308, 1545-2085},
	shorttitle = {Teams in {Organizations}},
	url = {https://www.annualreviews.org/content/journals/10.1146/annurev.psych.56.091103.070250},
	doi = {10.1146/annurev.psych.56.091103.070250},
	language = {en},
	number = {Volume 56, 2005},
	urldate = {2025-08-27},
	journal = {Annual Review of Psychology},
	author = {Ilgen, Daniel R. and Hollenbeck, John R. and Johnson, Michael and Jundt, Dustin},
	month = feb,
	year = {2005},
	note = {Publisher: Annual Reviews},
	pages = {517--543},
}

@article{schelble_lets_2022,
	title = {Let's {Think} {Together}! {Assessing} {Shared} {Mental} {Models}, {Performance}, and {Trust} in {Human}-{Agent} {Teams}},
	volume = {6},
	issn = {2573-0142},
	url = {https://dl.acm.org/doi/10.1145/3492832},
	doi = {10.1145/3492832},
	language = {en},
	number = {GROUP},
	urldate = {2025-08-28},
	journal = {Proceedings of the ACM on Human-Computer Interaction},
	author = {Schelble, Beau G. and Flathmann, Christopher and McNeese, Nathan J. and Freeman, Guo and Mallick, Rohit},
	month = jan,
	year = {2022},
	pages = {1--29},
}

@inproceedings{han_when_2024,
	address = {Honolulu HI USA},
	title = {When {Teams} {Embrace} {AI}: {Human} {Collaboration} {Strategies} in {Generative} {Prompting} in a {Creative} {Design} {Task}},
	isbn = {9798400703300},
	shorttitle = {When {Teams} {Embrace} {AI}},
	url = {https://dl.acm.org/doi/10.1145/3613904.3642133},
	doi = {10.1145/3613904.3642133},
	language = {en},
	urldate = {2025-08-28},
	booktitle = {Proceedings of the {CHI} {Conference} on {Human} {Factors} in {Computing} {Systems}},
	publisher = {ACM},
	author = {Han, Yuanning and Qiu, Ziyi and Cheng, Jiale and Lc, Ray},
	month = may,
	year = {2024},
	pages = {1--14},
}

@inproceedings{chiang_are_2023,
	address = {New York, NY, USA},
	series = {{CHI} '23},
	title = {Are {Two} {Heads} {Better} {Than} {One} in {AI}-{Assisted} {Decision} {Making}? {Comparing} the {Behavior} and {Performance} of {Groups} and {Individuals} in {Human}-{AI} {Collaborative} {Recidivism} {Risk} {Assessment}},
	isbn = {978-1-4503-9421-5},
	shorttitle = {Are {Two} {Heads} {Better} {Than} {One} in {AI}-{Assisted} {Decision} {Making}?},
	url = {https://dl.acm.org/doi/10.1145/3544548.3581015},
	doi = {10.1145/3544548.3581015},
	urldate = {2025-08-27},
	booktitle = {Proceedings of the 2023 {CHI} {Conference} on {Human} {Factors} in {Computing} {Systems}},
	publisher = {Association for Computing Machinery},
	author = {Chiang, Chun-Wei and Lu, Zhuoran and Li, Zhuoyan and Yin, Ming},
	month = apr,
	year = {2023},
	pages = {1--18},
}

@misc{li_generative_2024,
	title = {Generative {AI} {Enhances} {Team} {Performance} and {Reduces} {Need} for {Traditional} {Teams}},
	url = {http://arxiv.org/abs/2405.17924},
	doi = {10.48550/arXiv.2405.17924},
	urldate = {2025-08-28},
	publisher = {arXiv},
	author = {Li, Ning and Zhou, Huaikang and Mikel-Hong, Kris},
	month = may,
	year = {2024},
	note = {arXiv:2405.17924 [cs]},
}

@inproceedings{reicherts_ai_2025,
	address = {Yokohama Japan},
	title = {{AI}, {Help} {Me} {Think}—but for {Myself}: {Assisting} {People} in {Complex} {Decision}-{Making} by {Providing} {Different} {Kinds} of {Cognitive} {Support}},
	isbn = {9798400713941},
	shorttitle = {{AI}, {Help} {Me} {Think}—but for {Myself}},
	url = {https://dl.acm.org/doi/10.1145/3706598.3713295},
	doi = {10.1145/3706598.3713295},
	language = {en},
	urldate = {2025-08-28},
	booktitle = {Proceedings of the 2025 {CHI} {Conference} on {Human} {Factors} in {Computing} {Systems}},
	publisher = {ACM},
	author = {Reicherts, Leon and Zhang, Zelun Tony and Von Oswald, Elisabeth and Liu, Yuanting and Rogers, Yvonne and Hassib, Mariam},
	month = apr,
	year = {2025},
	pages = {1--19},
}

@misc{johnson_exploring_2025,
	title = {Exploring {Collaborative} {GenAI} {Agents} in {Synchronous} {Group} {Settings}: {Eliciting} {Team} {Perceptions} and {Design} {Considerations} for the {Future} of {Work}},
	shorttitle = {Exploring {Collaborative} {GenAI} {Agents} in {Synchronous} {Group} {Settings}},
	url = {http://arxiv.org/abs/2504.14779},
	doi = {10.48550/arXiv.2504.14779},
	urldate = {2025-08-29},
	publisher = {arXiv},
	author = {Johnson, Janet G. and Peralta, Macarena and Kaur, Mansanjam and Huang, Ruijie Sophia and Zhao, Sheng and Guan, Ruijia and Rajaram, Shwetha and Nebeling, Michael},
	month = apr,
	year = {2025},
	note = {arXiv:2504.14779 [cs]},
}

@inproceedings{shaer_ai-augmented_2024,
	address = {Honolulu HI USA},
	title = {{AI}-{Augmented} {Brainwriting}: {Investigating} the use of {LLMs} in group ideation},
	isbn = {9798400703300},
	shorttitle = {{AI}-{Augmented} {Brainwriting}},
	url = {https://dl.acm.org/doi/10.1145/3613904.3642414},
	doi = {10.1145/3613904.3642414},
	language = {en},
	urldate = {2025-08-31},
	booktitle = {Proceedings of the {CHI} {Conference} on {Human} {Factors} in {Computing} {Systems}},
	publisher = {ACM},
	author = {Shaer, Orit and Cooper, Angelora and Mokryn, Osnat and Kun, Andrew L and Ben Shoshan, Hagit},
	month = may,
	year = {2024},
	pages = {1--17},
}

@inproceedings{zhang_ladica_2025,
	address = {Yokohama Japan},
	title = {{LADICA}: {A} {Large} {Shared} {Display} {Interface} for {Generative} {AI} {Cognitive} {Assistance} in {Co}-located {Team} {Collaboration}},
	isbn = {9798400713941},
	shorttitle = {{LADICA}},
	url = {https://dl.acm.org/doi/10.1145/3706598.3713289},
	doi = {10.1145/3706598.3713289},
	language = {en},
	urldate = {2025-08-31},
	booktitle = {Proceedings of the 2025 {CHI} {Conference} on {Human} {Factors} in {Computing} {Systems}},
	publisher = {ACM},
	author = {Zhang, Zheng and Peng, Weirui and Chen, Xinyue and Cao, Luke and Li, Toby Jia-Jun},
	month = apr,
	year = {2025},
	pages = {1--22},
}

@article{feng_group_2025,
	title = {Group interaction patterns in generative {AI}-supported collaborative problem solving: {Network} analysis of the interactions among students and a {GAI} chatbot},
	volume = {56},
	issn = {1467-8535},
	shorttitle = {Group interaction patterns in generative {AI}-supported collaborative problem solving},
	url = {https://onlinelibrary.wiley.com/doi/abs/10.1111/bjet.13611},
	doi = {10.1111/bjet.13611},
	language = {en},
	number = {5},
	urldate = {2025-09-01},
	journal = {British Journal of Educational Technology},
	author = {Feng, Shihui},
	year = {2025},
	note = {\_eprint: https://bera-journals.onlinelibrary.wiley.com/doi/pdf/10.1111/bjet.13611},
	pages = {2125--2145},
}

@book{clark_supersizing_2010,
	title = {Supersizing the {Mind}: {Embodiment}, {Action}, and {Cognitive} {Extension}},
	isbn = {978-0-19-983104-3},
	shorttitle = {Supersizing the {Mind}},
	language = {en},
	publisher = {Oxford University Press},
	author = {Clark, Andy},
	month = dec,
	year = {2010},
}

@article{nussbaum_technology_2009,
	title = {Technology as small group face-to-face {Collaborative} {Scaffolding}},
	volume = {52},
	issn = {0360-1315},
	url = {https://www.sciencedirect.com/science/article/pii/S0360131508001073},
	doi = {10.1016/j.compedu.2008.07.005},
	number = {1},
	urldate = {2025-09-05},
	journal = {Computers \& Education},
	author = {Nussbaum, Miguel and Alvarez, Claudio and McFarlane, Angela and Gomez, Florencia and Claro, Susana and Radovic, Darinka},
	month = jan,
	year = {2009},
	pages = {147--153},
}

@article{memmert_towards_2023,
  title={Towards human-AI-collaboration in brainstorming: Empirical insights into the perception of working with a generative AI},
  author={Memmert, Lucas and Tavanapour, Navid},
  year={2023}
}

@article{wieland_electronic_2022,
	title = {Electronic {Brainstorming} {With} a {Chatbot} {Partner}: {A} {Good} {Idea} {Due} to {Increased} {Productivity} and {Idea} {Diversity}},
	volume = {5},
	issn = {2624-8212},
	shorttitle = {Electronic {Brainstorming} {With} a {Chatbot} {Partner}},
	url = {https://www.frontiersin.org/journals/artificial-intelligence/articles/10.3389/frai.2022.880673/full},
	doi = {10.3389/frai.2022.880673},
	language = {English},
	urldate = {2025-09-05},
	journal = {Frontiers in Artificial Intelligence},
	author = {Wieland, Britt and de Wit, Jan and de Rooij, Alwin},
	month = sep,
	year = {2022},
	note = {Publisher: Frontiers},
}

@article{braun_reflecting_2019,
	title = {Reflecting on reflexive thematic analysis},
	volume = {11},
	issn = {2159-676X},
	url = {https://doi.org/10.1080/2159676X.2019.1628806},
	doi = {10.1080/2159676X.2019.1628806},
	number = {4},
	urldate = {2025-09-06},
	journal = {Qualitative Research in Sport, Exercise and Health},
	author = {Braun, Virginia and Clarke, Victoria},
	month = aug,
	year = {2019},
	note = {Publisher: Routledge
\_eprint: https://doi.org/10.1080/2159676X.2019.1628806},
	pages = {589--597},
}

@article{sidji_human-ai_2024,
	title = {Human-{AI} {Collaboration} in {Cooperative} {Games}: {A} {Study} of {Playing} {Codenames} with an {LLM} {Assistant}},
	volume = {8},
	shorttitle = {Human-{AI} {Collaboration} in {Cooperative} {Games}},
	url = {https://dl.acm.org/doi/10.1145/3677081},
	doi = {10.1145/3677081},
	number = {CHI PLAY},
	urldate = {2025-09-06},
	journal = {Proc. ACM Hum.-Comput. Interact.},
	author = {Sidji, Matthew and Smith, Wally and Rogerson, Melissa J.},
	month = oct,
	year = {2024},
	pages = {316:1--316:25},
}

@article{woolley_evidence_2010,
	title = {Evidence for a {Collective} {Intelligence} {Factor} in the {Performance} of {Human} {Groups}},
	volume = {330},
	url = {https://www.science.org/doi/10.1126/science.1193147},
	doi = {10.1126/science.1193147},
	number = {6004},
	urldate = {2025-09-06},
	journal = {Science},
	author = {Woolley, Anita Williams and Chabris, Christopher F. and Pentland, Alex and Hashmi, Nada and Malone, Thomas W.},
	month = oct,
	year = {2010},
	note = {Publisher: American Association for the Advancement of Science},
	pages = {686--688},
}

@inproceedings{ma_are_2024,
	address = {New York, NY, USA},
	series = {{CHI} '24},
	title = {“{Are} {You} {Really} {Sure}?” {Understanding} the {Effects} of {Human} {Self}-{Confidence} {Calibration} in {AI}-{Assisted} {Decision} {Making}},
	isbn = {9798400703300},
	shorttitle = {“{Are} {You} {Really} {Sure}?},
	url = {https://dl.acm.org/doi/10.1145/3613904.3642671},
	doi = {10.1145/3613904.3642671},
	urldate = {2025-09-06},
	booktitle = {Proceedings of the 2024 {CHI} {Conference} on {Human} {Factors} in {Computing} {Systems}},
	publisher = {Association for Computing Machinery},
	author = {Ma, Shuai and Wang, Xinru and Lei, Ying and Shi, Chuhan and Yin, Ming and Ma, Xiaojuan},
	month = may,
	year = {2024},
	pages = {1--20},
}

@inproceedings{kahr_trust_2024,
	address = {New York, NY, USA},
	series = {{IUI} '24},
	title = {The {Trust} {Recovery} {Journey}. {The} {Effect} of {Timing} of {Errors} on the {Willingness} to {Follow} {AI} {Advice}.},
	isbn = {9798400705083},
	url = {https://dl.acm.org/doi/10.1145/3640543.3645167},
	doi = {10.1145/3640543.3645167},
	urldate = {2025-09-06},
	booktitle = {Proceedings of the 29th {International} {Conference} on {Intelligent} {User} {Interfaces}},
	publisher = {Association for Computing Machinery},
	author = {Kahr, Patricia K. and Rooks, Gerrit and Snijders, Chris and Willemsen, Martijn C.},
	month = apr,
	year = {2024},
	pages = {609--622},
}

@inproceedings{calisto_assertiveness-based_2023,
	address = {New York, NY, USA},
	series = {{CHI} '23},
	title = {Assertiveness-based {Agent} {Communication} for a {Personalized} {Medicine} on {Medical} {Imaging} {Diagnosis}},
	isbn = {978-1-4503-9421-5},
	url = {https://dl.acm.org/doi/10.1145/3544548.3580682},
	doi = {10.1145/3544548.3580682},
	urldate = {2025-09-06},
	booktitle = {Proceedings of the 2023 {CHI} {Conference} on {Human} {Factors} in {Computing} {Systems}},
	publisher = {Association for Computing Machinery},
	author = {Calisto, Francisco Maria and Fernandes, João and Morais, Margarida and Santiago, Carlos and Abrantes, João Maria and Nunes, Nuno and Nascimento, Jacinto C.},
	month = apr,
	year = {2023},
	pages = {1--20},
}

@inproceedings{shi_agent_2013,
	address = {New York, NY, USA},
	series = {{IUI} '13},
	title = {Agent metaphor for machine translation mediated communication},
	isbn = {978-1-4503-1965-2},
	url = {https://dl.acm.org/doi/10.1145/2449396.2449407},
	doi = {10.1145/2449396.2449407},
	urldate = {2025-09-06},
	booktitle = {Proceedings of the 2013 international conference on {Intelligent} user interfaces},
	publisher = {Association for Computing Machinery},
	author = {Shi, Chunqi and Lin, Donghui and Ishida, Toru},
	month = mar,
	year = {2013},
	pages = {67--74},
}

@article{zheng_roles_2019,
	title = {The {Roles} {Bots} {Play} in {Wikipedia}},
	volume = {3},
	issn = {2573-0142},
	url = {https://dl.acm.org/doi/10.1145/3359317},
	doi = {10.1145/3359317},
	language = {en},
	number = {CSCW},
	urldate = {2025-09-06},
	journal = {Proceedings of the ACM on Human-Computer Interaction},
	author = {Zheng, Lei (Nico) and Albano, Christopher M. and Vora, Neev M. and Mai, Feng and Nickerson, Jeffrey V.},
	month = nov,
	year = {2019},
	pages = {1--20},
}

@inproceedings{andrews_space_2010,
	address = {New York, NY, USA},
	series = {{CHI} '10},
	title = {Space to think: large high-resolution displays for sensemaking},
	isbn = {978-1-60558-929-9},
	shorttitle = {Space to think},
	url = {https://dl.acm.org/doi/10.1145/1753326.1753336},
	doi = {10.1145/1753326.1753336},
	urldate = {2025-09-06},
	booktitle = {Proceedings of the {SIGCHI} {Conference} on {Human} {Factors} in {Computing} {Systems}},
	publisher = {Association for Computing Machinery},
	author = {Andrews, Christopher and Endert, Alex and North, Chris},
	month = apr,
	year = {2010},
	pages = {55--64},
}

@article{olson_distance_2000,
	title = {Distance {Matters}},
	volume = {15},
	issn = {0737-0024},
	url = {https://doi.org/10.1207/S15327051HCI1523_4},
	doi = {10.1207/S15327051HCI1523_4},
	number = {2-3},
	urldate = {2025-09-06},
	journal = {Human–Computer Interaction},
	author = {Olson, Gary M. and Olson, Judith S.},
	month = sep,
	year = {2000},
	note = {Publisher: Taylor \& Francis
\_eprint: https://doi.org/10.1207/S15327051HCI1523\_4},
	pages = {139--178},
}

@article{hmelo-silver_facilitating_2008,
	title = {Facilitating {Collaborative} {Knowledge} {Building}},
	volume = {26},
	issn = {0737-0008},
	url = {https://doi.org/10.1080/07370000701798495},
	doi = {10.1080/07370000701798495},
	number = {1},
	urldate = {2025-09-06},
	journal = {Cognition and Instruction},
	author = {Hmelo-Silver, Cindy E. and Barrows, Howard S.},
	month = jan,
	year = {2008},
	note = {Publisher: Routledge
\_eprint: https://doi.org/10.1080/07370000701798495},
	pages = {48--94},
}

@article{wuchty_increasing_2007,
	title = {The {Increasing} {Dominance} of {Teams} in {Production} of {Knowledge}},
	volume = {316},
	url = {https://www.science.org/doi/full/10.1126/science.1136099},
	doi = {10.1126/science.1136099},
	number = {5827},
	urldate = {2025-09-06},
	journal = {Science},
	author = {Wuchty, Stefan and Jones, Benjamin F. and Uzzi, Brian},
	month = may,
	year = {2007},
	note = {Publisher: American Association for the Advancement of Science},
	pages = {1036--1039},
}

@article{fui-hoon_nah_generative_2023,
	title = {Generative {AI} and {ChatGPT}: {Applications}, challenges, and {AI}-human collaboration},
	volume = {25},
	issn = {1522-8053},
	shorttitle = {Generative {AI} and {ChatGPT}},
	url = {https://doi.org/10.1080/15228053.2023.2233814},
	doi = {10.1080/15228053.2023.2233814},
	number = {3},
	urldate = {2025-09-06},
	journal = {Journal of Information Technology Case and Application Research},
	author = {Fui-Hoon Nah, Fiona and Zheng, Ruilin and Cai, Jingyuan and Siau, Keng and Chen, Langtao},
	month = jul,
	year = {2023},
	note = {Publisher: Routledge
\_eprint: https://doi.org/10.1080/15228053.2023.2233814},
	pages = {277--304},
}

@misc{morris_design_2023,
	title = {The {Design} {Space} of {Generative} {Models}},
	url = {http://arxiv.org/abs/2304.10547},
	doi = {10.48550/arXiv.2304.10547},
	urldate = {2025-09-06},
	publisher = {arXiv},
	author = {Morris, Meredith Ringel and Cai, Carrie J. and Holbrook, Jess and Kulkarni, Chinmay and Terry, Michael},
	month = apr,
	year = {2023},
	note = {arXiv:2304.10547 [cs]},
}

@article{shi_understanding_2023,
	title = {Understanding {Design} {Collaboration} {Between} {Designers} and {Artificial} {Intelligence}: {A} {Systematic} {Literature} {Review}},
	volume = {7},
	shorttitle = {Understanding {Design} {Collaboration} {Between} {Designers} and {Artificial} {Intelligence}},
	url = {https://dl.acm.org/doi/10.1145/3610217},
	doi = {10.1145/3610217},
	number = {CSCW2},
	urldate = {2025-09-06},
	journal = {Proc. ACM Hum.-Comput. Interact.},
	author = {Shi, Yang and Gao, Tian and Jiao, Xiaohan and Cao, Nan},
	month = oct,
	year = {2023},
	pages = {368:1--368:35},
}

@inproceedings{van_den_broek_exploring_2024,
	address = {New York, NY, USA},
	series = {{PDC} '24},
	title = {Exploring the {Supportive} {Role} of {Artificial} {Intelligence} in {Participatory} {Design}: {A} {Systematic} {Review}},
	volume = {2},
	isbn = {9798400706547},
	shorttitle = {Exploring the {Supportive} {Role} of {Artificial} {Intelligence} in {Participatory} {Design}},
	url = {https://dl.acm.org/doi/10.1145/3661455.3669868},
	doi = {10.1145/3661455.3669868},
	urldate = {2025-09-06},
	booktitle = {Proceedings of the {Participatory} {Design} {Conference} 2024: {Exploratory} {Papers} and {Workshops} - {Volume} 2},
	publisher = {Association for Computing Machinery},
	author = {van den Broek, Simone and Sankaran, Supraja and de Wit, Jan and de Rooij, Alwin},
	month = aug,
	year = {2024},
	pages = {37--44},
}

@article{siemon_elaborating_2022,
	title = {Elaborating {Team} {Roles} for {Artificial} {Intelligence}-based {Teammates} in {Human}-{AI} {Collaboration}},
	volume = {31},
	copyright = {2022 The Author(s)},
	issn = {1572-9907},
	url = {https://link.springer.com/article/10.1007/s10726-022-09792-z},
	doi = {10.1007/s10726-022-09792-z},
	language = {en},
	number = {5},
	urldate = {2025-09-06},
	journal = {Group Decision and Negotiation},
	author = {Siemon, Dominik},
	month = oct,
	year = {2022},
	note = {Company: Springer
Distributor: Springer
Institution: Springer
Label: Springer
Publisher: Springer Netherlands},
	pages = {871--912},
}

@inproceedings{seymour_speculating_2024,
	address = {New York, NY, USA},
	series = {{CUI} '24},
	title = {Speculating {About} {Multi}-user {Conversational} {Interfaces} and {LLMs}: {What} {If} {Chatting} {Wasn}'t {So} {Lonely}?},
	isbn = {9798400705113},
	shorttitle = {Speculating {About} {Multi}-user {Conversational} {Interfaces} and {LLMs}},
	url = {https://dl.acm.org/doi/10.1145/3640794.3665888},
	doi = {10.1145/3640794.3665888},
	urldate = {2025-09-06},
	booktitle = {Proceedings of the 6th {ACM} {Conference} on {Conversational} {User} {Interfaces}},
	publisher = {Association for Computing Machinery},
	author = {Seymour, William and Rader, Emilee},
	month = jul,
	year = {2024},
	pages = {1--4},
}

@inproceedings{zamfirescu-pereira_why_2023,
	address = {Hamburg Germany},
	title = {Why {Johnny} {Can}’t {Prompt}: {How} {Non}-{AI} {Experts} {Try} (and {Fail}) to {Design} {LLM} {Prompts}},
	isbn = {978-1-4503-9421-5},
	shorttitle = {Why {Johnny} {Can}’t {Prompt}},
	url = {https://dl.acm.org/doi/10.1145/3544548.3581388},
	doi = {10.1145/3544548.3581388},
	language = {en},
	urldate = {2023-09-27},
	booktitle = {Proceedings of the 2023 {CHI} {Conference} on {Human} {Factors} in {Computing} {Systems}},
	publisher = {ACM},
	author = {Zamfirescu-Pereira, J.D. and Wong, Richmond Y. and Hartmann, Bjoern and Yang, Qian},
	month = apr,
	year = {2023},
	pages = {1--21},
}

@inproceedings{he_ai_2024,
	address = {New York, NY, USA},
	series = {{CHIWORK} '24},
	title = {{AI} and the {Future} of {Collaborative} {Work}: {Group} {Ideation} with an {LLM} in a {Virtual} {Canvas}},
	isbn = {9798400710179},
	shorttitle = {{AI} and the {Future} of {Collaborative} {Work}},
	url = {https://dl.acm.org/doi/10.1145/3663384.3663398},
	doi = {10.1145/3663384.3663398},
	urldate = {2024-09-03},
	booktitle = {Proceedings of the 3rd {Annual} {Meeting} of the {Symposium} on {Human}-{Computer} {Interaction} for {Work}},
	publisher = {Association for Computing Machinery},
	author = {He, Jessica and Houde, Stephanie and Gonzalez, Gabriel E. and Silva Moran, Darío Andrés and Ross, Steven I. and Muller, Michael and Weisz, Justin D.},
	month = jun,
	year = {2024},
	pages = {1--14},
}

@inproceedings{suh_ai_2021,
	address = {Yokohama Japan},
	title = {{AI} as {Social} {Glue}: {Uncovering} the {Roles} of {Deep} {Generative} {AI} during {Social} {Music} {Composition}},
	isbn = {978-1-4503-8096-6},
	shorttitle = {{AI} as {Social} {Glue}},
	url = {https://dl.acm.org/doi/10.1145/3411764.3445219},
	doi = {10.1145/3411764.3445219},
	language = {en},
	urldate = {2024-10-08},
	booktitle = {Proceedings of the 2021 {CHI} {Conference} on {Human} {Factors} in {Computing} {Systems}},
	publisher = {ACM},
	author = {Suh, Minhyang (Mia) and Youngblom, Emily and Terry, Michael and Cai, Carrie J},
	month = may,
	year = {2021},
	pages = {1--11},
}

@inproceedings{mukhopadhyay_osint_2025,
	address = {New York, NY, USA},
	series = {{CHI} '25},
	title = {{OSINT} {Clinic}: {Co}-designing {AI}-{Augmented} {Collaborative} {OSINT} {Investigations} for {Vulnerability} {Assessment}},
	isbn = {9798400713941},
	shorttitle = {{OSINT} {Clinic}},
	url = {https://dl.acm.org/doi/10.1145/3706598.3713283},
	doi = {10.1145/3706598.3713283},
	urldate = {2025-09-07},
	booktitle = {Proceedings of the 2025 {CHI} {Conference} on {Human} {Factors} in {Computing} {Systems}},
	publisher = {Association for Computing Machinery},
	author = {Mukhopadhyay, Anirban and Luther, Kurt},
	month = apr,
	year = {2025},
	pages = {1--22},
}

@inproceedings{liu_peergpt_2024,
	address = {New York, NY, USA},
	series = {{CHI} {EA} '24},
	title = {{PeerGPT}: {Probing} the {Roles} of {LLM}-based {Peer} {Agents} as {Team} {Moderators} and {Participants} in {Children}'s {Collaborative} {Learning}},
	isbn = {9798400703317},
	shorttitle = {{PeerGPT}},
	url = {https://doi.org/10.1145/3613905.3651008},
	doi = {10.1145/3613905.3651008},
	urldate = {2025-09-07},
	booktitle = {Extended {Abstracts} of the {CHI} {Conference} on {Human} {Factors} in {Computing} {Systems}},
	publisher = {Association for Computing Machinery},
	author = {Liu, Jiawen and Yao, Yuanyuan and An, Pengcheng and Wang, Qi},
	month = may,
	year = {2024},
	pages = {1--6},
}

@inproceedings{harteveld_teamwork_2019,
	address = {San Luis Obispo California USA},
	title = {Teamwork and adaptation in games ({TAG}): a survey to gauge teamwork},
	isbn = {978-1-4503-7217-6},
	shorttitle = {Teamwork and adaptation in games ({TAG})},
	url = {https://dl.acm.org/doi/10.1145/3337722.3337731},
	doi = {10.1145/3337722.3337731},
	language = {en},
	urldate = {2025-09-08},
	booktitle = {Proceedings of the 14th {International} {Conference} on the {Foundations} of {Digital} {Games}},
	publisher = {ACM},
	author = {Harteveld, Casper and Kleinman, Erica and Rizzo, Paola and Schouten, Dylan and Nguyen, Truong Huy and Liberty, Samuel and Kimbrough, Wade and Fombelle, Paul and El-Nasr, Magy Seif},
	month = aug,
	year = {2019},
	pages = {1--12},
}

@article{ellis_groupware_1991,
	title = {Groupware: some issues and experiences},
	volume = {34},
	issn = {0001-0782, 1557-7317},
	shorttitle = {Groupware},
	url = {https://dl.acm.org/doi/10.1145/99977.99987},
	doi = {10.1145/99977.99987},
	language = {en},
	number = {1},
	urldate = {2025-09-08},
	journal = {Communications of the ACM},
	author = {Ellis, Clarence A. and Gibbs, Simon J. and Rein, Gail},
	month = jan,
	year = {1991},
	pages = {39--58},
}

@article{nijstad_how_2006,
	title = {How the {Group} {Affects} the {Mind}: {A} {Cognitive} {Model} of {Idea} {Generation} in {Groups}},
	volume = {10},
	issn = {1088-8683},
	shorttitle = {How the {Group} {Affects} the {Mind}},
	url = {https://doi.org/10.1207/s15327957pspr1003_1},
	doi = {10.1207/s15327957pspr1003_1},
	number = {3},
	urldate = {2025-09-08},
	journal = {Personality and Social Psychology Review},
	author = {Nijstad, Bernard A. and Stroebe, Wolfgang},
	month = aug,
	year = {2006},
	note = {Publisher: SAGE Publications Inc},
	pages = {186--213},
}

@incollection{thompson_metacognition_2012,
	title = {Metacognition in {Teams} and {Organizations}},
	isbn = {978-0-203-86598-9},
	booktitle = {Social {Metacognition}},
	publisher = {Psychology Press},
	author = {Thompson, Leigh and Cohen, Taya R.},
	year = {2012},
	note = {Num Pages: 20},
}

@inproceedings{stuckel_effects_2008,
	address = {New York, NY, USA},
	series = {{CSCW} '08},
	title = {The effects of local lag on tightly-coupled interaction in distributed groupware},
	isbn = {978-1-60558-007-4},
	url = {https://dl.acm.org/doi/10.1145/1460563.1460635},
	doi = {10.1145/1460563.1460635},
	urldate = {2025-09-07},
	booktitle = {Proceedings of the 2008 {ACM} conference on {Computer} supported cooperative work},
	publisher = {Association for Computing Machinery},
	author = {Stuckel, Dane and Gutwin, Carl},
	month = nov,
	year = {2008},
	pages = {447--456},
}

@inproceedings{linebarger_benefits_2005,
	address = {New York, NY, USA},
	series = {{GROUP} '05},
	title = {Benefits of synchronous collaboration support for an application-centered analysis team working on complex problems: a case study},
	isbn = {978-1-59593-223-5},
	shorttitle = {Benefits of synchronous collaboration support for an application-centered analysis team working on complex problems},
	url = {https://dl.acm.org/doi/10.1145/1099203.1099211},
	doi = {10.1145/1099203.1099211},
	urldate = {2025-09-07},
	booktitle = {Proceedings of the 2005 {ACM} {International} {Conference} on {Supporting} {Group} {Work}},
	publisher = {Association for Computing Machinery},
	author = {Linebarger, John M. and Scholand, Andrew J. and Ehlen, Mark A. and Procopio, Michael J.},
	month = nov,
	year = {2005},
	pages = {51--60},
}

@article{kim_effect_2018,
	title = {The effect of metacognitive monitoring feedback on performance in a computer-based training simulation},
	volume = {67},
	issn = {0003-6870},
	url = {https://www.sciencedirect.com/science/article/pii/S0003687017302272},
	doi = {10.1016/j.apergo.2017.10.006},
	urldate = {2025-09-08},
	journal = {Applied Ergonomics},
	author = {Kim, Jung Hyup},
	month = feb,
	year = {2018},
	pages = {193--202},
}

@article{nonose_effect_2012,
	title = {The {Effect} of {Metacognition} in {Cooperation} on {Team} {Behaviors}},
	language = {en},
	author = {Nonose, Kohei and Kanno, Taro and Furuta, Kazuo},
	year = {2012},
}

@article{wiltshire_training_2014,
	title = {Training for {Collaborative} {Problem} {Solving}: {Improving} {Team} {Process} and {Performance} through {Metacognitive} {Prompting}},
	volume = {58},
	issn = {1071-1813},
	shorttitle = {Training for {Collaborative} {Problem} {Solving}},
	url = {https://doi.org/10.1177/1541931214581241},
	doi = {10.1177/1541931214581241},
	number = {1},
	urldate = {2025-09-08},
	journal = {Proceedings of the Human Factors and Ergonomics Society Annual Meeting},
	author = {Wiltshire, Travis J. and Rosch, Kelly and Fiorella, Logan and Fiore, Stephen M.},
	month = sep,
	year = {2014},
	note = {Publisher: SAGE Publications Inc},
	pages = {1154--1158},
}

@article{cui_ai-enhanced_2024,
	title = {{AI}-enhanced collective intelligence},
	volume = {5},
	issn = {26663899},
	url = {https://linkinghub.elsevier.com/retrieve/pii/S2666389924002332},
	doi = {10.1016/j.patter.2024.101074},
	language = {en},
	number = {11},
	urldate = {2025-09-08},
	journal = {Patterns},
	author = {Cui, Hao and Yasseri, Taha},
	month = nov,
	year = {2024},
	pages = {101074},
}

@inproceedings{shin_integrating_2023,
	address = {Hamburg Germany},
	title = {Integrating {AI} in {Human}-{Human} {Collaborative} {Ideation}},
	isbn = {978-1-4503-9422-2},
	url = {https://dl.acm.org/doi/10.1145/3544549.3573802},
	doi = {10.1145/3544549.3573802},
	language = {en},
	urldate = {2025-09-08},
	booktitle = {Extended {Abstracts} of the 2023 {CHI} {Conference} on {Human} {Factors} in {Computing} {Systems}},
	publisher = {ACM},
	author = {Shin, Joon Gi and Koch, Janin and Lucero, Andrés and Dalsgaard, Peter and Mackay, Wendy E.},
	month = apr,
	year = {2023},
	pages = {1--5},
}

@article{koch_imagesense_2020,
	title = {{ImageSense}: {An} {Intelligent} {Collaborative} {Ideation} {Tool} to {Support} {Diverse} {Human}-{Computer} {Partnerships}},
	volume = {4},
	shorttitle = {{ImageSense}},
	url = {https://hal.science/hal-02867303},
	doi = {10.1145/3392850},
	number = {CSCW1},
	urldate = {2025-09-08},
	journal = {Proceedings of the ACM on Human-Computer Interaction},
	author = {Koch, Janin and Taffin, Nicolas and Beaudouin-Lafon, Michel and Laine, Markku and Lucero, Andrés and Mackay, Wendy},
	month = may,
	year = {2020},
	note = {Publisher: Association for Computing Machinery (ACM)},
	pages = {1--27},
}

@inproceedings{salikutluk_evaluation_2024,
	address = {Honolulu HI USA},
	title = {An {Evaluation} of {Situational} {Autonomy} for {Human}-{AI} {Collaboration} in a {Shared} {Workspace} {Setting}},
	isbn = {9798400703300},
	url = {https://dl.acm.org/doi/10.1145/3613904.3642564},
	doi = {10.1145/3613904.3642564},
	language = {en},
	urldate = {2025-09-08},
	booktitle = {Proceedings of the {CHI} {Conference} on {Human} {Factors} in {Computing} {Systems}},
	publisher = {ACM},
	author = {Salikutluk, Vildan and Schöpper, Janik and Herbert, Franziska and Scheuermann, Katrin and Frodl, Eric and Balfanz, Dirk and Jäkel, Frank and Koert, Dorothea},
	month = may,
	year = {2024},
	pages = {1--17},
}

@inproceedings{verheijden_collaborative_2023,
	address = {Hamburg Germany},
	title = {Collaborative {Diffusion}: {Boosting} {Designerly} {Co}-{Creation} with {Generative} {AI}},
	isbn = {978-1-4503-9422-2},
	shorttitle = {Collaborative {Diffusion}},
	url = {https://dl.acm.org/doi/10.1145/3544549.3585680},
	doi = {10.1145/3544549.3585680},
	language = {en},
	urldate = {2025-09-08},
	booktitle = {Extended {Abstracts} of the 2023 {CHI} {Conference} on {Human} {Factors} in {Computing} {Systems}},
	publisher = {ACM},
	author = {Verheijden, Mathias Peter and Funk, Mathias},
	month = apr,
	year = {2023},
	pages = {1--8},
}

@inproceedings{park_generative_2023,
	address = {San Francisco CA USA},
	title = {Generative {Agents}: {Interactive} {Simulacra} of {Human} {Behavior}},
	isbn = {9798400701320},
	shorttitle = {Generative {Agents}},
	url = {https://dl.acm.org/doi/10.1145/3586183.3606763},
	doi = {10.1145/3586183.3606763},
	language = {en},
	urldate = {2025-09-08},
	booktitle = {Proceedings of the 36th {Annual} {ACM} {Symposium} on {User} {Interface} {Software} and {Technology}},
	publisher = {ACM},
	author = {Park, Joon Sung and O'Brien, Joseph and Cai, Carrie Jun and Morris, Meredith Ringel and Liang, Percy and Bernstein, Michael S.},
	month = oct,
	year = {2023},
	pages = {1--22},
}

@inproceedings{park_social_2022,
	address = {Bend OR USA},
	title = {Social {Simulacra}: {Creating} {Populated} {Prototypes} for {Social} {Computing} {Systems}},
	isbn = {978-1-4503-9320-1},
	shorttitle = {Social {Simulacra}},
	url = {https://dl.acm.org/doi/10.1145/3526113.3545616},
	doi = {10.1145/3526113.3545616},
	language = {en},
	urldate = {2025-09-08},
	booktitle = {Proceedings of the 35th {Annual} {ACM} {Symposium} on {User} {Interface} {Software} and {Technology}},
	publisher = {ACM},
	author = {Park, Joon Sung and Popowski, Lindsay and Cai, Carrie and Morris, Meredith Ringel and Liang, Percy and Bernstein, Michael S.},
	month = oct,
	year = {2022},
	pages = {1--18},
}

@article{wu_agent_2019,
	title = {Agent, {Gatekeeper}, {Drug} {Dealer}: {How} {Content} {Creators} {Craft} {Algorithmic} {Personas}},
	volume = {3},
	issn = {2573-0142},
	shorttitle = {Agent, {Gatekeeper}, {Drug} {Dealer}},
	url = {https://dl.acm.org/doi/10.1145/3359321},
	doi = {10.1145/3359321},
	language = {en},
	number = {CSCW},
	urldate = {2025-09-08},
	journal = {Proceedings of the ACM on Human-Computer Interaction},
	author = {Wu, Eva Yiwei and Pedersen, Emily and Salehi, Niloufar},
	month = nov,
	year = {2019},
	pages = {1--27},
}

@article{robbins_title_nodate,
	title = {Title: {Essentials} of {Organizational} {Behavior}, 11th edition},
	language = {en},
	author = {Robbins, Stephen P and Judge, Timothy A},
}

@article{stevens_knowledge_1994,
	title = {The {Knowledge}, {Skill}, and {Ability} {Requirements} for {Teamwork}: {Implications} for {Human} {Resource} {Management}},
	volume = {20},
	issn = {0149-2063},
	shorttitle = {The {Knowledge}, {Skill}, and {Ability} {Requirements} for {Teamwork}},
	url = {https://doi.org/10.1177/014920639402000210},
	doi = {10.1177/014920639402000210},
	number = {2},
	urldate = {2025-09-08},
	journal = {Journal of Management},
	author = {Stevens, Michael J. and Campion, Michael A.},
	month = apr,
	year = {1994},
	note = {Publisher: SAGE Publications Inc},
	pages = {503--530},
}

@article{hubert_current_2024,
	title = {The current state of artificial intelligence generative language models is more creative than humans on divergent thinking tasks},
	volume = {14},
	issn = {2045-2322},
	url = {https://www.nature.com/articles/s41598-024-53303-w},
	doi = {10.1038/s41598-024-53303-w},
	language = {en},
	number = {1},
	urldate = {2025-09-08},
	journal = {Scientific Reports},
	author = {Hubert, Kent F. and Awa, Kim N. and Zabelina, Darya L.},
	month = feb,
	year = {2024},
	pages = {3440},
}

@article{gao_coaicoder_2024,
	title = {{CoAIcoder}: {Examining} the {Effectiveness} of {AI}-assisted {Human}-to-{Human} {Collaboration} in {Qualitative} {Analysis}},
	volume = {31},
	issn = {1073-0516, 1557-7325},
	shorttitle = {{CoAIcoder}},
	url = {https://dl.acm.org/doi/10.1145/3617362},
	doi = {10.1145/3617362},
	language = {en},
	number = {1},
	urldate = {2025-09-08},
	journal = {ACM Transactions on Computer-Human Interaction},
	author = {Gao, Jie and Choo, Kenny Tsu Wei and Cao, Junming and Lee, Roy Ka-Wei and Perrault, Simon},
	month = feb,
	year = {2024},
	pages = {1--38},
}

@article{deng_crossgai_2024,
	title = {{CrossGAI}: {A} {Cross}-{Device} {Generative} {AI} {Framework} for {Collaborative} {Fashion} {Design}},
	volume = {8},
	issn = {2474-9567},
	shorttitle = {{CrossGAI}},
	url = {https://dl.acm.org/doi/10.1145/3643542},
	doi = {10.1145/3643542},
	language = {en},
	number = {1},
	urldate = {2025-09-08},
	journal = {Proceedings of the ACM on Interactive, Mobile, Wearable and Ubiquitous Technologies},
	author = {Deng, Hanhui and Jiang, Jianan and Yu, Zhiwang and Ouyang, Jinhui and Wu, Di},
	month = mar,
	year = {2024},
	pages = {1--27},
}

@article{duan2025gender,
  title={Gender Stereotypes toward Non-gendered Generative AI: The Role of Gendered Expertise and Gendered Linguistic Cues},
  author={Duan, Wen and McNeese, Nathan and Li, Lingyuan},
  journal={Proceedings of the ACM on Human-Computer Interaction},
  volume={9},
  number={1},
  pages={1--35},
  year={2025},
  publisher={ACM New York, NY, USA}
}

@article{munyaka2023decision,
author = {Munyaka, Imani and Ashktorab, Zahra and Dugan, Casey and Johnson, J. and Pan, Qian},
title = {Decision Making Strategies and Team Efficacy in Human-AI Teams},
year = {2023},
issue_date = {April 2023},
publisher = {Association for Computing Machinery},
address = {New York, NY, USA},
volume = {7},
number = {CSCW1},
url = {https://doi.org/10.1145/3579476},
doi = {10.1145/3579476},
journal = {Proc. ACM Hum.-Comput. Interact.},
month = apr,
articleno = {43},
numpages = {24}
}

@inproceedings{wobbrock_aligned_2011,
	address = {New York, NY, USA},
	series = {{CHI} '11},
	title = {The aligned rank transform for nonparametric factorial analyses using only anova procedures},
	isbn = {978-1-4503-0228-9},
	url = {https://doi.org/10.1145/1978942.1978963},
	doi = {10.1145/1978942.1978963},
	urldate = {2025-11-27},
	booktitle = {Proceedings of the {SIGCHI} {Conference} on {Human} {Factors} in {Computing} {Systems}},
	publisher = {Association for Computing Machinery},
	author = {Wobbrock, Jacob O. and Findlater, Leah and Gergle, Darren and Higgins, James J.},
	month = may,
	year = {2011},
	pages = {143--146},
}

@inproceedings{elkin_aligned_2021,
	address = {New York, NY, USA},
	series = {{UIST} '21},
	title = {An {Aligned} {Rank} {Transform} {Procedure} for {Multifactor} {Contrast} {Tests}},
	isbn = {978-1-4503-8635-7},
	url = {https://dl.acm.org/doi/10.1145/3472749.3474784},
	doi = {10.1145/3472749.3474784},
	urldate = {2025-11-27},
	booktitle = {The 34th {Annual} {ACM} {Symposium} on {User} {Interface} {Software} and {Technology}},
	publisher = {Association for Computing Machinery},
	author = {Elkin, Lisa A. and Kay, Matthew and Higgins, James J. and Wobbrock, Jacob O.},
	month = oct,
	year = {2021},
	pages = {754--768},
}

@article{xu2023effectiveness,
  title={The effectiveness of collaborative problem solving in promoting students’ critical thinking: A meta-analysis based on empirical literature},
  author={Xu, Enwei and Wang, Wei and Wang, Qingxia},
  journal={Humanities and Social Sciences Communications},
  volume={10},
  number={1},
  pages={1--11},
  year={2023},
  publisher={Palgrave}
}

@article{tegos2015promoting,
  title={Promoting academically productive talk with conversational agent interventions in collaborative learning settings},
  author={Tegos, Stergios and Demetriadis, Stavros and Karakostas, Anastasios},
  journal={Computers \& Education},
  volume={87},
  pages={309--325},
  year={2015},
  publisher={Elsevier}
}

@inproceedings{xu2017new,
  title={A new chatbot for customer service on social media},
  author={Xu, Anbang and Liu, Zhe and Guo, Yufan and Sinha, Vibha and Akkiraju, Rama},
  booktitle={Proceedings of the 2017 CHI conference on human factors in computing systems},
  pages={3506--3510},
  year={2017}
}

@inproceedings{porcheron2017animals,
  title={" Do Animals Have Accents?" Talking with Agents in Multi-Party Conversation},
  author={Porcheron, Martin and Fischer, Joel E and Sharples, Sarah},
  booktitle={Proceedings of the 2017 ACM conference on computer supported cooperative work and social computing},
  pages={207--219},
  year={2017}
}

@inproceedings{dohsaka2009effects,
  title={Effects of conversational agents on human communication in thought-evoking multi-party dialogues},
  author={Dohsaka, Kohji and Asai, Ryota and Higashinaka, Ryuichiro and Minami, Yasuhiro and Maeda, Eisaku},
  booktitle={Proceedings of the SIGDIAL 2009 Conference},
  pages={217--224},
  year={2009}
}

@article{dyke2013enhancing,
  title={Enhancing scientific reasoning and discussion with conversational agents},
  author={Dyke, Gregory and Adamson, David and Howley, Iris and Ros{\'e}, Carolyn Penstein},
  journal={IEEE Transactions on Learning Technologies},
  volume={6},
  number={3},
  pages={240--247},
  year={2013},
  publisher={IEEE}
}

@article{hagemann2017complex,
  title={Complex problem solving in teams: the impact of collective orientation on team process demands},
  author={Hagemann, Vera and Kluge, Annette},
  journal={Frontiers in psychology},
  volume={8},
  pages={1730},
  year={2017},
  publisher={Frontiers Media SA}
}

@article{bain2022whisperx,
  title={WhisperX: Time-Accurate Speech Transcription of Long-Form Audio},
  author={Bain, Max and Huh, Jaesung and Han, Tengda and Zisserman, Andrew},
  journal={INTERSPEECH 2023},
  year={2023}
}

@inproceedings{fotaris2019escape,
  title={Escape rooms for learning: A systematic review},
  author={Fotaris, Panagiotis and Mastoras, Theodoros},
  booktitle={Proceedings of the European Conference on Games Based Learning},
  volume={2019},
  number={1},
  pages={235--243},
  year={2019}
}

@inproceedings{zhou2024understanding,
  title={Understanding nonlinear collaboration between human and AI agents: A co-design framework for creative design},
  author={Zhou, Jiayi and Li, Renzhong and Tang, Junxiu and Tang, Tan and Li, Haotian and Cui, Weiwei and Wu, Yingcai},
  booktitle={Proceedings of the 2024 CHI conference on human factors in computing systems},
  pages={1--16},
  year={2024}
}

@inproceedings{hwang2021ideabot,
  title={IdeaBot: investigating social facilitation in human-machine team creativity},
  author={Hwang, Angel Hsing-Chi and Won, Andrea Stevenson},
  booktitle={Proceedings of the 2021 CHI conference on human factors in computing systems},
  pages={1--16},
  year={2021}
}

@inproceedings{bittner_where_2019,
	title = {Where is the {Bot} in our {Team}? {Toward} a {Taxonomy} of {Design} {Option} {Combinations} for {Conversational} {Agents} in {Collaborative} {Work}},
	shorttitle = {Where is the {Bot} in our {Team}?},
	url = {http://hdl.handle.net/10125/59469},
	doi = {10.24251/HICSS.2019.035},
	language = {en},
	urldate = {2025-05-29},
	author = {Bittner, Eva and Oeste-Reiß, Sarah and Leimeister, Jan Marco},
	year = {2019},
}

@article{flathmann_empirically_2024,
	title = {Empirically {Understanding} the {Potential} {Impacts} and {Process} of {Social} {Influence} in {Human}-{AI} {Teams}},
	volume = {8},
	url = {https://dl.acm.org/doi/10.1145/3637326},
	doi = {10.1145/3637326},
	number = {CSCW1},
	urldate = {2025-05-29},
	journal = {Proc. ACM Hum.-Comput. Interact.},
	author = {Flathmann, Christopher and Duan, Wen and Mcneese, Nathan J. and Hauptman, Allyson and Zhang, Rui},
	month = apr,
	year = {2024},
	pages = {49:1--49:32},
}

@article{zhang_investigating_2023,
	title = {Investigating {AI} {Teammate} {Communication} {Strategies} and {Their} {Impact} in {Human}-{AI} {Teams} for {Effective} {Teamwork}},
	volume = {7},
	issn = {2573-0142},
	url = {https://dl.acm.org/doi/10.1145/3610072},
	doi = {10.1145/3610072},
	language = {en},
	number = {CSCW2},
	urldate = {2025-05-29},
	journal = {Proceedings of the ACM on Human-Computer Interaction},
	author = {Zhang, Rui and Duan, Wen and Flathmann, Christopher and McNeese, Nathan and Freeman, Guo and Williams, Alyssa},
	month = sep,
	year = {2023},
	pages = {1--31},
}

@misc{wang_adaptive_2025,
	title = {Adaptive {Human}-{Agent} {Teaming}: {A} {Review} of {Empirical} {Studies} from the {Process} {Dynamics} {Perspective}},
	shorttitle = {Adaptive {Human}-{Agent} {Teaming}},
	url = {http://arxiv.org/abs/2504.10918},
	doi = {10.48550/arXiv.2504.10918},
	urldate = {2025-05-29},
	publisher = {arXiv},
	author = {Wang, Mengyao and Wu, Jiayun and Ma, Shuai and Li, Nuo and Zhang, Peng and Gu, Ning and Lu, Tun},
	month = apr,
	year = {2025},
	note = {arXiv:2504.10918 [cs]},
}

@inproceedings{chiang_enhancing_2024,
	address = {Greenville SC USA},
	title = {Enhancing {AI}-{Assisted} {Group} {Decision} {Making} through {LLM}-{Powered} {Devil}'s {Advocate}},
	isbn = {979-8-4007-0508-3},
	url = {https://dl.acm.org/doi/10.1145/3640543.3645199},
	doi = {10.1145/3640543.3645199},
	language = {en},
	urldate = {2025-05-29},
	booktitle = {Proceedings of the 29th {International} {Conference} on {Intelligent} {User} {Interfaces}},
	publisher = {ACM},
	author = {Chiang, Chun-Wei and Lu, Zhuoran and Li, Zhuoyan and Yin, Ming},
	month = mar,
	year = {2024},
	pages = {103--119},
}

@inproceedings{houde_controlling_2025,
	title = {Controlling {AI} {Agent} {Participation} in {Group} {Conversations}: {A} {Human}-{Centered} {Approach}},
	shorttitle = {Controlling {AI} {Agent} {Participation} in {Group} {Conversations}},
	url = {http://arxiv.org/abs/2501.17258},
	doi = {10.1145/3708359.3712089},
	urldate = {2025-05-29},
	booktitle = {Proceedings of the 30th {International} {Conference} on {Intelligent} {User} {Interfaces}},
	author = {Houde, Stephanie and Brimijoin, Kristina and Muller, Michael and Ross, Steven I. and Moran, Dario Andres Silva and Gonzalez, Gabriel Enrique and Kunde, Siya and Foreman, Morgan A. and Weisz, Justin D.},
	month = mar,
	year = {2025},
	note = {arXiv:2501.17258 [cs]},
	pages = {390--408},
}

@misc{noauthor_acorn_nodate,
	title = {Acorn {Cottage}},
	url = {https://www.quarantini.space/ac-joining-instructions},
	language = {en-US},
	urldate = {2025-08-25},
        year = {2025},
	journal = {Quarantini Boredom Escapes},
}

@misc{team_enchambered_nodate,
	title = {Alone {Together:} {Enchambered} {Escape} {Room}},
	url = {https://www.enchambered.com/puzzles/alone-together/},
	language = {en-US},
	urldate = {2025-08-25},
        year = {2025},
	journal = {Enchambered: Sacramento Escape Room},
}

@inproceedings{kleinman_untapped_2024,
	address = {New York, NY, USA},
	series = {{CHI} {PLAY} {Companion} '24},
	title = {The {Untapped} {Potential} of {Escape} {Rooms} as {Gamified} {Research} {Environments}},
	isbn = {979-8-4007-0692-9},
	url = {https://dl.acm.org/doi/10.1145/3665463.3678865},
	doi = {10.1145/3665463.3678865},
	urldate = {2025-06-02},
	booktitle = {Companion {Proceedings} of the 2024 {Annual} {Symposium} on {Computer}-{Human} {Interaction} in {Play}},
	publisher = {Association for Computing Machinery},
	author = {Kleinman, Erica and Harteveld, Casper},
	month = oct,
	year = {2024},
	pages = {276--278},
}

@misc{mccandless_liberating_2020,
	title = {Liberating {Structures}: {Change} {Methods} for {Everybody} {Every} {Day}},
	shorttitle = {Liberating {Structures}},
	url = {https://keithmccandless.medium.com/liberating-structures-change-methods-for-everybody-every-day-648e9c0d04a7},
	language = {en},
	urldate = {2025-06-19},
	journal = {Medium},
	author = {McCandless, Keith},
	month = sep,
	year = {2020},
}

@inproceedings{kraus_improving_2023,
	title = {Improving {Proactive} {Dialog} {Agents} {Using} {Socially}-{Aware} {Reinforcement} {Learning}},
	url = {http://arxiv.org/abs/2211.15359},
	doi = {10.1145/3565472.3595611},
	urldate = {2025-08-14},
	booktitle = {Proceedings of the 31st {ACM} {Conference} on {User} {Modeling}, {Adaptation} and {Personalization}},
	author = {Kraus, Matthias and Wagner, Nicolas and Riekenbrauck, Ron and Minker, Wolfgang},
	month = jun,
	year = {2023},
	note = {arXiv:2211.15359 [cs]},
	pages = {146--155},
}

@article{Resick2010,
  author    = {Resick, Christian J. and Dickson, Marcus W. and Mitchelson, Jacqueline K. and Allison, Laura K. and Clark, M. Anne},
  title     = {Team composition, cognition, and effectiveness: Examining mental model similarity and accuracy},
  journal   = {Group Dynamics: Theory, Research, and Practice},
  year      = {2010},
  volume    = {14},
  number    = {2},
  pages     = {174--191},
  doi       = {10.1037/a0018444}
}

@incollection{Hart1988,
  author    = {Hart, Sandra G. and Staveland, Lowell E.},
  title     = {Development of NASA-TLX (Task Load Index): Results of empirical and theoretical research},
  booktitle = {Advances in Psychology},
  editor    = {Hancock, Peter A. and Meshkati, Najmedin},
  volume    = {52},
  pages     = {139--183},
  publisher = {North-Holland},
  year      = {1988},
  doi       = {10.1016/S0166-4115(08)62386-9}
}

@article{Bendell2025,
  author    = {Bendell, R. and Williams, J. and Fiore, Stephen M. and Jentsch, F.},
  title     = {Artificial social intelligence in teamwork: how team traits influence human-AI dynamics in complex tasks},
  journal   = {Frontiers in Robotics and AI},
  year      = {2025},
  volume    = {12},
  pages     = {1487883},
  doi       = {10.3389/frobt.2025.1487883},
  pmid      = {40034799},
  pmcid     = {PMC11873349},
  publisher = {Frontiers Media SA}
}

@article{braun2006using,
  title={Using thematic analysis in psychology},
  author={Braun, Virginia and Clarke, Victoria},
  journal={Qualitative research in psychology},
  volume={3},
  number={2},
  pages={77--101},
  year={2006},
  publisher={Taylor \& Francis}
}

@article{mcdonald2019reliability,
  author       = {Nora McDonald and Sarita Schoenebeck and Andrea Forte},
  title        = {Reliability and Inter-rater Reliability in Qualitative Research: Norms and Guidelines for {CSCW} and {HCI} Practice},
  journal      = {Proceedings of the ACM on Human-Computer Interaction},
  volume       = {3},
  number       = {CSCW},
  articleno    = {72},
  year         = {2019},
  month        = {nov},
  pages        = {72:1--72:23},
  doi          = {10.1145/3359174},
  url          = {https://doi.org/10.1145/3359174},
  publisher    = {Association for Computing Machinery}
}

@inproceedings{Vakeva2025,
  author    = {Jaakko V{\"a}kev{\"a} and Perttu H{\"a}m{\"a}l{\"a}inen and Janne Lindqvist},
  title     = {{``Don't You Dare Go Hollow'': How Dark Souls Helps Players Cope with Depression, a Thematic Analysis of Reddit Discussions}},
  booktitle = {Proceedings of the 2025 CHI Conference on Human Factors in Computing Systems (CHI '25)},
  year      = {2025},
  publisher = {Association for Computing Machinery},
  address   = {New York, NY, USA},
  pages     = {458:1--458:20},
  articleno = {458},
  numpages  = {20},
  doi       = {10.1145/3706598.3714075},
  url       = {https://doi.org/10.1145/3706598.3714075}
}

@inproceedings{earle-randell_how_2025,
	title = {How {Virtual} {Agents} {Can} {Shape} {Human}-{Human} {Collaboration}: {A} {Systematic} {Review}},
	isbn = {978-3-031-98420-4},
	shorttitle = {How {Virtual} {Agents} {Can} {Shape} {Human}-{Human} {Collaboration}},
	url = {https://link.springer.com/chapter/10.1007/978-3-031-98420-4_33},
	doi = {10.1007/978-3-031-98420-4_33},
	language = {en},
	urldate = {2025-08-23},
	booktitle = {Artificial {Intelligence} in {Education}},
	publisher = {Springer, Cham},
	author = {Earle-Randell, Toni V. and Zhang, Shan and Schroeder, Noah and Boyer, Kristy E. and Dorley, Emmanuel},
	year = {2025},
	note = {ISSN: 1611-3349},
	pages = {468--486},
}

@inproceedings{hutchinson2003technology,
  title={Technology probes: inspiring design for and with families},
  author={Hutchinson, Hilary and Mackay, Wendy and Westerlund, Bo and Bederson, Benjamin B and Druin, Allison and Plaisant, Catherine and Beaudouin-Lafon, Michel and Conversy, St{\'e}phane and Evans, Helen and Hansen, Heiko and others},
  booktitle={Proceedings of the SIGCHI conference on Human factors in computing systems},
  pages={17--24},
  year={2003}
}

@inproceedings{kuang_enhancing_2024,
	address = {New York, NY, USA},
	series = {{CHI} '24},
	title = {Enhancing {UX} {Evaluation} {Through} {Collaboration} with {Conversational} {AI} {Assistants}: {Effects} of {Proactive} {Dialogue} and {Timing}},
	isbn = {9798400703300},
	shorttitle = {Enhancing {UX} {Evaluation} {Through} {Collaboration} with {Conversational} {AI} {Assistants}},
	url = {https://dl.acm.org/doi/10.1145/3613904.3642168},
	doi = {10.1145/3613904.3642168},
	urldate = {2025-08-23},
	booktitle = {Proceedings of the 2024 {CHI} {Conference} on {Human} {Factors} in {Computing} {Systems}},
	publisher = {Association for Computing Machinery},
	author = {Kuang, Emily and Li, Minghao and Fan, Mingming and Shinohara, Kristen},
	month = may,
	year = {2024},
	pages = {1--16},
}

@String{Computing = "Computing" }

@String{Computer = "{IEEE} Computer" }

@String{Springer = "Springer-Verlag" }

\appendix

\section{Appendix: Generative AI Usage}
We used ChatGPT to 1) generate the initial captions and descriptions for the images, 2) for polishing the quality of text, and 3) format the tables.

\section{Appendix: Screenshot of Agents Embedded in Puzzle Screens}
\label{sec:screenshot}
\begin{figure*}[ht]
    \centering
    \includegraphics[width=.85\textwidth]{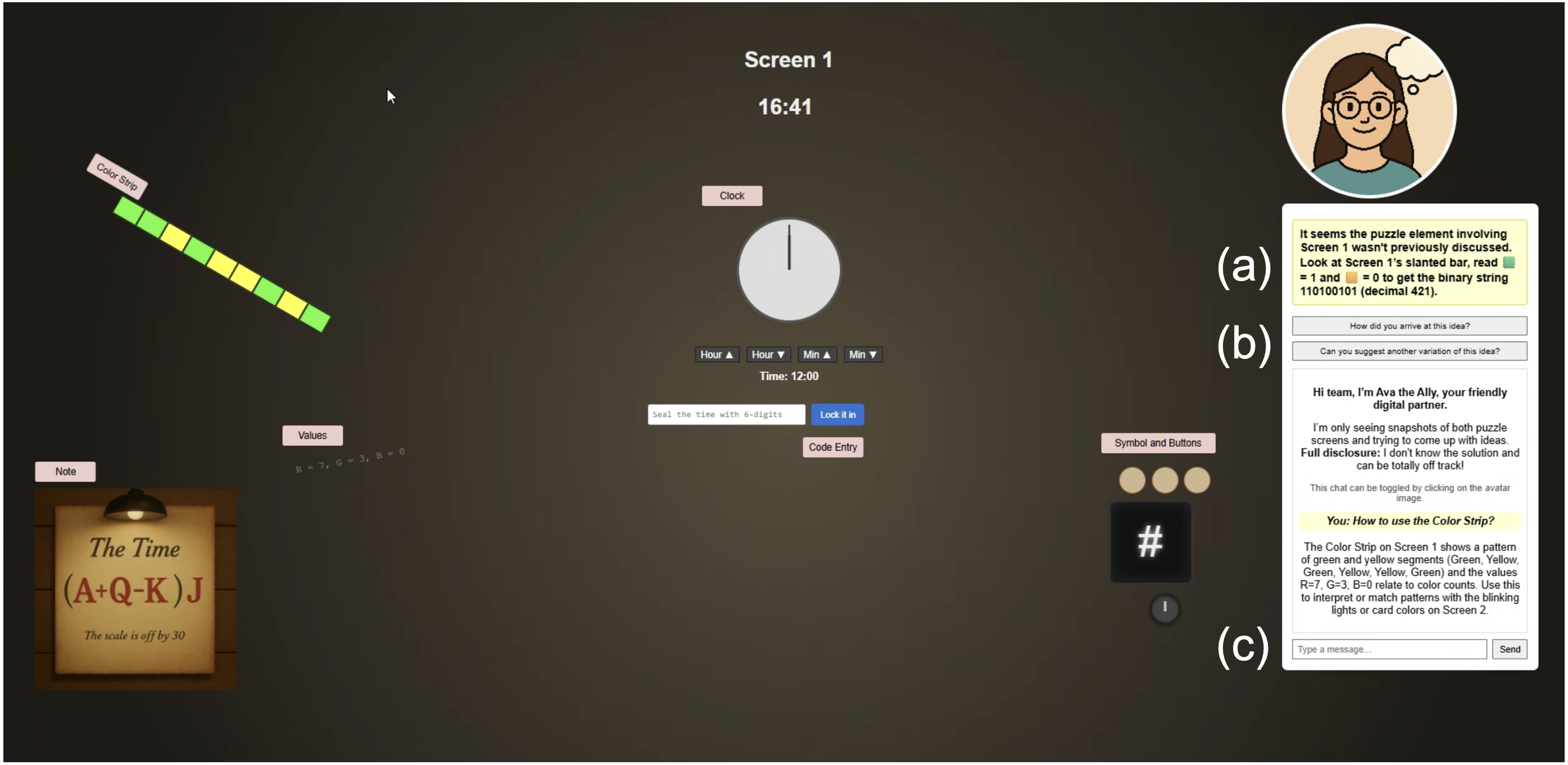}
    \caption{Screenshot of Screen 1 of Puzzle 1 with the peer agent condition.}
    \label{fig:peer_shot}
    \Description{Shows a digital puzzle interface labeled Screen 1 with a countdown timer at the top right displaying 16:41. On the left side, a diagonal Color Strip alternates between green and yellow squares. Beneath it is a Note that reads, ``The Time (A+Q–K) J. The scale is off by 30,'' along with a Values label indicating R = 7, G = 3, B = 0. At the center is a large Clock set to 12:00 with dropdown controls for hours and minutes, a field for entering a 6-digit time, and a blue ``Lock it in'' button. To the right is the avatar of Ava the Ally, the digital partner, with a thought bubble above her head. Ava provides puzzle hints in a yellow-highlighted box: ``It seems the puzzle element involving Screen 1 wasn't previously discussed. Look at Screen 1's slanted bar, read green = 1 and yellow = 0 to get the binary string 110100101 (decimal 421).'' Below are interactive prompts such as ``How did you arrive at this idea?'' and ``Can you suggest another variation of this idea?'' followed by additional text where Ava explains her role, noting that she only sees puzzle snapshots and may not always be correct. The chat area continues with Ava analyzing the Color Strip, describing its sequence of green and yellow segments, explaining how the RGB values may map onto counts, and suggesting connections to blinking lights or card colors on Screen 2. At the lower right, a Symbol and Buttons panel shows three brown circles above a black square containing a white hash (#) symbol, resembling an input or code mechanism.}
\end{figure*}

\begin{figure*}[ht]
    \centering
    \includegraphics[width=.85\textwidth]{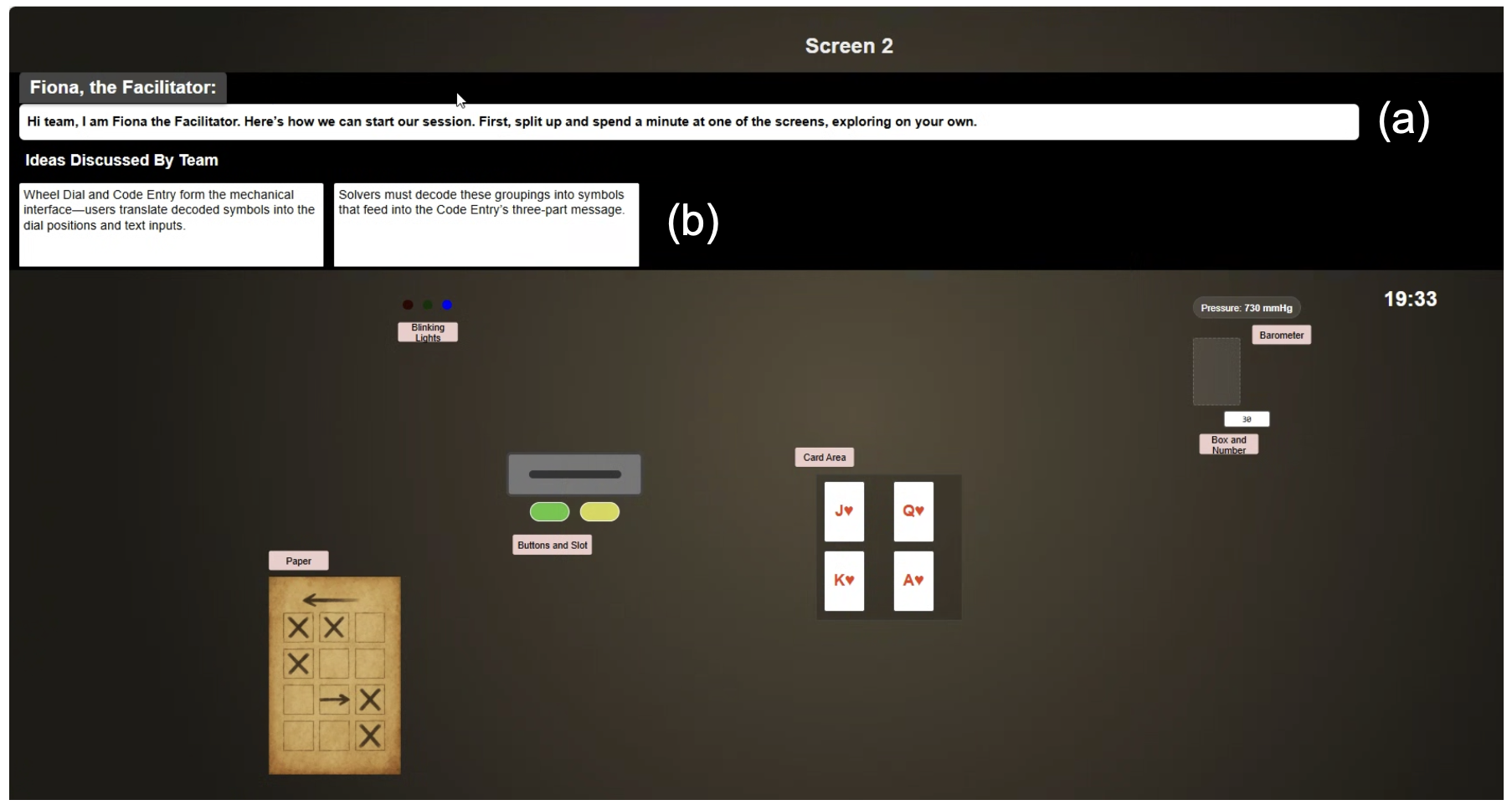}
    \caption{Screenshot of Screen 2 of Puzzle 1 with the facilitator agent condition.}
    \label{fig:facilitator_shot}
    \Description{Shows a digital puzzle interface labeled Screen 2 with a countdown timer at the top right displaying 19:33. At the top, a black banner introduces Fiona, the facilitator, who gives instructions in a scripted message: ``Hi team, I am Fiona the facilitator. Here's how we can start our session. First, split up and spend a minute at one of the screens, exploring on your own.'' Below this, a section titled Ideas Discussed By Group contains two white text boxes, one explaining that the Wheel Dial and Code Entry form the mechanical interface where users translate decoded symbols into dial positions and text inputs, and another stating that solvers must decode groupings into symbols for the three-part message. The puzzle area displays several interactive components including a Paper showing a grid of Xs with arrows, a set of green and yellow Buttons with a slot, three Blinking Lights labeled red, green, and blue, a Card Area showing four playing cards (Jack, Queen, King, and Ace of hearts), and a Barometer displaying pressure at 730 mmHg above a Box and Number labeled ``30.'' The layout resembles a digital escape-room puzzle combining symbolic, numerical, and mechanical inputs.}
    
\end{figure*}

Figures \ref{fig:peer_shot} and \ref{fig:facilitator_shot} show representative screenshots of the facilitator and peer AI agents as embedded within the puzzle screens used in the study.

\section{Appendix: Prompts used in the design of the facilitator and peer agent}
\label{appendix_prompts}
The prompts used in the design of the facilitator and peer agents, including prompts for summary generation, proactive idea generation, contextualization, and chat responses are detailed here.

\subsection{Prompt to generate summaries for the facilitator}
\label{facilitator_prompt}
You are an expert facilitator who turns raw transcripts of in-person group discussions into tightly focused, puzzle-element–driven summaries. When given a transcript of the team discussion and access to the on-screen labels/images, follow this two-step process: Step 1: Based only on the transcript (do not use the images), identify and summarize the most explicitly mentioned puzzle solution ideas. The ideas should not be facilitator's advice, which talks about the team structure and reminders. Summarize each idea in one short sentence. Step 2: Now consider the puzzle-element labels in the screenshots. From the ideas you got from the transcript, create two coherent summaries that are most closely associated with those puzzle elements. Summarize each in two very short sentences and return them as a numbered list separated by a line break. Provide no additional commentary or analysis.

\subsection{Prompt to generate proactive peer thoughts}
\label{peer_thoughts}
Based on the two screens with elements to solve a puzzle, come up with short and succinct ideas (6) to brainstorm possible solutions. Show how elements from the two screens are connected.

\subsection{Prompt to contextualize the peer thoughts based on ongoing team discussion}
\label{peer_context}

Provide a succinct contextualized version of this thought. Structure the response as one short sentence to contextualize based on what the transcript summary says about the puzzle element mentioned in the thought; if it wasn't discussed, say so. Then share the thought without any additional commentary. Always share ideas with uncertainty---not solutions. No fluff. Keep response to 2 short sentences.

\subsection{Prompt to generate peer response to user queries over chat}
\label{peer_chat}

You are Ava, a peer sharing ideas on the puzzle. You're looking at a two-screen puzzle and should respond to user queries based on them. The puzzle is split across both screens. Some more information about the interactions possible with the screens: … Provide the response as a succinct summary (2 lines) based on the query details.

\section{Appendix: Group Interview Guide}
\label{Interview_questions}
Thank you for participating in the study. We will now move on the the group interview. We would like for you as a group to discuss the different AI features, and how it impacted your performance as a team and the team processes, such as, communication, coordination, and planning. We will go over each feature, and there are no right or wrong answers. So please share your experience freely, which will help us think about the next steps on how to design AI support for collaborative problem-solving tasks. 

\textit{Questions for facilitator agent:}
\begin{enumerate}
    \item Can you describe your team's performance in this puzzle? How did the facilitator features, like suggesting workflows and providing summaries, impact your performance and mental workload? 
    \item Can you describe your team collaboration, including communication, planning, and coordination, while solving this puzzle with the facilitator AI?
    \item How did you incorporate these features in your teamwork? Were the features helpful?
\end{enumerate}

\textit{Questions for peer agent:}
\begin{enumerate}
    \item Can you describe your team's performance in this puzzle? How did the peer AI features like proactive thoughts and the chat impact your performance and mental workload?
    \item Can you describe your team collaboration, including communication, planning, and coordination while solving this puzzle with the peer AI?
    \item How did you incorporate these features in your teamwork? Were the features helpful?
\end{enumerate}

\textit{Questions for No AI:}
\begin{enumerate}
\item Can you describe your team's performance on this puzzle? How mentally demanding was the task?
\item Can you describe how your team communicated during this session?
\item Can you also talk about the planning and coordination aspects?
\end{enumerate}

\section{Appendix: Survey Measures}
\label{Surveys}

\subsection{AI Perception Survey}
Participants rated their perceptions of the AI agent using the 5-item AI Perception Scale \cite{Bendell2025}, on a 5-point Likert scale (1 = Strongly Disagree, 5 = Strongly Agree). Using our data, we obtained $\alpha$ = .83. Items included: 

\begin{enumerate}
    \item The AI agent's recommendation improved our team score
    \item The AI agent's recommendations improved our team coordination
    \item I felt comfortable depending on the AI agent 
    \item I understand why the AI agent made its recommendations
    \item I think the AI agent is trustworthy
\end{enumerate}

\subsection{Perceived Coordination Scale}
The Perceived Coordination Scale was adapted from Tesluk and Mathieu \cite{tesluk1999overcoming} and demonstrated high internal consistency in our sample ($\alpha$ = .87). Items were rated on a 5-point scale (1 = Strongly Disagree, 5 = Strongly Agree) and included:
\begin{enumerate}
    \item People on my team helped each other out when needed
    \item We all cooperated to get the work done 
    \item On my team, people shared their knowledge with each other 
    \item On the whole, the members of this team all did their fair share of the work
    \item My team coordinated activities to make this run smoothly
\end{enumerate}

\subsection{NASA-TLX}
\begin{enumerate}
    \item How mentally demanding was the task?
    \item How physically demanding was the task?
    \item How hurried or rushed was the pace of the task?
    \item How successful were you in accomplishing what you were asked to do? 
    \item How hard did you have to work to accomplish your level of performance?
    \item How insecure, discouraged, irritated, stressed, and annoyed were you? 
\end{enumerate}

\section{Appendix: Quantitative Results}
This appendix presents the complete quantitative results for all outcome measures analyzed in the study. The tables report ART ANOVA statistics, estimated marginal means (EMMs), and post-hoc contrasts for task performance, perceived coordination, workload, and AI perception measures.
\label{quants}
\begin{table}[ht]
\centering
\caption{ART ANOVA results, estimated marginal means (EMMs), and post-hoc contrasts for puzzle performance, perceived coordination, and NASA--TLX workload.}
\label{tab:art_results_primary}
\resizebox{\columnwidth}{!}{%
\begin{tabular}{lcccc}
\toprule
\textbf{Outcome / Effect} & \textbf{$F$} & \textbf{$Df$} & \textbf{$Df(res)$} & \textbf{$p$} \\
\midrule

\multicolumn{5}{l}{\textbf{Puzzle Performance (Score)}} \\
Condition & 13.33 & 2 & 6.00 & \textbf{.006}** \\
Puzzle & 16.66 & 2 & 6.00 & \textbf{.004}** \\
Condition $\times$ Puzzle & 3.78 & 4 & 4.79 & .093 \\[4pt]

\multicolumn{5}{l}{\textit{EMMs (Condition): NoAI = 9.33,\; Peer = 4.67,\; Facilitator = 14.50}} \\[2pt]
\multicolumn{5}{l}{\textit{Posthoc Pairwise Comparisons (Holm-adjusted, Condition):}} \\
\multicolumn{5}{l}{NoAI -- Peer: $\beta$ = 4.67,\; SE = 1.91,\; $t(6)=2.45$,\; \textbf{$p=.070$},\; $d = 1.41$} \\
\multicolumn{5}{l}{NoAI -- Facilitator: $\beta$ = --5.17,\; SE = 1.91,\; $t(6)=-2.71$,\; \textbf{$p=.070$},\; $d = -1.57$} \\
\multicolumn{5}{l}{Peer -- Facilitator: $\beta$ = --9.83,\; SE = 1.91,\; $t(6)=-5.16$,\; \textbf{p=.006**},\; $d = -2.98$} \\[6pt]

\multicolumn{5}{l}{\textit{EMMs (Puzzle): P1 = 15.50,\; P2 = 6.50,\; P3 = 6.50}} \\
\multicolumn{5}{l}{\textit{Posthoc Pairwise Comparisons (Holm-adjusted, Puzzle):}} \\
\multicolumn{5}{l}{P1 -- P2: $\beta$ = 9,\; SE = 1.80,\; $t(6)=5.00$,\; \textbf{p=.007}**,\; $d = 2.89$} \\
\multicolumn{5}{l}{P1 -- P3: $\beta$ = 9,\; SE = 1.80,\; $t(6)=5.00$,\; \textbf{p=.007}**,\; $d = 2.89$} \\
\multicolumn{5}{l}{P2 -- P3: $\beta$ = 0,\; SE = 1.80,\; $t(6)=0.00$,\; $p=1.000$,\; $d = 0.00$} \\

\midrule
\multicolumn{5}{l}{\textbf{Perceived Team Coordination}} \\
Condition & 0.22 & 2 & 6.00 & .812 \\
Puzzle & 1.01 & 2 & 6.00 & .418 \\
Condition $\times$ Puzzle & 1.08 & 4 & 4.79 & .459 \\[4pt]
\multicolumn{5}{l}{EMMs (Condition): NoAI = 9.17,\; Peer = 9.17,\; Facilitator = 10.17} \\
\multicolumn{5}{l}{EMMs (Puzzle): P1 = 10.83,\; P2 = 9.00,\; P3 = 8.67} \\

\midrule
\multicolumn{5}{l}{\textbf{Workload (NASA--TLX)}} \\
Condition & 5.75 & 2 & 60.00 & \textbf{.005}** \\
Puzzle & 16.78 & 2 & 60.00 & \textbf{<.001}*** \\
Condition $\times$ Puzzle & 0.57 & 4 & 7.24 & .696 \\ [4pt]

\multicolumn{5}{l}{\textit{EMMs (Condition): Facilitator = 30.4,\; NoAI = 31.6,\; Peer = 47.5}} \\
\multicolumn{5}{l}{\textit{Contrasts (Holm-adjusted):}} \\
\multicolumn{5}{l}{Facilitator -- NoAI: $\beta$ = --1.21,\; SE = 5.63,\; $t(60)= -0.22$,\; $p=.831$,\; $d=-0.06$} \\
\multicolumn{5}{l}{Facilitator -- Peer: $\beta$ = --17.10,\; SE = 5.63,\; $t(60)= -3.04$,\; $\textbf{p=.011*}$,\; $d=-0.88$} \\
\multicolumn{5}{l}{NoAI -- Peer: $\beta$ = --15.90,\; SE = 5.63,\; $t(60)= -2.82$,\; $\textbf{p=.013*}$,\; $d=-0.82$} \\[4pt]

\multicolumn{5}{l}{\textit{EMMs (Puzzle): P1 = 19.7,\; P2 = 44.1,\; P3 = 45.7}} \\
\multicolumn{5}{l}{\textit{Contrasts (Holm-adjusted):}} \\
\multicolumn{5}{l}{P1 -- P2: $\beta$ = --24.42,\; SE = 5.03,\; $t(60)= -4.86$,\; \textbf{p< .001}***,\; $d=-1.40$} \\
\multicolumn{5}{l}{P1 -- P3: $\beta$ = --25.96,\; SE = 5.03,\; $t(60)= -5.16$,\; \textbf{p< .001}***,\; $d=-1.49$} \\
\multicolumn{5}{l}{P2 -- P3: $\beta$ = --1.54,\; SE = 5.03,\; $t(60)= -0.31$,\; $p=.760$,\; $d=-0.09$} \\

\bottomrule
\end{tabular}
}
{\centering \textit{Note.} $^{***} p < .001$;\; $^{**} p < .01$;\; $^{*} p < .05$. \par}

\label{tab:art_perf}
\end{table}

\begin{table}[ht]
\centering
\caption{ART ANOVA results, estimated marginal means (EMMs), and post-hoc contrasts for AI perception measures (Peer and Facilitator conditions only).}
\label{tab:art_results_ai_perception}
\resizebox{\columnwidth}{!}{%
\begin{tabular}{lcccc}
\toprule
\textbf{AI Perception Measure / Effect} & \textbf{$F$} & \textbf{$Df$} & \textbf{$Df(res)$} & \textbf{$p$} \\
\midrule

\multicolumn{5}{l}{\textit{\textbf{``The AI agent’s recommendation improved our team score”}}} \\
Condition & 5.33 & 1 & 3.00 & .104 \\
Puzzle & 3.27 & 2 & 4.80 & .127 \\
Condition $\times$ Puzzle & 2.61 & 2 & 4.80 & .171 \\[4pt]

\multicolumn{5}{l}{EMMs (Condition): Peer = 8.83,\; Facilitator = 4.17} \\
\multicolumn{5}{l}{EMMs (Puzzle): P1 = 6.00,\; P2 = 3.75,\; P3 = 9.75} \\

\midrule
\multicolumn{5}{l}{\textit{\textbf{“The AI agent’s recommendations improved our team coordination”}}} \\
Condition & 122.50 & 1 & 3.00 & \textbf{.002}** \\
Puzzle & 8.87 & 2 & 5.69 & \textbf{.018}* \\
Condition $\times$ Puzzle & 1.20 & 2 & 4.50 & .383 \\[4pt]

\multicolumn{5}{l}{\textit{EMMs (Condition): Peer = 9.42,\; Facilitator = 3.58}} \\
\multicolumn{5}{l}{\textit{Posthoc (Condition):}} \\
\multicolumn{5}{l}{Peer -- Facilitator: est = 5.83,\; SE = 0.527,\; $t(3)=11.07$,\; \textbf{p=.002**},\; $d=6.39$} \\[4pt]

\multicolumn{5}{l}{\textit{EMMs (Puzzle): P1 = 7.75,\; P2 = 2.50,\; P3 = 9.25}} \\
\multicolumn{5}{l}{\textit{Posthoc (Puzzle; Holm-adjusted):}} \\
\multicolumn{5}{l}{P1 -- P2: $\beta$ = 5.25,\; SE = 1.68,\; $t(5.69)=3.12$,\; \textbf{p=.044*},\; $d=2.43$} \\
\multicolumn{5}{l}{P1 -- P3: $\beta$ = --1.50,\; SE = 1.68,\; $t(5.69)=-0.89$,\; $p=.409$,\; $d=-0.69$} \\
\multicolumn{5}{l}{P2 -- P3: $\beta$ = --6.75,\; SE = 1.68,\; $t(5.69)=-4.01$,\; \textbf{p=.024*},\; $d=-3.13$} \\

\midrule
\multicolumn{5}{l}{\textit{\textbf{``I felt comfortable depending on the AI agent”}}} \\
Condition & 0.71 & 1 & 3.00 & .462 \\
Puzzle & 0.94 & 2 & 4.80 & .452 \\
Condition $\times$ Puzzle & 2.35 & 2 & 4.80 & .194 \\[4pt]
\multicolumn{5}{l}{EMMs (Condition): Peer = 7.50,\; Facilitator = 5.50} \\
\multicolumn{5}{l}{EMMs (Puzzle): P1 = 7.25,\; P2 = 4.25,\; P3 = 8.00} \\

\midrule
\multicolumn{5}{l}{\textit{\textbf{``I understand why the AI agent made its recommendations”}}} \\
Condition & 0.31 & 1 & 3.00 & .619 \\
Puzzle & 1.32 & 2 & 5.56 & .339 \\
Condition $\times$ Puzzle & 0.67 & 2 & 4.80 & .552 \\[4pt]
\multicolumn{5}{l}{EMMs (Condition): Peer = 7.25,\; Facilitator = 5.75} \\
\multicolumn{5}{l}{EMMs (Puzzle): P1 = 8.88,\; P2 = 6.12,\; P3 = 4.50} \\

\midrule
\multicolumn{5}{l}{\textit{\textbf{``I think the AI agent is trustworthy”}}} \\
Condition & 0.06 & 1 & 3.00 & .827 \\
Puzzle & 0.63 & 2 & 4.80 & .570 \\
Condition $\times$ Puzzle & 1.27 & 2 & 4.80 & .360 \\[4pt]
\multicolumn{5}{l}{EMMs (Condition): Peer = 6.17,\; Facilitator = 6.83} \\
\multicolumn{5}{l}{EMMs (Puzzle): P1 = 5.75,\; P2 = 5.25,\; P3 = 8.50} \\

\bottomrule 
\end{tabular}
}
{\centering \textit{Note.} $^{***} p < .001$;\; $^{**} p < .01$;\; $^{*} p < .05$. \par}

\label{tab:art_perception}
\end{table}

\section{Appendix: Exploratory Factor Analysis}
\label{factor}
This appendix provides additional details from our Exploratory Factor Analysis (EFA), including standardized factor loadings, Bartlett’s Test of Sphericity, and the variance explained by each extracted factor.

\subsection{Model Fit Indices}
The root mean square of the residuals (RMSR) was 0.08. \\
The Tucker--Lewis Index (TLI) of factoring reliability was 0.983. \\
The RMSEA was 0.012 with a 90\% confidence interval of [0, 0.088].

\begin{table}[ht]
\centering
\caption{Standardized factor loadings from the three-factor EFA (ML extraction, oblimin rotation).}
\label{tab:efa-loadings}
\begin{tabular}{p{0.52\linewidth}ccc}
\toprule
\textbf{Item} & \textbf{ML1} & \textbf{ML2} & \textbf{ML3} \\
\midrule
People on my team helped each other out when needed & 0.90 & 0.03 & 0.05 \\
We all cooperated to get the work done & 0.81 & 0.06 & 0.07 \\
People shared their knowledge with each other & 0.52 & -0.17 & 0.18 \\
Members did their fair share of the work & 0.55 & -0.21 & 0.22 \\
My team coordinated activities to make this run smoothly & 0.84 & 0.01 & -0.15 \\
\midrule
AI agent's recommendation improved our team score & -0.02 & 0.89 & -0.03 \\
AI agent's recommendations improved our team coordination & 0.04 & 0.84 & 0.07 \\
I felt comfortable depending on the AI agent & 0.08 & 0.86 & 0.06 \\
I understand why the AI agent made its recommendations & -0.20 & 0.22 & -0.18 \\
I think the AI agent is trustworthy & -0.20 & 0.60 & -0.19 \\
\midrule
How mentally demanding was the task? & -0.14 & 0.01 & 0.60 \\
How physically demanding was the task? & -0.18 & -0.12 & 0.12 \\
How hurried or rushed was the pace of the task? & -0.17 & -0.04 & 0.47 \\
How successful were you in accomplishing what you were asked to do? & -0.23 & 0.21 & 0.44 \\
How hard did you have to work to accomplish your performance? & 0.17 & -0.01 & 0.77 \\
How insecure, discouraged, irritated, stressed, and annoyed were you? & -0.53 & -0.02 & 0.36 \\
\bottomrule
\end{tabular}
\end{table}

\begin{table}[ht]
\centering
\caption{Bartlett’s Test of Sphericity evaluates whether the correlation matrix is significantly different from the identity matrix, indicating sufficient inter-item correlations for EFA. A significant result supports the use of factor analysis for these data.}
\label{tab:bartlett}
\begin{tabular}{lc}
\toprule
\textbf{Statistic} & \textbf{Value} \\
\midrule
$\chi^2$ & 536.14 \\
df & 120 \\
$p$ & $< .001$ \\
\bottomrule
\end{tabular}
\end{table}

\begin{table}[ht]
\centering
\caption{Summary of factor extraction results from the maximum-likelihood EFA with oblimin rotation. ML1, ML2, and ML3 represent the extracted latent factors, corresponding respectively to (1) \textit{Perceived Coordination}, (2) \textit{AI Perception}, and (3) \textit{Workload}. The table reports the variance explained by each factor.}
\label{tab:fa-summary}
\begin{tabular}{lccc}
\toprule
 & \textbf{ML1} & \textbf{ML2} & \textbf{ML3} \\
\midrule
SS Loadings & 3.31 & 2.84 & 1.70 \\
Proportion Variance & 0.21 & 0.18 & 0.11 \\
Cumulative Variance & 0.21 & 0.38 & 0.49 \\
Proportion Explained & 0.42 & 0.36 & 0.22 \\
Cumulative Proportion & 0.42 & 0.78 & 1.00 \\
\bottomrule
\end{tabular}
\end{table}

\begin{table}[t]
\centering
\caption{Factor correlations for the three-factor EFA solution using oblimin rotation. ML1, ML2, and ML3 represent the extracted latent factors, corresponding respectively to (1) \textit{Perceived Coordination}, (2) \textit{AI Perception}, and (3) \textit{Workload}. }
\label{tab:fa-correlations}
\begin{tabular}{lccc}
\toprule
 & \textbf{ML1} & \textbf{ML2} & \textbf{ML3} \\
\midrule
ML1 & 1.00 & -0.19 & 0.11 \\
ML2 & -0.19 & 1.00 & -0.14 \\
ML3 & 0.11 & -0.14 & 1.00 \\
\bottomrule
\end{tabular}

\end{table}

\clearpage
\end{document}